\documentclass[traditabstract]{aa} 
\usepackage{graphicx}
\usepackage{natbib}
\usepackage{psfig}
\usepackage{txfonts}
\newcommand{\ergs}{\>{\rm erg}\,{\rm s}^{-1}}

\defcitealias{capetti2015b}{Paper~I}
\defcitealias{fossati1998}{F98}
\begin{document}

   \title{Testing the blazar sequence with the least luminous BL Lacs.}

   \subtitle{}

   \author{C. M. Raiteri\inst{1}
          \and
          A. Capetti\inst{1}
          }

   \institute{INAF-Osservatorio Astrofisico di Torino, via Osservatorio 20, I-10025 Pino Torinese, Italy}

   \date{}

  \abstract {In a previous paper, we proposed a new method to select low-power
    BL Lacs (LPBLs) based on  mid-infrared emission and flux contrast
    through the Ca II spectral break; that study led to the selection of a
    complete sample formed by 34 LPBLs with $0.05<z \le 0.15$ and radio
    luminosities spanning the range $\log L_{\rm r} = 39.2$--41.5 [$\rm erg \,
      s^{-1}$]. We now assemble the broadband spectral energy distributions
    (SEDs) of these sources to investigate their nature and compare them with
    brighter BL~Lacs. We find that the ratios between the X-ray and radio
    luminosities range from $\sim 20$ to $\sim 30000$ and that the synchrotron
    peak frequencies span a wide energy interval, from $\log \rm \, \nu_{\rm
      peak} \sim 13.5$ to $\sim 20$ [Hz]. This indicates a broad variety of SED
    shapes and a mixture of BL Lac flavors. Indeed, although the majority of our LPBLs are
    high-energy peaked BL Lacs (HBLs), we find that a quarter of them are low-energy peaked
    BL Lacs (LBLs), despite the fact that the sample is biased against the
    selection of LBLs. The analysis of the median LPBL SED confirms 
    disagreement with the blazar sequence at low radio luminosities. Furthermore, if we limit the sample to the LBLs subsample, we find that
    their median SED shape is essentially indistinguishable from that of the
    most luminous BL Lacs. 
    We conclude that the observed radio power is not the main
    driving parameter of the multiwavelength properties of BL Lacs.}

   \maketitle

\section{Introduction}

Blazars are active galactic nuclei (AGNs) showing extreme properties.  Their
flux variability at all wavelengths,  high and variable polarization, apparent
superluminal motion of the radio knots, and high brightness temperatures are all
signatures of relativistic motion in a plasma jet that is oriented at a small angle
with respect to the line of sight. The consequent Doppler beaming of the
nonthermal jet emission usually makes this  emission dominate over the thermal emission
from both the AGN nucleus (accretion disk, broad and narrow line regions,
dusty torus) and the host galaxy. Therefore, the study of blazar emission is
the most efficient tool to unveil the nature of extragalactic jets. 

Blazars include flat-spectrum radio quasars (FSRQs) and BL Lac objects. 
The classical distinction between them is based on the
strength of the emission lines, whose rest-frame equivalent width (EW) must be
smaller than 5 \AA\ to define a BL Lac \citep{sti91}.  Another quantity
used to identify blazars is the 
value of the flux density contrast across the Ca II spectral break \citep{sto91}, which
is reduced with respect to that of quiescent galaxies in case of a nonthermal contribution.

BL Lacs were further divided into radio-selected BL Lacs (RBL) and X-ray
selected BL Lacs (XBL) until \citet{pad95} proposed a more physical
classification into high-energy cutoff BL Lacs (HBL) and low-energy cutoff BL
Lacs (LBL), setting the dividing line between the two classes at $F_{\rm x}/F_{\rm r} \sim
200$, where $F_{\rm x}$ is the 0.3--3.5 keV flux and $F_{\rm r}= \nu F_\nu$ at 5 GHz.
In the usual $\log \nu \, F_\nu$ versus $\log \nu$ representation, the
blazar spectral energy distribution (SED) shows two main bumps corresponding
to the synchrotron (from radio to UV or X-rays) and inverse-Compton (from X to $\gamma$ rays) 
jet emission contributions. In the LBL SED, the synchrotron peak falls in the infrared--optical band, 
while it is in the UV--X-rays in HBL\footnote{An alternative definition names low-, intermediate-, and high-synchrotron peaked (LSP, ISP, and HSP) BL Lacs as those sources with the synchrotron peak frequency $<10^{14}$, 
between $10^{14}$ and $10^{15}$, and $>10^{15} \rm \, Hz$, respectively \citep{abdo2010}.}. 
The operation of new TeV Telescopes, such as the
{\em Major Atmospheric Gamma Imaging Cherenkov Telescopes} ({\em MAGIC}), the
{\em High Energy Stereoscopic System} ({\em H.E.S.S.}) and the
{\em Very Energetic Radiation Imaging Telescope Array System} ({\em VERITAS}),
and the planning of more
sophisticated TeV instruments, such as the Cherenkov Telescope Array
\citep[CTA;][]{acharya2013} has  called attention to the extremely high-energy peaked BL Lacs (EHBL).  Following \citet{bonnoli2015}, they are
characterized by $F_{\rm x}/F_{\rm r} \ga 10^{4}$, where $F_{\rm x}$ is the X-ray flux in the
0.1-2.4 keV band, and $F_{\rm r}= \nu F_\nu$ at 1.4 GHz.

Fossati et al.\ (1998; hereafter F98) proposed a unifying view of the blazar SED behavior. These authors suggested that when the radio power
decreases the synchrotron peak frequency shifts to higher energies and the
Compton dominance, i.e.,\ the ratio between the inverse-Compton and
synchrotron peak luminosities, decreases.  In this way, FSRQs, LBLs, and HBLs form
a continuous spectral sequence, which is the so-called ``blazar sequence''.  This
picture was subsequently confirmed by \citet{donato2001}, who also included hard X-ray data in the
analysis. 
However, some authors have criticized this view.
On one side, the existence of high-power HBL has been claimed \citep[see, e.g.,][]{padovani2003,gio12a}.
On the other side, LBL of low power have been found in deeper radio-selected samples \citep{padovani2003,caccianiga2004,anton2005}.
These authors argued that the blazar sequence of \citetalias{fossati1998} may be the result of selection biases because the high-power sources were mostly radio-selected objects, while the low-power sources were mainly X-ray selected objects \citep[see also][]{landt2008}. 
Moreover, \citet{nieppola2008} suggested that the anticorrelation between synchrotron peak frequency and luminosity disappears when correcting for the relativistic Doppler boosting effect.
The blazar sequence has then been revisited by \citet{ghisellini2008}, who substituted the dependence of the jet emission properties on the jet power alone with a dependence on two parameters, namely the black hole mass and accretion rate.
The accretion rate has also been invoked by \citet{meyer2011} as a second parameter needed to define the SED features.

   \begin{figure}
   \resizebox{\hsize}{!}{\includegraphics{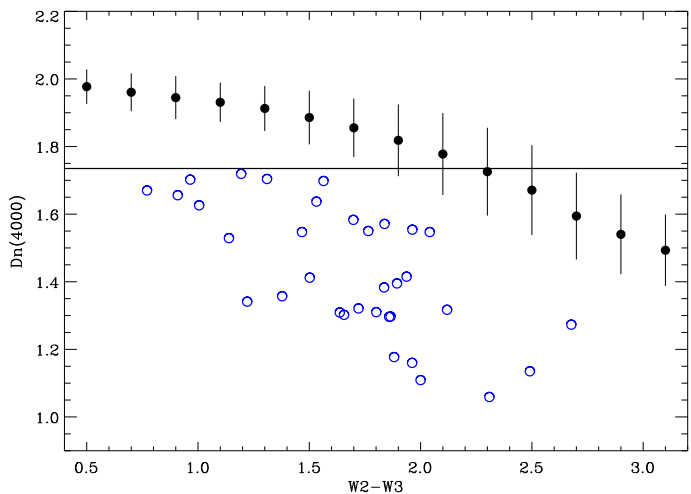}}
\caption{The LPBLs identified in \citetalias{capetti2015b} (blue circles). They lie in the ``forbidden zone" of the Dn(4000) versus the W2-W3 diagram, which is defined by the Dn(4000) limit obtained by requiring that the nonthermal jet contribution is at least 1/3 of the host galaxy one at 3900 \AA, and by the 1-sigma lower limit to the average relation holding for the 21065 bright nearby galaxies (black dots; see \citetalias{capetti2015b} for details).
}
\label{uvmir}
\end{figure}

The least luminous blazars are extremely interesting because they enable us to
explore the behavior of the blazars properties at very low radio power and they probe the jet formation and emission processes at the lowest levels of accretion.
These sources allow us to test the low-power end of the blazar sequence, which predicts that
they are all HBLs. 
A major difficulty in finding low-luminosity BL Lacs comes from the dilution of
the jet emission by the host starlight in the optical and near-IR
bands. In \citet[][hereafter Paper I]{capetti2015b}, we proposed a new method to
select low-power BL Lacs (LPBLs) based on a new diagnostic plane that includes
the W2$-$W3 color from the {\em Wide-Field Infrared Survey Explorer} ({\em
  WISE}) all-sky survey (AllWISE) and the Dn(4000) index derived from the Digital
Sky Survey (SDSS). We isolated 36 LPBLs candidates up to redshift 0.15. Their
radio luminosity at 1.4 GHz goes from $\log L_{\rm r} = 39.2$ to 41.5
[$\rm erg \, s^{-1}$]. By carefully considering the completeness of our
sample, we analyzed the BL Lac radio luminosity function (RLF), finding a
dramatic paucity of LPBLs with respect to the extrapolation of the RLF toward
low power. We thus suggested that a sharp break is present in their RLF,
located at $\log L_{\rm r} \sim 41$ [$\rm erg \, s^{-1}$]. This implies that
the LPBLs we consider represent the very end of the luminosity function of
this class of objects and, therefore, they represent the ideal test bed for a
better understanding of the nature of relativistic jets at low power.

In this paper, we analyze  the LPBLs of \citetalias{capetti2015b} in detail to
explore their nature. 
In Sect.\ \ref{sample} and \ref{data}, we present the LPBLs sample and the collected
multiwavelength data. These are used in Sect.\ \ref{nature} to estimate their  
broadband spectral indices and to build their SEDs. 
The SED properties are analyzed in the context of the blazar sequence
in Sect.\ \ref{sequence}, while in Sect.\ \ref{beam} we discuss the effect of 
Doppler beaming. Our summary and conclusions are given in
Sect.\ \ref{summary}.

Throughout the paper we adopt a cosmology with $H_0=70 \, \rm km \, s^{-1} \,
Mpc^{-1}$, $\Omega_{\rm M}=0.29$, and $\Omega_\Lambda=0.71$.  When we speak of
luminosities $L_{\rm b}$ in a given band b (where b = r, i, o, x, $\gamma$ for
radio, infrared, optical, X-ray, and $\gamma$-ray, respectively) we mean
$L_{\rm b} = 4 \, \pi \, D^2 \, \nu \, F_\nu$, where $D$ is the luminosity
distance and $F_\nu$ the flux density at the frequency $\nu$, corrected for
the Galactic absorption, when necessary, and $k$-corrected. The parameter $F_\nu$ is assumed
to have a power-law dependence on frequency, $F_\nu \sim \nu^{-\alpha}$, where
$\alpha$ is the energy spectral index. This is linked to the photon spectral
index $\Gamma$ through $\Gamma=\alpha+1$.
We also use the term ``color'' to indicate a broadband spectral index.

\section{The LPBLs sample}
\label{sample}

In \citetalias{capetti2015b} we selected LPBLs according to a new
diagnostic plane that includes the W2$-$W3 color from AllWISE and the Dn(4000)
index\footnote{The Dn(4000) index represents the ratio between the flux densities on the red (4000--4100 \AA) and blue (3850--3950 \AA) side of the Ca~II spectral break \citep{balogh1999}.} derived from the SDSS. 
In this plane, LPBLs
populate the so-called
forbidden zone, a region that is scarcely populated by other sources; see Fig.\ \ref{uvmir}. 
The forbidden zone is circumscribed, on the one hand, by a limit on the Dn(4000) and is obtained by requiring that the ratio $f$ between the nonthermal jet flux to the host galaxy flux at 3900 \AA\ is at least 1/3, and, on the other hand, by the Dn(4000) versus W2$-$W3 relationship that holds for a sample of 21065 bright nearby galaxies. 
Starting from the radio-selected
sample by \citet{best2012} and filtering objects with small emission line EW
($\lesssim 5 $\AA\ in the rest frame), we isolated 36 LPBL candidates up to redshift 0.15. Their
radio luminosity at 1.4 GHz spans the range $\log L_{\rm r} = 39.2$--41.5
[$\rm erg \, s^{-1}$].

The name itself, forbidden zone, was conceived to mean that we are not expecting a significant number of contaminants there. 
Indeed, in \citetalias{capetti2015b}, we were able to identify just one class of contaminants, the E+A galaxies, which we discarded by looking at their $\rm H\delta$ absorption.
The further selection of BL Lac candidates based on small EW also excludes possible contaminants from most of the AGN classes.
A preliminary analysis of the selected LPBLs properties gave strength to the idea that they are genuine BL Lacs.
All but one\footnote{The only exception is ID=7186, with $\sigma = 139 \, \rm km
\, s^{-1}$ and $\log (M_{\rm BH} / M_\odot) = 7.5$.} LPBL are hosted in massive
elliptical galaxies with stellar velocity dispersion $\sigma > 160 \, \rm km
\, s^{-1}$, leading to black hole masses $\log (M_{\rm BH} / M_\odot) > 7.8$ \citep{tre02}, i.e.,\ above the threshold to produce a radio-loud AGN \citep{chiaberge2011}.
Moreover, all LPBLs have radio morphologies 
dominated by a compact core in the FIRST images, and in some cases they also show a one-sided jet or a halo.
All these properties are typical of BL Lacs. 
Furthermore, the values of their radio-optical spectral index overlap with those characterizing BL Lacs.

Among the 36 LPBLs identified in \citetalias{capetti2015b}, 35 objects are in the
redshift range $0.05 < z \le 0.15$ and, as discussed in that paper, form a
complete sample.  Among them, we realized that in the object with ID=9591 the
radio source is actually not associated with the optical galaxy. Indeed, the
offset between the radio and optical sources is 4$\farcs$8. In all other
objects the offset is significantly smaller with a distribution having a
median value of only 0$\farcs$3. We then remove it from the sample. Therefore, we
are left with 34 objects, which are listed in Table \ref{lumi}.

\section{Data selection}
\label{data}
In our search for multiwavelength data for our 34 LPBLs, we focused on those bands that are more suitable to study the jet emission and on those catalogs/archives that are able to provide the largest coverage of our sample.
The resulting broadband SEDs are shown in the Appendix and two examples are shown in Fig.\ \ref{sedexample}.
Data downloaded from the Asi Science Data
Center\footnote{\tt http://tools.asdc.asi.it/} (ASDC) are added ``in background" for completeness.
From inspection of the SEDs it is clear that
because of the host contamination, the spectral bands in which we 
can measure the jet emission are significantly less in LPBLs than
in brighter blazars. Indeed, the host represents the bulk of
the emission in the region spanning from the mid-infrared to the optical. As
described in more detail below, the method used for our BL Lac selection enables us
to isolate the jet emission in two bands 
located at the boundaries of this region.  
In the following, we discuss the data used to analyze the jet emission in the various bands.

   \begin{figure*}
    \vspace{0.5cm}
    \centerline{
    \psfig{figure=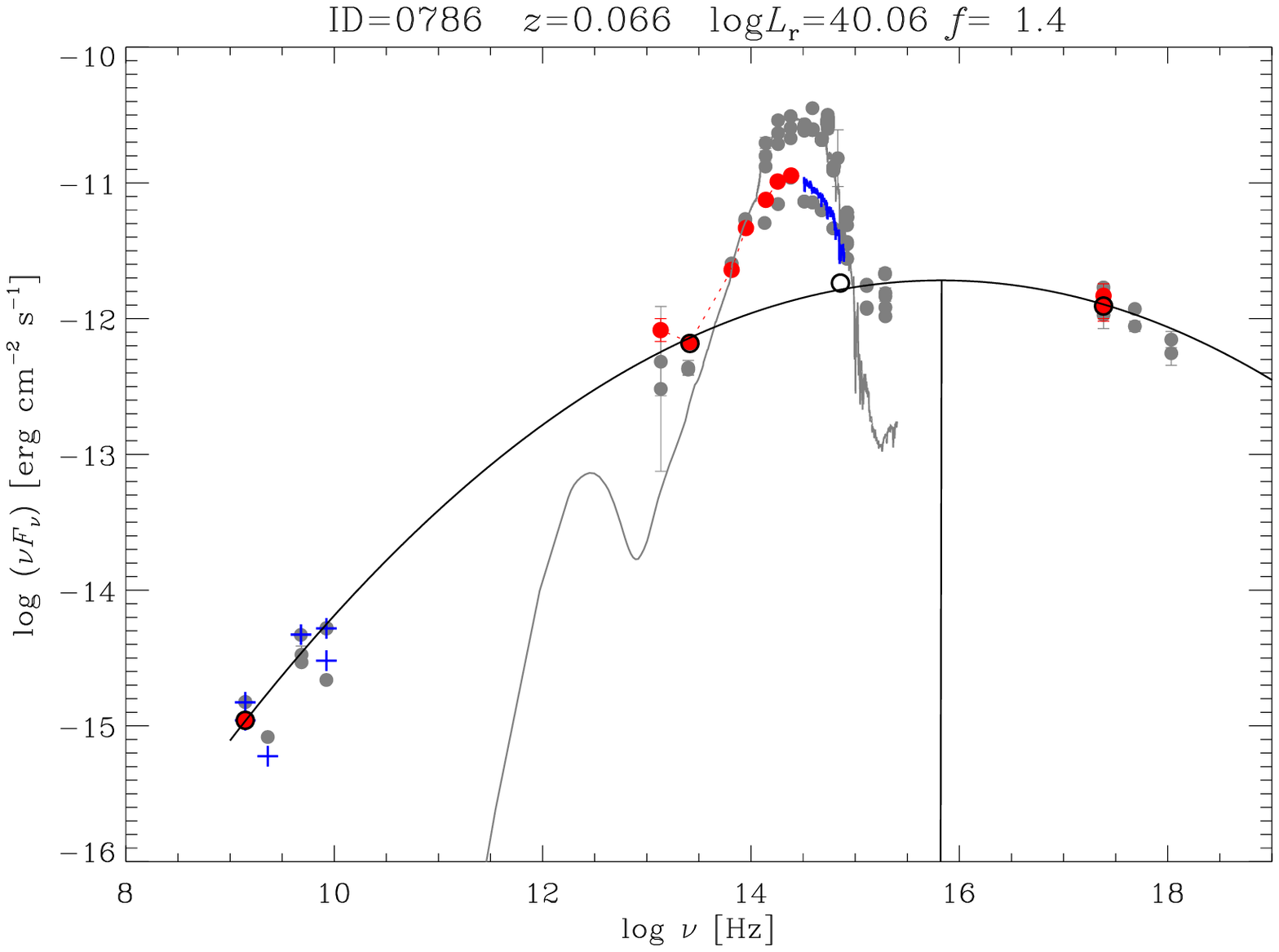,width=.50\linewidth}
    \psfig{figure=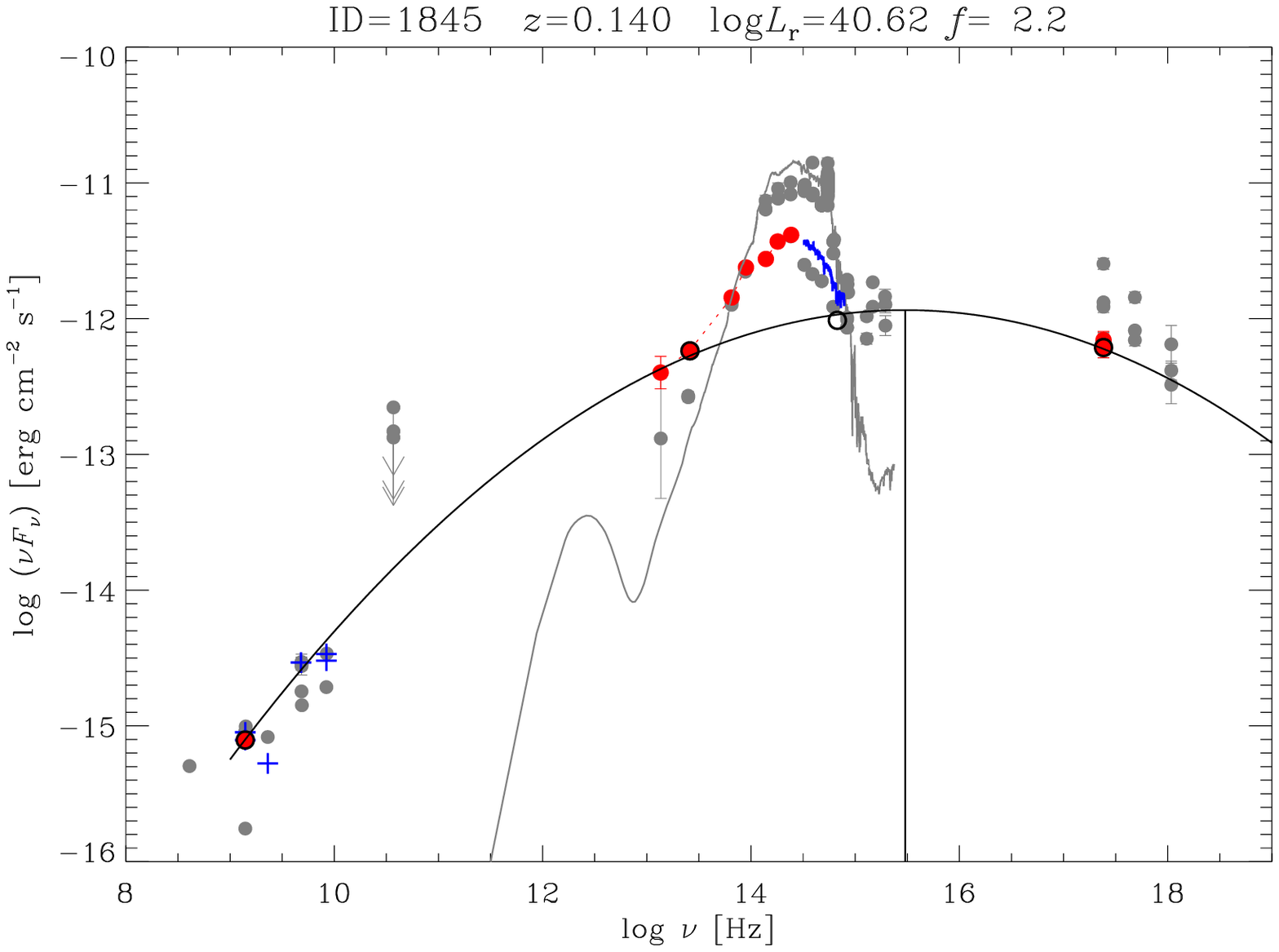,width=.50\linewidth}
    }
    \caption{Examples of the SED of two LPBLs, namely ID=786 and ID=1845. We show
      data analyzed by us (colored symbols) as well as data downloaded from the ASDC (grey dots); 
      see Fig.\ \ref{sed_z1a} for a full description.
      The near-IR, optical, and X-ray data have been corrected for Galactic absorption.
      We overplot the logarithmic parabolic fit through the data points at 1.4 GHz, 
      $\rm 12 \, \mu m$, rest-frame 3900 \AA, and 1 keV (black empty circles), where the infrared and optical fluxes have been corrected for the host galaxy contribution.
The grey line represents an elliptical galaxy template normalized to the W2 flux density.}
    \label{sedexample}
   \end{figure*}

\begin{table*}
\caption{The 34 BL Lac objects identified by \citet{capetti2015b} and their $k$-corrected luminosities in the various bands.}
\label{lumi}
\centering                   
\begin{tabular}{rrrccccrcc}
\hline\hline
ID & R.A.  & Dec.  & $z$ & $\log L_{\rm r}$ & $\log L_{\rm i}$ & $\log L_{\rm o}$ & $\log L_{\rm x}$ & $\log L_\gamma$ & $\Gamma_\gamma$\\
   & [deg] & [deg] &     & [$\ergs$]        & [$\ergs$]        & [$\ergs$]        &  [$\ergs$]       & [$\ergs$]       &                \\
\hline
   507 & 118.654457 &  39.17994 & 0.096 & 40.20 & 42.99 & 43.56 & 42.85    &       &            \\
   786 & 193.445877 &   3.44177 & 0.066 & 40.06 & 42.82 & 43.31 & 43.07    & 43.00 & 1.84 (0.10)\\
  1087 & 234.945602 &   3.47208 & 0.131 & 39.86 & 42.58 & 42.94 & 42.49    &       &            \\
  1103 & 164.657227 &  56.46977 & 0.143 & 41.23 & 44.26 & 44.58 & 43.95    & 44.42 & 1.95 (0.03)\\ 
  1315 &  14.083661 &$-$9.60826 & 0.103 & 40.58 & 43.39 & 43.68 & 43.89    & 43.22 & 1.79 (0.12)\\ 
  1768 & 201.445267 &   5.41502 & 0.135 & 40.05 & 42.74 & 43.06 & 42.90    &       &            \\ 
  1805 & 119.060333 &  27.40993 & 0.140 & 40.17 & 42.73 & 42.86 & $<$42.85 &       &            \\
  1845 & 163.433868 &  49.49889 & 0.140 & 40.62 & 43.44 & 43.77 & 43.43    & 43.53 & 1.80 (0.10)\\ 
  1903 & 121.805740 &  34.49784 & 0.139 & 40.36 & 43.07 & 43.07 & $<$42.89 &       &            \\ 
  1906 & 122.412010 &  34.92701 & 0.082 & 40.56 & 42.98 & 43.27 & 43.66    & 42.74 & 1.67 (0.13)\\
  2061 & 119.695808 &  27.08766 & 0.099 & 40.34 & 43.46 & 43.26 & $<$42.55 & 43.28 & 2.13 (0.12)\\
  2083 & 132.650848 &  34.92296 & 0.145 & 40.39 & 43.29 & 43.78 & 43.31    & 43.25 & 1.92 (0.20)\\ 
  2281 & 144.188004 &   5.15749 & 0.131 & 40.29 & 43.25 & 43.24 & $<$42.82 &       &            \\ 
  2341 & 151.793533 &  50.39902 & 0.133 & 40.30 & 42.54 & 42.92 & $<$42.59 &       &            \\
  2440 & 212.955994 &  52.81670 & 0.076 & 40.81 & 42.87 & 42.83 & 42.65    & 42.76 & 2.56 (0.23)\\ 
  3091 & 216.876160 &  54.15659 & 0.106 & 40.06 & 43.15 & 42.97 & 42.81    &       &            \\ 
  3403 & 217.135864 &  42.67252 & 0.129 & 40.43 & 43.38 & 44.03 & 44.91    & 43.39 & 1.57 (0.09)\\ 
  3591 & 172.926163 &  47.00241 & 0.126 & 40.87 & 42.88 & 43.14 & 42.94    &       &            \\ 
  3951 & 183.795746 &   7.53463 & 0.136 & 40.75 & 43.32 & 43.76 & 43.93    &       &            \\ 
  3958 & 185.383606 &   8.36228 & 0.132 & 40.98 & 42.72 & 43.29 & 43.45    &       &            \\ 
  4022 & 233.009293 &  30.27470 & 0.065 & 39.95 & 42.77 & 43.02 & 43.38    & 42.63 & 1.77 (0.13)\\
  4156 & 196.580185 &  11.22771 & 0.086 & 40.48 & 42.75 & 42.73 & $<$42.32 &       &            \\ 
  4601 & 199.005905 &   8.58712 & 0.051 & 39.19 & 42.28 & 42.37 & $<$41.86 &       &            \\ 
  4756 & 229.690521 &   6.23225 & 0.102 & 40.89 & 43.13 & 43.60 & 44.00    &       &            \\
  5076 & 180.764603 &  60.52198 & 0.065 & 40.39 & 43.64 & 43.88 & 42.68    & 43.16 & 2.21 (0.08)\\
  5997 & 127.270111 &  17.90440 & 0.089 & 40.81 & 42.95 & 43.22 & 43.30    &       &            \\
  6152 & 162.411652 &  27.70362 & 0.144 & 40.08 & 42.67 & 43.14 & 42.95    &       &            \\ 
  6428 & 169.276047 &  20.23538 & 0.138 & 40.92 & 43.66 & 44.20 & 45.01    & 44.03 & 1.87 (0.05)\\ 
  6943 & 194.383087 &  24.21118 & 0.140 & 39.89 & 42.78 & 43.57 & 44.34    &       &            \\ 
  6982 & 212.616913 &  14.64450 & 0.144 & 41.53 & 43.41 & 43.57 & 42.79    &       &            \\ 
  7186 & 211.548355 &  22.31630 & 0.128 & 39.70 & 42.53 & 42.73 & 42.31    &       &            \\
  7223 & 223.784271 &  19.33760 & 0.115 & 39.73 & 42.80 & 42.91 & 43.46    &       &            \\ 
  9640 & 227.671326 &  33.58465 & 0.114 & 39.33 & 42.92 & 42.88 & 43.68    &       &            \\
 10537 & 233.696716 &  37.26515 & 0.143 & 40.23 & 43.19 & 43.90 & 42.87    & 43.54 & 2.11 (0.12)\\
\hline\hline
\end{tabular}

\medskip
Column description: (1) ID number; (2 and 3) coordinates in degrees; (4) redshift, $k$-corrected luminosity at (5) 5 GHz; (6) 12 $\mu$m; (7) 3900 \AA; (8) 1 keV,
(9) 1 Gev; 3FGL $\gamma$-ray spectral index (with error). 
\end{table*}

\subsection{Radio}
Our 34 LPBLs were selected from the sample of 18286 radio sources built by
\citet{best2012}. We use
radio data from the Faint Images of the Radio
Sky at Twenty-cm (FIRST) survey. Therefore, our reference radio frequency is 1.4
GHz. 

In general, there are many data available in the radio band. 
In particular, for 22 LPBLs the Green Bank survey \citep{gregory1991} provides information at 4.8 GHz.
The average radio spectral index between 1.4 and 4.8 GHz is 0.02 with a standard deviation of 0.07. This confirms that we are dealing with objects with flat radio spectra, as expected for BL Lacs.
The scatter of radio data for the same source is likely due to the jet emission variability and to different spatial resolution. Only a few objects have millimeter observations and these are not further
considered.
   
\subsection{Infrared}
In the infrared we use the all-sky survey made by the {\em WISE} satellite \citep{wri10},
which provides magnitudes in the four filters W1 ($\rm 3.4 \mu m$), W2 ($\rm
4.6 \mu m$), W3 ($\rm 12 \mu m$), and W4 ($\rm 22 \mu m$). Cross-match with the AllWISE catalog\footnote{\tt
  http://wise2.ipac.caltech.edu/docs/release/allwise/} was performed with a 3
arcsec radius \citepalias[see][]{capetti2015b}.

We choose $\rm 12 \mu m$ as the reference wavelength in the infrared to
reduce the strong contamination by the host galaxy light and, at the same time,
have accurate data. We calculate the host contribution at the W3 wavelength by
normalizing the SWIRE template of a 13 Gyr  old elliptical galaxy\footnote{\tt
  http://www.iasf-milano.inaf.it/$\sim$polletta/templates/
  swire\_templates.html} \citep{pol07} to the W2 value (see Fig.\ \ref{sedexample}). This contribution is
then subtracted to the AllWISE W3 flux density to get the jet emission at that
wavelength. This correction is likely overestimated, since it assumes that the
whole emission in W2 is from the host galaxy; however, it has a negligible
effect on the W3 value because of the steep fall of the galaxy emission going
toward the far infrared. Indeed, the host contribution in W3 is only about 1\% of the jet contribution.

\subsection{Optical}
We calculate the jet flux density $F_{\rm o}$ at 3900 \AA\ rest frame from
the Dn(4000) values derived from the SDSS spectra by the group from the Max
Planck Institute for Astrophysics and Johns Hopkins
University\footnote{\tt http://www.mpa-garching.mpg.de/SDSS/}
\citep{bri04,tre04},
$$F_{\rm o}={\rm {D^0_n - Dn(4000)} \over {D^0_n-1}} F_{\rm SDSS}$$
where $F_{\rm SDSS}$ is the SDSS flux density at 3900
\AA\ (reddening-corrected), and ${\rm D^0_n}$ is the Dn(4000) value for
quiescent galaxies, which slightly changes with redshift
\citepalias[see][]{capetti2015b}.

\subsection{X-rays}

Most of our sources have X-ray flux measurements.
In principle, these could also be affected by the emission from the host hot
coronae. However, the LPBL X-ray luminosity usually exceeds those measured from the
galactic components, which are generally smaller than $10^{42}$ $\ergs$ \citep{fabbiano1992}.

\begin{figure}
   \resizebox{\hsize}{!}{\includegraphics{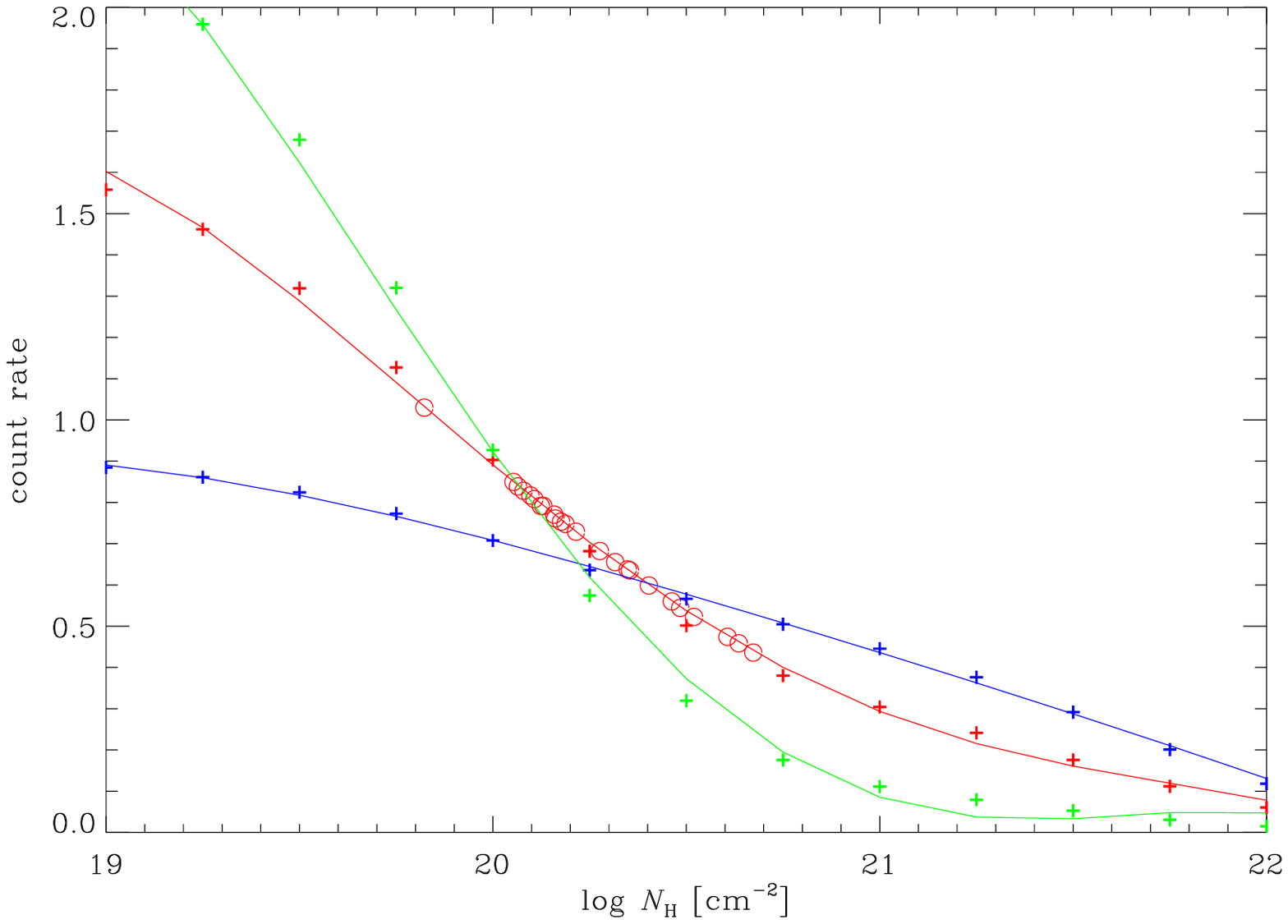}}
      \caption{The {\em ROSAT} count rate corresponding to an unabsorbed flux
        of $10^{-11} \rm \, erg \, cm^{-2} \, s^{-1}$ versus the Galactic
        hydrogen column density $N_{\rm H}$ for the three values of the X-ray
        photon spectral index $\Gamma_{\rm x}=1$ (blue), 2 (red), and 3 (green). Plus
        signs indicate the values derived with WebPIMMS, and solid curves
        represent polynomial fits through them. Open circles correspond to the
        $N_{\rm H}$ values of our LPBLs.}
         \label{nh_rass}
   \end{figure}

Flux densities at 1 keV were derived from the {\em ROSAT} All Sky
Survey (RASS), which provides the source count rate in the 0.1--2.4 keV energy
range. Counts-to-flux conversion factors are given for three possible photon
spectral indices $\Gamma_{\rm x}=1,2,3$.  Adopting a search radius of 30
arcsec, we found seven sources in the RASS Faint Source Catalog\footnote{\tt
  http://www.xray.mpe.mpg.de/rosat/survey/rass-fsc/} and 17 in the Bright
Source catalog\footnote{\tt
  http://www.xray.mpe.mpg.de/rosat/survey/rass-bsc/\\main/cat.html}
\citep{voges1999}. Since in general we do not know the spectral slope of each
source, we consider all the three possibilities and correct for Galactic
absorption using
WebPIMMS\footnote{https://heasarc.gsfc.nasa.gov/cgi-bin/Tools/w3pimms/w3pimms.pl}. Figure
\ref{nh_rass} shows the {\em ROSAT} count rates corresponding to an un-absorbed
flux of $10^{-11} \rm \, erg \, cm^{-2} \, s^{-1}$ as a function of $N_{\rm
  H}$ for the three spectral indices. The spread of the count rate due to
different slopes is minimum for $\log N_{\rm H} = 20.2$--20.3 [$\rm cm^{-2}$],
and grows for both increasing and decreasing Galactic absorption.  We fit the
three relationships with fourth-order polynomial curves and apply them to find the
de-absorbed flux corresponding to our LPBLs count rates. 

For the following
analysis, we use the 1 keV de-absorbed flux densities obtained with $\Gamma_{\rm
  x}=2$. However, given the distribution of column densities for our LPBLs, since all but one (ID=1103) of them is $20.0 \lesssim \log N_{\rm H} \lesssim 20.7$ [$\rm cm^{-2}$], the flux
measurements obtained by adopting $\Gamma_{\rm x}=1$ or 3 differ by at
most 40\% from the adopted values.

For the ten sources lacking RASS data, we searched for observations with other
X-ray satellites.  We derive 1 keV flux densities of ID=3591 from the seven-year {\em Swift}-XRT point source catalog \citep[1SWXRT;][]{delia2013}.  In
the case of ID=7186, we use the 3XMM-DR5 catalog\footnote{\tt
  http://xmmssc.irap.omp.eu/Catalogue/3XMM-DR5/\\3XMM\_DR5.html}, containing
source detections drawn from 13 years of {\em XMM-Newton} observations
\citep{rosen2015}.   We use {\em Chandra} data reported by the
ASDC for ID=1087.

Taking into account that the faintest LPBL in the RASS has a count rate of 0.025
counts/s, we assume an upper limit of 0.02 counts/s for the remaining seven sources without X-ray detection. This corresponds to a flux of $1.12 \, 10^{-13} \, \rm erg \, cm^{-2} \, s^{-1}$ for a
$\Gamma_{\rm x}=2$ spectral index, and implies 6--10 counts for typical
exposure times of 300--500 s. The actual flux upper limits are obtained by
correcting the limit to the count rate for Galactic absorption toward each
individual source.

\subsection{$\gamma$-rays}
The {\em Fermi} Large Area Telescope Third Source Catalog
\citep[3FGL;][]{acero2015} includes the first four years of data from the mission.  We found 13 {\em Fermi} counterparts to our objects, up to
a maximum distance of $\sim 400$ arcsec.  In the catalog, the spectra of all these objects
have been fitted with a power law, which is plotted in the SEDs reported in the Appendix with its uncertainties on both the flux and spectral index (giving rise to a ``butterfly"). 
According to the 3FGL, only one object, ID=3403,
has a possible association with a TeV source.

\bigskip

In summary, to analyze the jet emission in LPBLs, we can rely on
data in the radio, mid-infrared, optical, X rays (for most of them), and $\gamma$ rays (for some of them).

Even if all objects span a small range of redshift, we need to perform
$k$-correction to derive rest-frame luminosities (see Table 1) and colors. When
$k$-correcting (or transforming flux densities from one frequency to another
inside the same band), we need spectral indices.  In the radio we assume
$\alpha_{\rm r}=0$, while at $\gamma$ rays we use the specific index of each source.
In more detail, to calculate the rest-frame 1 GeV luminosity $L_\gamma$ reported in Table
\ref{lumi}, we transformed the flux density at the pivot energy (energy at
which the error is minimum) into flux density at 1 GeV and then $k$-corrected,
 in both cases, with the specific spectral index of each
source given in the 3FGL.
For the other bands we follow \citetalias{fossati1998}.  Since we do not know a
priori what type of BL Lac objects we are dealing with, we use average values
between those corresponding to the Slew and 1 Jy samples, namely,
$\alpha_{\rm i}=0.70$, $\alpha_{\rm o}=0.94$, $\alpha_{\rm x}=1.325$.
We stress that because of the limited range of redshift of the LPBLs sample, the
corrections are very small and the precise values adopted do not affect significantly
our analysis.

\section{The nature of LPBLs}
\label{nature}
One useful quantity to establish the flavor of BL Lacs is the ratio of the
X-ray to radio fluxes. As mentioned in the Introduction, according to
\citet{pad95} and \citet{bonnoli2015}, this ratio should be greater than 200
and 10000 to define an HBL and EHBL, respectively. Equivalently,
\citetalias{fossati1998} defined HBL as those objects with $\alpha_{\rm rx} \la 0.75$,
where the LBL has $\alpha_{\rm rx} \ga 0.75$.

We calculate broadband spectral indices for our LPBLs after dereddening and $k$-correction. To ease the comparison with previous results, radio fluxes were converted from 1.4 to 5 GHz and optical fluxes from 3900 to 5500 \AA.

In Table \ref{tab2} we list the ratio of the X-ray to the radio luminosity and
the radio-X-ray spectral index. Based on these parameters, which give the same indication,
seven objects are LBL,
19 are HBL, and four are EHBL.  For the remaining four objects, all without an X-ray
detection, the resulting lower limit on $\alpha_{\rm rx}$ does not lead to a secure
classification.

As mentioned in the Introduction, we are dealing with sources whose flux is variable at all wavelengths with
timescales ranging from minutes to months. We are using catalog data, which are not taken simultaneously, and this may 
cause some distortion of the SED. 
However, while the individual classifications must be taken with some caution, we 
do not expect that variability can alter the conclusions drawn from a sizeable sample of objects.

\begin{figure}
\resizebox{\hsize}{!}{\includegraphics{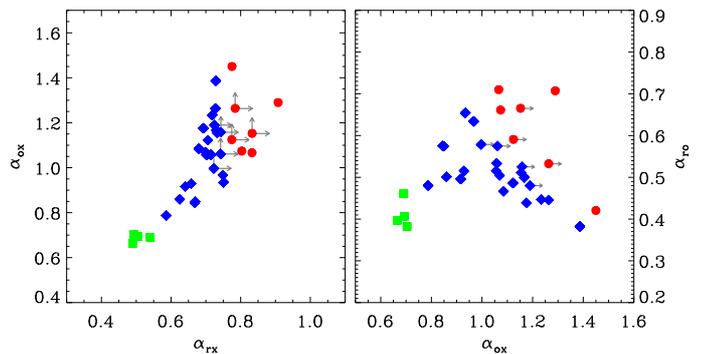}}
\caption{Broadband spectral indices for the objects in our LPBL sample, separating LBL (red circles) from HBL (blue diamonds) and EHBL (green squares)}
\label{donato}
\end{figure}

In Fig.\ \ref{donato} we show the optical-X-ray versus
radio-X-ray and the radio-optical versus optical-X-ray spectral indices for the three BL Lac classes. 
They are in good agreement with the results obtained by \citet{donato2001}, because the different BL Lac classes cover the same range of values. The positive dependence of
$\alpha_{\rm ox}$ on $\alpha_{\rm rx}$ is a signature of the SED curvature
that increases from EHBL toward LBL. This effect can be studied in a more
quantitative way by reproducing the SEDs with a log-parabolic model.

\begin{figure}
 \resizebox{\hsize}{!}{\includegraphics{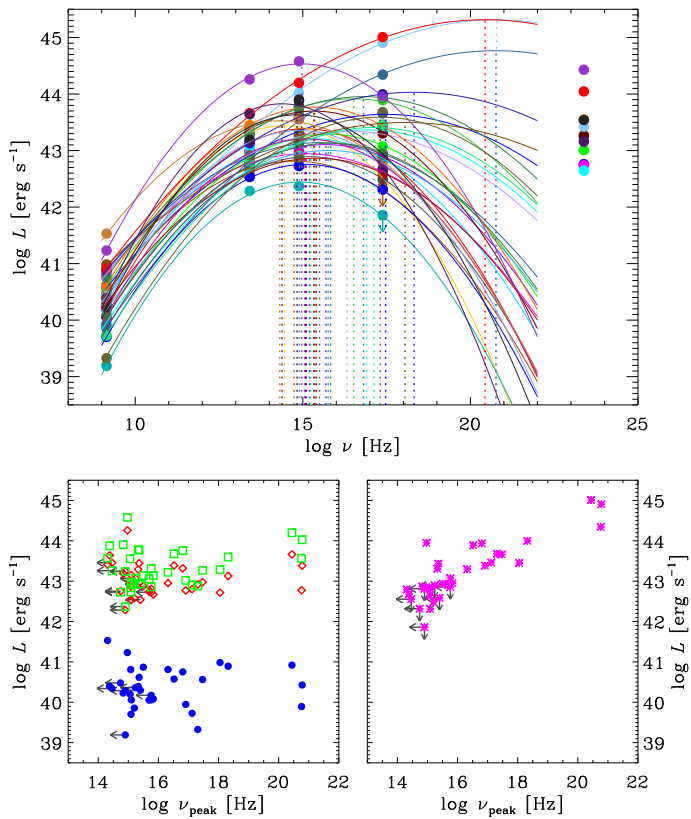}}
\caption{Upper panel: radio-to-$\gamma$-ray SEDs for the 34 objects analyzed in
  this paper (colored dots).  Solid lines represent parabolic fits to the
  radio--X-ray data. Vertical dotted lines identify the parabola apices, which
  give an indication of the synchrotron peak position. For the objects without
  X-ray detection, the peak location should be considered an upper limit.
  Lower panels: luminosities versus the synchrotron peak frequency. Blue dots,
  red diamonds, green squares, and pink asterisks refer to
  radio, infrared, optical, and X-ray bands, respectively. In all panels, upper
  limits due to the lack of detection in the X-ray band are indicated with arrows.}
\label{sedall}
\end{figure}

In Fig.\ \ref{sedall} we show the radio, infrared, optical, X-ray, and
$\gamma$-ray luminosities of our LPBLs (see Table \ref{lumi}).  The
radio--X-ray SEDs are fitted with a log-parabolic curve, whose apex roughly
represents the synchrotron peak.  Although in principle this should be
determined through a synchrotron emission model, the uncertainty derived from
the choice of the model parameters makes a simple parabolic fit equally valid for our purposes.
In the lower panel, luminosities in the various bands are plotted as a function
of the synchrotron peak frequency $\nu_{\rm peak}$.  There is no significant
trend of $\nu_{\rm  peak}$ with $L_{\rm r}$, $L_{\rm i}$, or $L_{\rm o}$. 
Conversely, $\nu_{\rm  peak}$ clearly increases with $L_{\rm x}$. 
This is the natural consequence
of the large spread in the X-ray luminosities of the various BL Lac classes. 

   \begin{figure}
   \resizebox{\hsize}{!}{\includegraphics{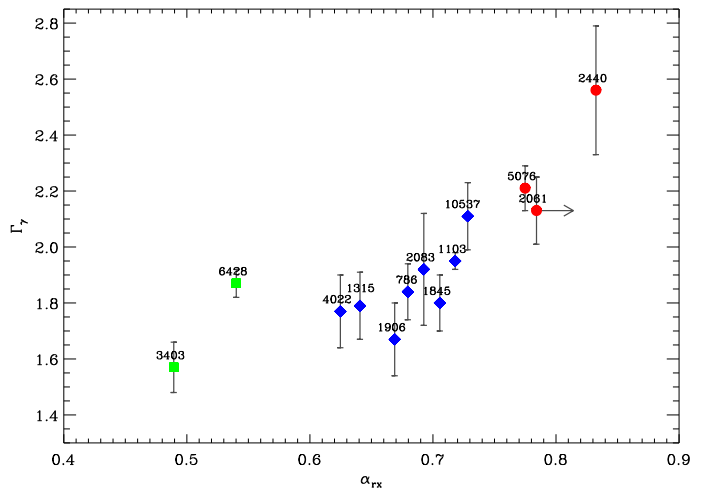}}
      \caption{ $\gamma$-ray photon spectral index $\Gamma_\gamma$ versus the radio-X-ray energy spectral index $\alpha_{\rm rx}$. Red dots represent LBL, blue diamonds HBL, and green squares EHBL. }
         \label{gaga}
   \end{figure}

Our estimate of the synchrotron peak frequency ranges from $\log \nu = 14.3$ to 20.8 (see Table \ref{tab2}),
i.e.,\ from the infrared to hard X-rays, confirming that the sample
includes both LBL as well as HBL and EHBL. For the sources not detected in
X-rays, the peak location must be considered as an upper limit. By adopting $\log \nu_{\rm peak} = 15$ as
a dividing value between LBL and HBL, overall we
recover a classification in concordance with that based on the X-rays/radio
comparison. There are only three exceptions (indicated with an asterisk in
Table \ref{tab2}).  For these objects we look for further data that can
constrain their nature. The X-ray spectrum of ID=3591 is soft (XRT data from
the ASDC) and this would define it as an HBL.   We
have $\gamma$-ray information from {\em Fermi} for the other two objects.  In Figure \ref{gaga} we plot
$\Gamma_\gamma$ versus $\alpha_{\rm rx}$, highlighting the various BL Lac
classes.  As expected, $\Gamma_\gamma$ increases with $\alpha_{\rm rx}$ and
all HBL have $\Gamma_\gamma<2$ but one, ID=10537, whose nature thus remains
unclear.  The LBL definition of ID=2440 is instead confirmed by its very soft
$\gamma$-ray spectrum.
Furthermore, for two LPBLs not classified with the X-rays/radio method,
the peak location suggests a definition as LBL (these objects are denoted
as ``LBL?'' in Table \ref{tab2}).

However, in LBLs the X-ray emission is generally dominated by the inverse-Compton
contribution.  Therefore, a fit to the synchrotron bump including the
X-ray emission overestimates $\nu_{\rm peak}$. We tested the
relevance of this effect by fitting the radio-to-optical SED of our
sources, and indeed we found that in some cases the three-point fit gives a  
$\nu_{\rm peak}$ that is significantly lower than the four-point fit, and as
low as $\log \nu_{\rm peak} \sim 13.5$ (see Fig.\ \ref{sedexample2})\footnote{
To make a comparison with the SEDs in the Appendix easier, here the SEDs are plotted in the observed $\log \nu \, F_\nu$ diagram. As a consequence, their $\nu_{\rm peak}$, even when corrected for redshift, may be slightly different from the values reported in Table \ref{tab2}.}. 
In particular, the $\nu_{\rm peak}$ shift is significant for most objects without an X-ray detection and, in general, means that the four-point fit overestimates the synchrotron contribution to the X-ray flux.

At the opposite end of the  $\nu_{\rm peak}$ distribution,  we
  obtained formal values as high as 10$^{20.8}$ Hz for the EHBLs, which are  not 
well constrained by the available measurements,however.
In two of such sources the SED shown in the Appendix suggests a peak location between 
$10^{18}$--$10^{19}$ Hz. 

BL~Lacs drawn from flux limited radio samples are known to
include mostly LBL \citep[e.g.,][]{padovani2003}. Conversely, the LPBL sample
considered here, although initially selected in the radio band,
is biased {\it against} the selection of LBL. Indeed, one of
the requirements is a significant dilution of the host emission by the
nonthermal emission across the Ca~II break; sources peaking at low
energies, well before the optical band, are less likely to be
selected. Indeed, as shown by Fig.\ 15 in \citetalias{capetti2015b}, the selection
function of LPBL has a cut-off at a luminosity that shifts toward a higher
power when $\alpha_{\rm ro}$ increases, thus disfavoring the detection of LBL
with respect to HBL.

In summary, our classification of LPBLs in their different flavors shows that
while the bulk of our LPBL sample is formed by HBLs (or EHBLs),
despite the negative bias in our sample about one quarter of them are LBLs.

According to \citetalias{fossati1998}, all the sources with $\log L_{\rm r} <
42$ at 5 GHz have $\log \nu_{\rm peak} > 15.5$. When correcting for the
difference in cosmology, the above luminosity limit translates into $\log
L_{\rm r} < 41.2$ at 1.4 GHz.  All our LPBLs but two fulfill this requirement,
but they are characterized by peak frequencies that, for the objects that lack an X-ray
detection, may reach $\log \nu_{\rm peak} \sim 13.5$.  We recall that
\citet{padovani2003} found six BL Lacs under the above luminosity limit with
peak frequencies extending down to $\sim 10^{13} \rm \, Hz$ and seven were found by
\citet{anton2005} down to $\sim 10^{14} \rm \, Hz$.  In both cases, the
minimum 1.4 GHz luminosity of the objects is $\log L_{\rm r} \sim 40.2$, so
our analysis confirms and extends  the results of these previous studies to lower luminosities.

\begin{table}
\caption{Classification of the LPBLs into the various BL~Lacs groups based on
  the ratio between X-ray to radio luminosities or, equivalently, to the
  radio-X-ray spectral index. We also report the synchrotron peak frequency obtained from a
  log-parabolic fit to the $k$-corrected radio, infrared, optical, and X-ray luminosities. 
Objects with discordant classification are indicated with an asterisk.
Notes on individual sources are given in the Appendix.}
\label{tab2}
\centering                   
\begin{tabular}{rrrrll}
\hline\hline
ID       & $L_{\rm x}/L_{\rm r}$ & $\alpha_{\rm rx}$ & $\nu_{\rm peak}$ & Class & Notes\\
\hline
     507 &      445   &    0.73 &    15.1 & HBL & Y \\
     786 &      1036  &    0.68 &    15.8 & HBL & Y \\
    1087 &      426   &    0.73 &    15.2 & HBL &   \\
    1103 &      524   &    0.72 &    15.0 & HBL & Y \\
    1315 &      2061  &    0.64 &    16.5 & HBL & Y \\
    1768 &      701   &    0.70 &    15.7 & HBL &   \\
    1805 &   $<$482   & $>$0.72 & $<$15.8 & ?   &   \\
    1845 &      654   &    0.71 &    15.4 & HBL & Y \\
    1903 &   $<$341   & $>$0.74 & $<$15.2 & ?   &   \\
    1906 &      1251  &    0.67 &    17.5 & HBL & Y \\
    2061 &   $<$163   & $>$0.78 & $<$14.5 & LBL & Y \\
    2083 &      823   &    0.69 &    15.3 & HBL &   \\
    2281 &   $<$336   & $>$0.74 & $<$14.9 & LBL? & Y \\
    2341 &   $<$194   & $>$0.77 & $<$15.4 & LBL  &   \\
    2440 &      69    &    0.83 &    15.1 & LBL* & Y \\
    3091 &      561   &    0.71 &    15.1 & HBL  & Y \\
    3403 &      29999 &    0.49 &    20.8 & EHBL & Y \\
    3591 &      117   &    0.80 &    15.5 & LBL* & Y \\
    3951 &      1520  &    0.66 &    16.8 & HBL  &   \\
    3958 &      294   &    0.75 &    18.1 & HBL  &   \\
    4022 &      2733  &    0.62 &    16.9 & HBL  & Y \\
    4156 &   $<$ 69   & $>$0.83 & $<$14.7 & LBL  & Y \\
    4601 &   $<$468   & $>$0.72 & $<$14.9 & LBL? & Y \\
    4756 &      1271  &    0.67 &    18.3 & HBL  & Y \\
    5076 &      192   &    0.77 &    14.4 & LBL  & Y \\
    5997 &      305   &    0.75 &    16.3 & HBL  &   \\
    6152 &      732   &    0.70 &    15.8 & HBL  &   \\
    6428 &      12207 &    0.54 &    20.4 & EHBL & Y \\
    6943 &      28108 &    0.49 &    20.8 & EHBL & Y \\
    6982 &      18    &    0.91 &    14.3 & LBL  &   \\
    7186 &      404   &    0.73 &    15.1 & HBL  & Y \\
    7223 &      5415  &    0.59 &    17.1 & HBL  &   \\
    9640 &      22528 &    0.51 &    17.3 & EHBL & Y \\
   10537 &      438   &    0.73 &    14.8 & HBL* & Y \\
\hline\hline
\end{tabular}
\end{table}

   \begin{figure*}
    \vspace{0.5cm}
    \centerline{
    \psfig{figure=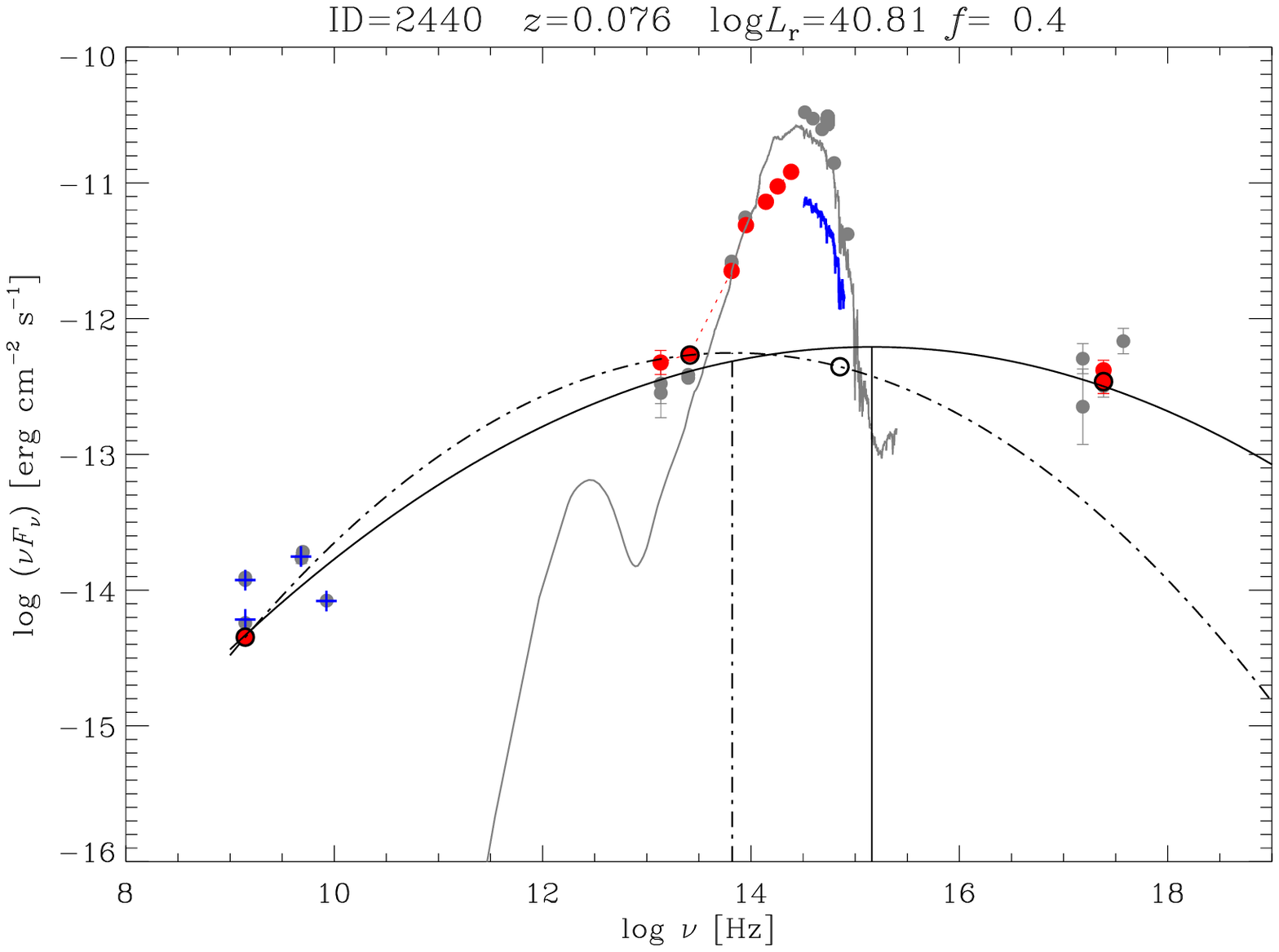,width=.50\linewidth}
    \psfig{figure=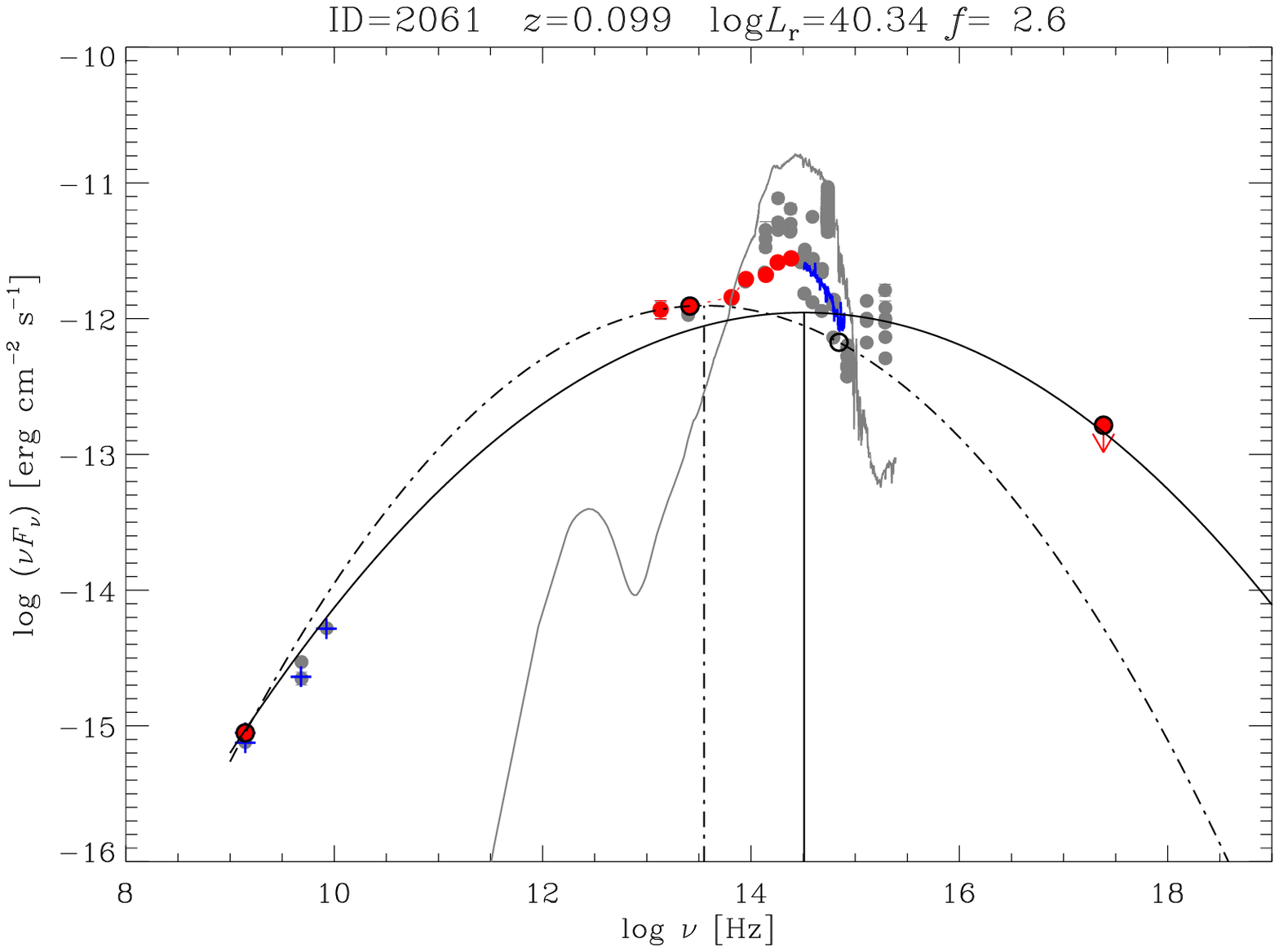,width=.50\linewidth}
    }
    \caption{Examples of the SED of two LPBLs, namely ID=2440 and ID=2061. The
      solid line represents a logarithmic parabolic fit, including the radio,
      infrared, optical, and X-ray data points, which does not match the
      infrared and optical data. These are instead well reproduced by a
      3-point fit excluding the X-ray datum (dot-dashed line). 
The case shown in the left panel 
      indicates that the X-ray emission is likely dominated by the
      inverse-Compton component; that in the right panel highlights how 
      the upper limit to the X-ray flux leads to a large overestimate 
      of the synchrotron peak frequency. The grey line represents an elliptical galaxy template normalized in W2 and the grey dots show data collected from the ASDC.}
    \label{sedexample2}
   \end{figure*}

\section{The blazar sequence}
\label{sequence}

\begin{figure}
\sidecaption
\resizebox{\hsize}{!}{\includegraphics{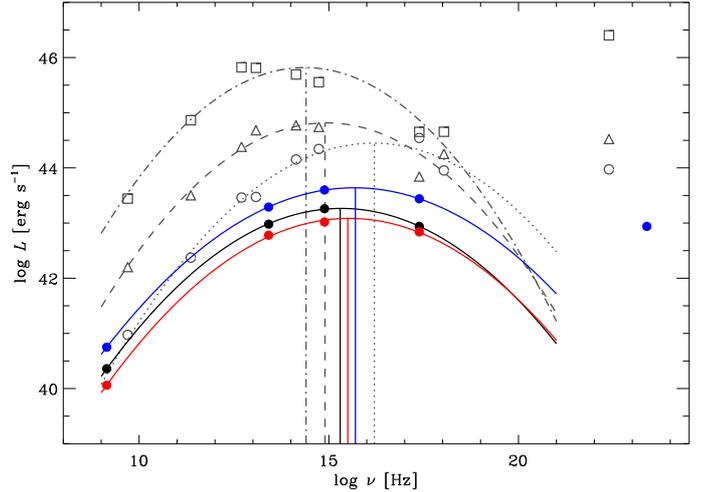}}\caption{Median SED for the 34 objects analyzed in this paper (black dots). The red and blue dots
  represent the SEDs obtained by splitting the sample into two groups of increasing radio
  luminosities (see text and Tab.\ \ref{tab3}). Gray symbols refer
  to the SEDs of the three faintest subsamples in the \citetalias{fossati1998} work,
  complemented by the hard X-ray data points from  \citet{donato2001}. The curves are
  log-parabolic fits to the radio-to-X-ray median luminosities with the
  peak indicated with vertical lines.
}
\label{mediane}
\end{figure}

As mentioned in the Introduction, \citetalias{fossati1998} identified a continuous
trend of the blazar SEDs, where BL Lacs of lower radio power exhibit higher
synchrotron and inverse-Compton peak frequencies and lower Compton dominance
with respect to the higher power objects. In their work, the faintest SED has
a median 5 GHz luminosity\footnote{We have lowered by a factor 1.85 and 1.76 the
  points of \citetalias{fossati1998} and \citet{donato2001}, respectively, to account for
  the differences in the choice of cosmological parameters.} of $\log L_{\rm r} = 40.97$, and according to their
model the peak synchrotron frequency is around $\log \nu= 16.5$ [Hz].
Subsequently, \citet{donato2001} added a hard X-ray point to the
\citetalias{fossati1998} SEDs and slightly modified the model, so that the peak
frequency of the faintest SED shifted to $\log \nu \sim 16$. If instead we fit
their radio-to-hard-X points with a simple log-parabolic curve, we get a peak
frequency of $\log \nu = 16.2$. 
The data points of the faintest three SED in these works are plotted in
Fig.\ \ref{mediane} and some of their properties are reported in Table \ref{tab3}. 

Following \citetalias{fossati1998} we consider the median luminosities in the
various bands of all 34 LPBLs, obtaining the values tabulated in Table
\ref{tab3} and plotted in Fig.\ \ref{mediane}. 
Building median SEDs has the advantage of reducing the effects of variability 
when nonsimultaneous data are used.

The presence of upper limits to the X-ray fluxes does not affect
the estimate of the median luminosity in this band because they are all
located within the faintest half of the distribution. For this reason, the
median is well defined.\footnote{This result does not apply to the least
  luminous LPBLs subsample discussed below. The presence of upper limits
  within the 50\% percentile leads to an uncertainty in the median of 0.2
  dex.}  The radio luminosity of the faintest median SED in \citetalias{fossati1998} and \citet{donato2001}
 is similar to that of our sample of LPBLs.  However, the two SEDs diverge at increasing frequencies,
reaching a difference of a factor $\sim 40$ at 1 keV.  Indeed, a log-parabolic
fit gives a peak frequency at $\log \nu = 15.3$ for the LPBLs, which is substantially lower than that
found by \citetalias{fossati1998} and Donato et al.  
This highlights that our sample is very different from that of the faintest objects analyzed by \citetalias{fossati1998} because it includes a much higher fraction of LBLs. 

We can further split our sample into groups of radio luminosity. We define
two subsamples of equal size. The median luminosities are
reported in Table \ref{tab3} and shown in Fig.\ \ref{mediane}.
The median $\gamma$-ray luminosity is only defined for the brighter sample.
There are no significant differences in both shape and peak frequency between these two SEDs, 
despite the difference in radio luminosity.
Moreover, the corresponding $\nu_{\rm peak}$ lies in between the peaks of the two least 
luminous groups considered by \citetalias{fossati1998}.
Apparently, the trend of increasing $\nu_{\rm peak}$ 
at decreasing $\log L_{\rm r}$ does not apply to the least luminous BL~Lacs.

\begin{figure}
\resizebox{\hsize}{!}{\includegraphics{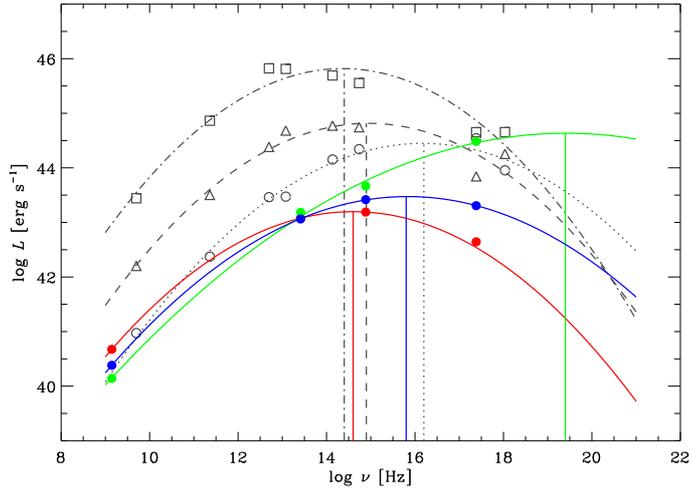}}
\caption{The same as Fig.\ \ref{mediane}, but now different colors refer to the three
flavors of BL Lac objects (red=LBL, blue=HBL, green=EHBL).
}
\label{classi}
\end{figure}

Because of the different flavor of BL Lacs in our sample, the above described median SEDs are the result of a mixture of sources with different properties.
Therefore, we calculate the median SEDs for LBL, HBL, and EHBL separately.
The result is shown in Fig.\ \ref{classi}, where the parabolic fits are made through the radio-to-X-ray data for the HBL and EHBL cases, while for the LBL we do not consider the X-ray point for the above discussed reasons. Aside from the obvious shift of $\nu_{\rm peak}$, there are two notable features. The first is the small difference in radio luminosity among the three classes, where LBL (EHBL) are only a factor of two brighter (fainter) than HBL. The second is the similarity of our LBL SED shape with that of the LBL in \citetalias{fossati1998}, which have radio luminosities 30--600 times higher.

\begin{table}
\caption{Properties of the median BL~Lacs SED}
\label{tab3}
\centering                   
\begin{tabular}{lrccr}
\hline\hline
Selection & $N$ &  $\log L_{\rm 5 GHz}$ & $\log L_{\rm 1 keV }$ & $\log \nu_{\rm peak}$ \\
\hline
\multicolumn{5}{c}{Fossati et al. 1998}\\
\hline
$\log L_{\rm r}=43$--44 &    17  &  43.44   & 44.65 & 14.4 \\
$\log L_{\rm r}=42$--43 &    10  &  42.20   & 43.84 & 14.9 \\
$\log L_{\rm r}<42$  &    38  &  40.97   & 44.54 & 16.2 \\
\hline
\hline
\multicolumn{5}{c}{This work}\\
\hline                            
all      & 34  &  40.91   & 42.94 & 15.3 \\
$\log L_{\rm r}>40.9$  & 17  &  41.30   & 43.44 & 15.7 \\
$\log L_{\rm r}<40.9$  & 17  &  40.61   & 42.84 & 15.5 \\
\hline                            
LBL      &  7  &  40.67   & 42.64 & 14.6 \\
HBL      & 19  &  40.38   & 43.31 & 15.8 \\
EHBL     &  4  &  40.14   & 44.48 & 19.4 \\
\hline
\end{tabular}

\medskip
Column description: (1) Group description, (2) number of objects in the radio band, (3) and (4)
median radio and X-rays luminosities, (4) synchrotron peak frequency. 
\end{table}

\section{The beaming effect}
\label{beam}
Up to now we have not considered the fact that blazars are beamed sources with
their luminosities enhanced by relativistic Doppler boosting, which also
blueshifts the synchrotron peak frequency. The relationship between the
emitted luminosities $L'$ and their beamed counterparts is $L=\delta^4 \, L'$,
where $\delta$ is the Doppler factor\footnote{The Doppler
  factor is defined as $\delta=[\Gamma_{\rm b} \, (1-\beta \cos \theta)]^{-1}$,
  where $\beta$ is the ratio of the plasma velocity to the speed of light,
  $\theta$ the viewing angle of the emitting plasma, and $\Gamma_{\rm
    b}=(1-\beta)^{-1/2}$ the plasma bulk Lorentz factor.}. Analogously, the
frequencies transform as $\nu= \delta \, \nu'$.

\citet{nieppola2008} considered Doppler factors derived from the study of the total flux density
variability and found that they anticorrelate with the peak frequency. These
authors suggested that the anticorrelation between $L_{\rm r}$ and $\nu_{\rm
  peak}$ becomes a positive correlation when correcting for beaming.
\citet{nieppola2008} used Doppler factors inferred from the analysis of radio
data. However, the jet radiation at different frequencies is likely emitted
from different regions along the jet, each with its own Doppler factor.
Indeed, the usually long (weeks to months) delay of the radio variations after
the higher energy variations (in particular optical and $\gamma$-ray), and the longer
radio variability timescales \citep[e.g.,][]{rai08a,vil09a} suggest that the
radio emission comes from a more external and extended jet zone than that
producing the high-frequency synchrotron radiation. Taking into account that
the Doppler factor $\delta$ depends on both the bulk Lorentz factor and the
viewing angle, accelerating/decelerating processes and/or bends in the jet
structure \citep[e.g.,][]{lister2001,lister2013} would cause a spread of
$\delta$ values within the same source in different bands. In turn, this would
produce a distortion of the observed SED shape with respect to the intrinsic SED
shape.

\section{Summary and conclusions}
\label{summary}
We explored the properties of the SED of a complete sample of 34 BL~Lacs of
low power. These objects have been selected in the redshift range $0.05 < z <
0.15$ in \citetalias{capetti2015b} based on a new diagnostic plane that includes the
W2$-$W3 color from WISE and the Dn(4000) index derived from the SDSS.

We collected multiwavelength data focusing on those bands that allow us to study the jet emission. 
We have a complete coverage
at the frequencies used for the sample selection (1.4 GHz, 12 $\mu$m, and rest-frame 3900
\AA). X-ray fluxes are available for most of the sources (27 out of 34), while
only 13 of them are detected at $\gamma$-rays. 

We classified the LPBL based on the ratio between radio and X-rays
luminosities or, equivalently, on the radio-X-ray spectral index, finding that seven of them are LBL, 19 are HBL, and four are
EHBL. The lack of X-rays data prevent us from obtaining a robust classification for
the last four objects. 

We derived an estimate of the synchrotron peak frequency from each SED through a
log-parabolic fit to the $k$-corrected jet luminosities.  Overall, the mix of
BL~Lac types is confirmed by the location of the peak frequency, ranging from
$\log \nu = 14.3$ to 20.8, i.e.,\ from the infrared to the hard X-rays. However, 
the SEDs of some LPBLs are better described by a three-point fit (excluding the
X-ray data).
For these sources, the peak frequency shifts at much lower energies,
as low as $\log \nu_{\rm peak} \sim 13.5$, which implies that the X-ray emission also receives a 
contribution from the inverse-Compton process, as expected in LBLs.
 LPBLs with synchrotron
peaks located at even lower energies would be very difficult to find with our method. 
Indeed,
the nonthermal contribution to the optical flux might be insufficient to
lower the Dn(4000) and the lines EW might not be significantly reduced with
respect to the parent population.

We estimated the median SED for the sample as a whole and for two
subsamples based on different thresholds of radio luminosity. 
The advantage of this approach is to mitigate the effects of variability 
when nonsimultaneous data are used, as in our case.
We found that the peak frequencies obtained with such an analysis all fall at $\log \nu_{\rm
  peak} \sim 15.5$ Hz. These values are lower than that characterizing
the least luminous group of BL~Lacs considered by
\citetalias{fossati1998} that have similar radio luminosity of our LPBLs. 
In our sample, the
sources classified as LBL, HBL, or EHBL do not differ significantly in radio
luminosity. The peak frequency of the median SED of LBLs
is located at $\log \nu_{\rm peak} \sim 14.6$, consistent with the value
derived by \citetalias{fossati1998} for much more luminous BL~Lacs.

Apparently,  there is no clear connection between the
SED shape and the radio power within the LPBL sample. At least at low luminosity, BL~Lacs do not
populate any preferred region in the $\log L_{\rm r}$ versus $\nu_{\rm peak}$ 
plane and the trend of increasing $\nu_{\rm peak}$ at decreasing $L_{\rm r}$ does
not extend to the least luminous BL~Lacs. The reason is the presence
of a significant fraction of LBL objects. This is remarkable when considering that
our LPBL sample is biased {\it against} the selection of LBLs.

Furthermore, the SED of the LBLs included in our sample is very similar to
that of radio-selected BL~Lacs that are up to two-three orders of magnitude more
luminous. These results cast serious doubts on the idea that the radio
luminosity is the main driving parameter of the multiwavelength properties of
BL~Lacs. 

In contrast, there is a clear positive dependence of $\nu_{\rm peak}$ on $\log
L_{\rm x}$. This implies that X-ray flux-limited samples necessarily lead to
the selection of objects with synchrotron peaks at higher energies. Previous
works had already suggested that the blazar sequence is an artifact due to
biases in the samples used by \citetalias{fossati1998}. The present analysis
confirms this result with a much larger sample of LPBLs reaching
luminosities one order of magnitude lower.

Finally, the shape of the intrinsic SED can be distorted by relativistic beaming in case there are accelerating/decelerating processes in the jet and/or bends in the jet structure.
This suggests caution in interpreting the observed spread in the spectral behavior of BL Lacs.

\begin{acknowledgements}
Part of this work is based on archival data, software, or online services provided by the ASI Science Data Center (ASDC). 

This research has made use of data obtained from the 3XMM XMM-Newton serendipitous source catalog compiled by the ten institutes of the XMM-Newton Survey Science Centre selected by ESA.

Funding for SDSS-III has been provided by the Alfred P. Sloan Foundation, the
Participating Institutions, the National Science Foundation, and the
U.S. Department of Energy Office of Science. The SDSS-III website is
http://www.sdss3.org/.  SDSS-III is managed by the Astrophysical Research
Consortium for the Participating Institutions of the SDSS-III Collaboration,
including the University of Arizona, the Brazilian Participation Group,
Brookhaven National Laboratory, University of Cambridge, Carnegie Mellon
University, University of Florida, the French Participation Group, the German
Participation Group, Harvard University, the Instituto de Astrofisica de
Canarias, the Michigan State/Notre Dame/JINA Participation Group, Johns
Hopkins University, Lawrence Berkeley National Laboratory, Max Planck
Institute for Astrophysics, Max Planck Institute for Extraterrestrial Physics,
New Mexico State University, New York University, Ohio State University,
Pennsylvania State University, University of Portsmouth, Princeton University,
the Spanish Participation Group, University of Tokyo, University of Utah,
Vanderbilt University, University of Virginia, University of Washington, and
Yale University.

This publication makes use of data products from the Two Micron All Sky Survey, which is a joint project of the University of Massachusetts and the Infrared Processing and Analysis Center/California Institute of Technology, funded by the National Aeronautics and Space Administration and the National Science Foundation.

\end{acknowledgements}

\clearpage
\appendix
\section{Spectral energy distributions}
In Figure \ref{sed_z1a}, we show the broadband SEDs of our
sources.  They contain the FIRST as well as other radio data (NVSS; Green Bank, VLBI); 
the AllWISE flux densities in the W1, W2, W3, and
W4 bands, the $J, H, K$ flux densities from the 2MASS; the SDSS spectrum; the RASS and
3XMM X-ray data, and the 3FGL $\gamma$-ray fluxes.  The near-infrared,
optical, and X-ray data have been corrected for Galactic absorption.

The RASS 0.1--2.4 fluxes have been converted into 1 keV flux densities with
all the three possible spectral indices $\Gamma_{\rm x}=1,2,3$ to quantify the uncertainties.  
At $\gamma$ rays, we show the 3FGL data for the various
energy bins, the flux density at the pivot energy, and the butterfly resulting
from the power-law spectral fit.

We also highlight the reference radio (1.4 GHz), infrared (W3 band), optical
(3900 \AA\ rest-frame), X-ray (1 keV), and $\gamma$-ray (1 GeV) jet flux
densities (see Section 3).  In the W3 band, the jet flux has been corrected
for the host contamination estimated from an elliptical galaxy template
normalized in the W2 band.  The optical jet emission has been derived from the
Dn(4000).  For the sources for which no RASS detection was available, we give
upper limits obtained by assuming a {\em ROSAT} count rate of 0.02 counts/sec.
The 1 GeV flux density was obtained from the flux density at the pivot
energy.
We also plot data downloaded from the ASDC.

\subsection{Notes on individual sources}

\begin{description}

\item[ID=507] The soft X-ray spectrum from {\em XMM-Newton} confirms its HBL nature.

\item[ID=786] Its HBL flavor is verified by the soft X-ray and hard $\gamma$-ray spectra.

\item[ID=1103] The soft X-ray and mildly hard $\gamma$-ray spectra agree with an HBL classification.

\item[ID=1315] Its HBL nature is confirmed by the soft X-ray and hard $\gamma$-ray spectra.

\item[ID=1845] Its HBL flavor is verified by the soft X-ray and hard $\gamma$-ray spectra.

\item[ID=1906] Its HBL nature is confirmed by the soft X-ray and hard $\gamma$-ray spectra.

\item[ID=2061] LBL with no detection in X-rays. With respect to the four-point fit to the SED, the three-point fit shifts $\log \nu_{\rm peak}$ from $<14.5$ to $13.7$.

\item[ID=2281] LBL with no detection in X-rays; $\log \nu_{\rm peak}$ decreases from $<14.9$ to 14.1, changing the SED fit from four to three points. 

\item[ID=2440] Its LBL definition is confirmed by its hard X-ray and soft $\gamma$-ray spectra, which also suggests a Compton
  dominance. Indeed, $\log \nu_{\rm peak}$ goes from $<15.1$ in the four-point fit to the SED to 14.0 in the three-point fit.

\item[ID=3091] We classified it as HBL, however, $\log \nu_{\rm peak}$ shifts from $15.1$ in a four-point to $13.7$ in a three-point SED fit. 
It may be that the SDSS spectrum met the source in a particularly low optical state.

\item[ID=3403] This is the famous EHBL 1ES 1426+426, which has been extensively studied in X-rays.
Indeed, its SED reveals an extremely bright X-ray flux and an
exceptionally hard $\gamma$-ray spectrum, clearly identifying the source as an
EHBL.
The value $\log \nu_{\rm peak}=20.8$ we found is clearly overestimated; the
{\em XMM-Newton} and ASDC data allow us to position it at $\log \nu_{\rm peak} \sim 17$--18. 

\item[ID=3591] Its LBL classification based on the radio and X-ray brightness is contradicted
by the high synchrotron peak frequency as well as a soft X-ray (XRT) and a flat UV (UVOT+GALEX) spectrum (ASDC data).
The 5 GHz flux density calculated from its 1.4 GHz FIRST value
adopting $\alpha_{\rm r}=0$ is much higher than the other radio data, so that
the wrong LBL definition is likely due to an underestimate of its $L_{\rm x}/L_{\rm r}$ and overestimate of its $\alpha_{\rm rx}$.  

\item[ID=4022] Its HBL nature is confirmed by the high X-ray flux with a
soft X-ray and a hard $\gamma$-ray spectrum.

\item[ID=4156] LBL without X-ray detection with $\log \nu_{\rm peak}$ shifting from $<14.7$ in a four-point to 14.1 in a three-point SED fit.

\item[ID=4601] LBL without X-ray detection, with $\log \nu_{\rm peak}$ decreasing from $<14.9$ to 14.4 when changing the SED fit from four to three points.

\item[ID=4756] The soft X-ray spectrum from {\em XMM-Newton} agrees with its HBL classification. 

\item[ID=5076] Its LBL flavor is confirmed by the low X-ray flux compared to the infrared and
optical fluxes and by its soft $\gamma$-ray spectrum.

\item[ID=6428] EHBL with a clearly overestimated $\log \nu_{\rm peak}=20.4$. The ASDC data show that its X-ray flux is extremely variable and that {\em ROSAT} found the source in a very bright state.  Nonetheless,  when considering the other
X-ray data also, the source still appears as an HBL, which is proved by the mildly hard $\gamma$-ray spectrum. 

\item[ID=6943] EHBL with a clearly overestimated $\log \nu_{\rm peak}=20.8$.
The {\em XMM-Newton} data confirm the X-ray brightness given by the RASS and add spectral information in favor of its EHBL
nature, suggesting a $\log \nu_{\rm peak}$ of 17--18. 

\item[ID=7186] Its HBL classification is strengthened by the soft {\em XMM-Newton} X-ray spectrum.

\item[ID=9640] A high X-ray state compared to the infrared
and optical states, coupled with the hard X-ray spectrum detected by {\em XMM-Newton}, confirm its EHBL nature and suggest 
that the synchrotron peak falls at $\log \nu > 18$.

\item[ID=10537] The flavor of this BL Lac is difficult to assess. Both the four-point and three-point SED fits fail to give a satisfactory reproduction of the data.
The high optical flux suggests an HBL classification, which is supported by a possible soft X-ray spectrum, but contradicted by the mildly soft $\gamma$-ray spectrum.

\end{description}

\begin{figure*}
\centerline{\psfig{figure=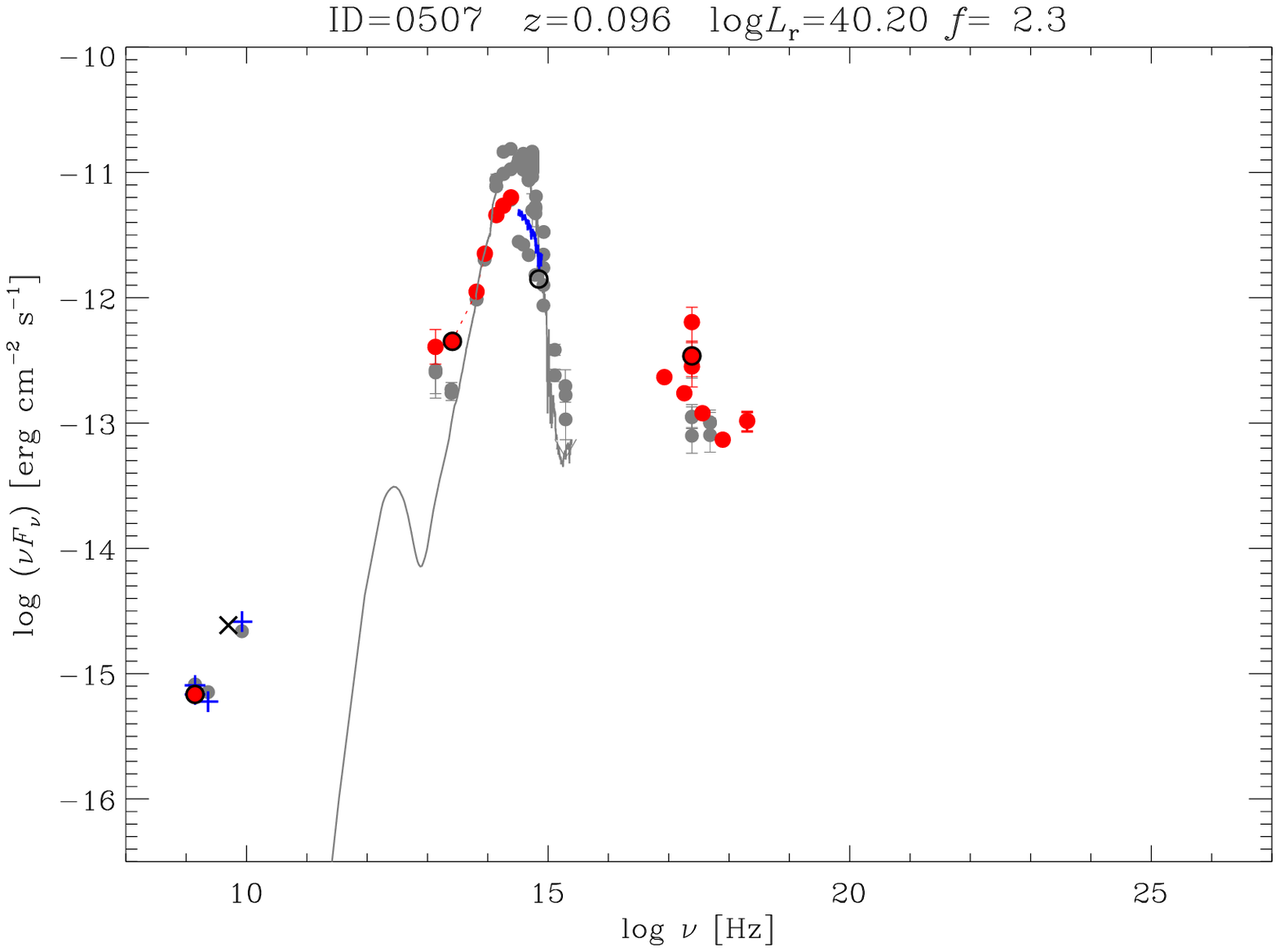,width=0.50\linewidth}
\psfig{figure=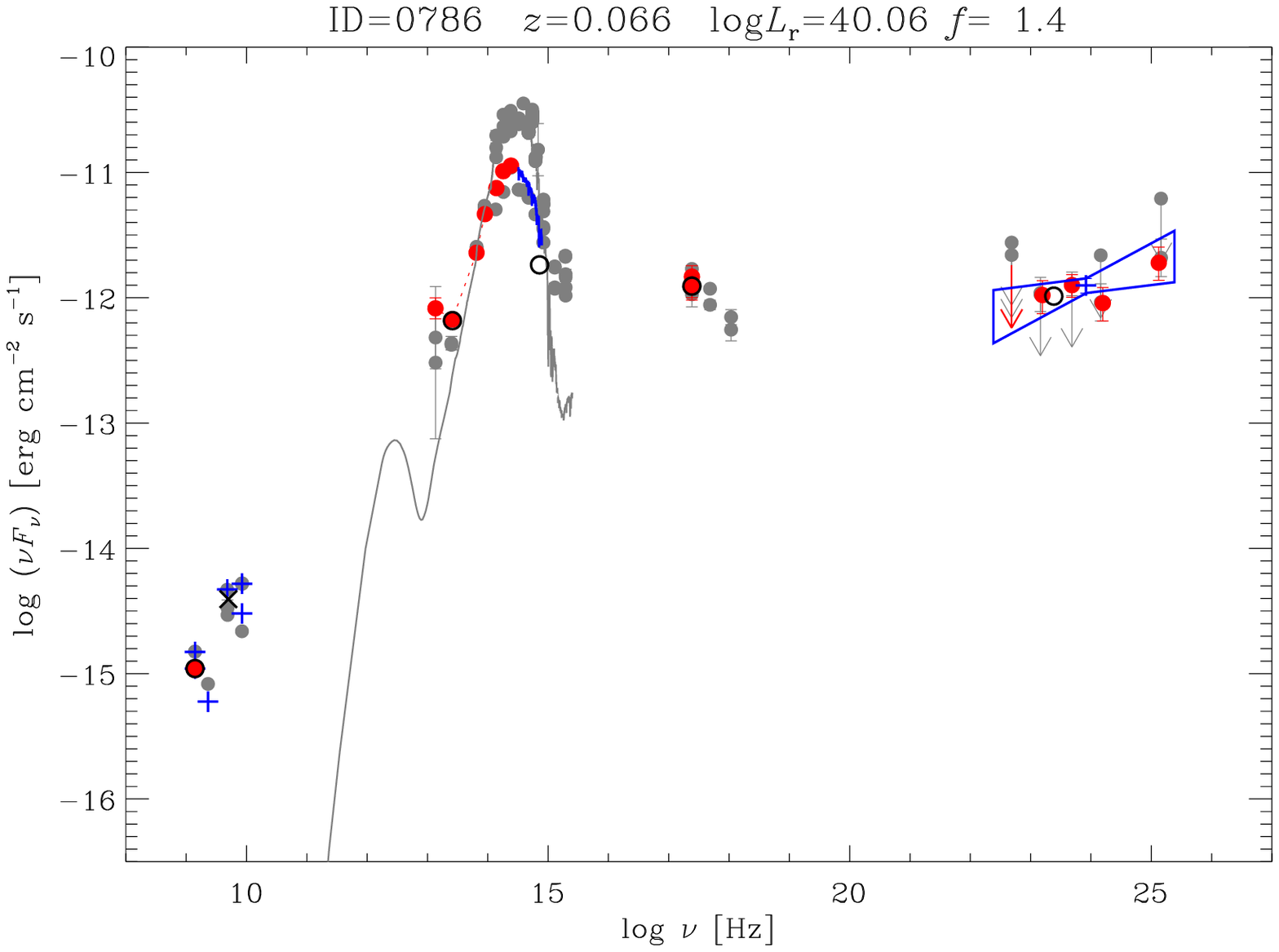,width=0.50\linewidth}}
\vspace{0.2cm}
\centerline{\psfig{figure=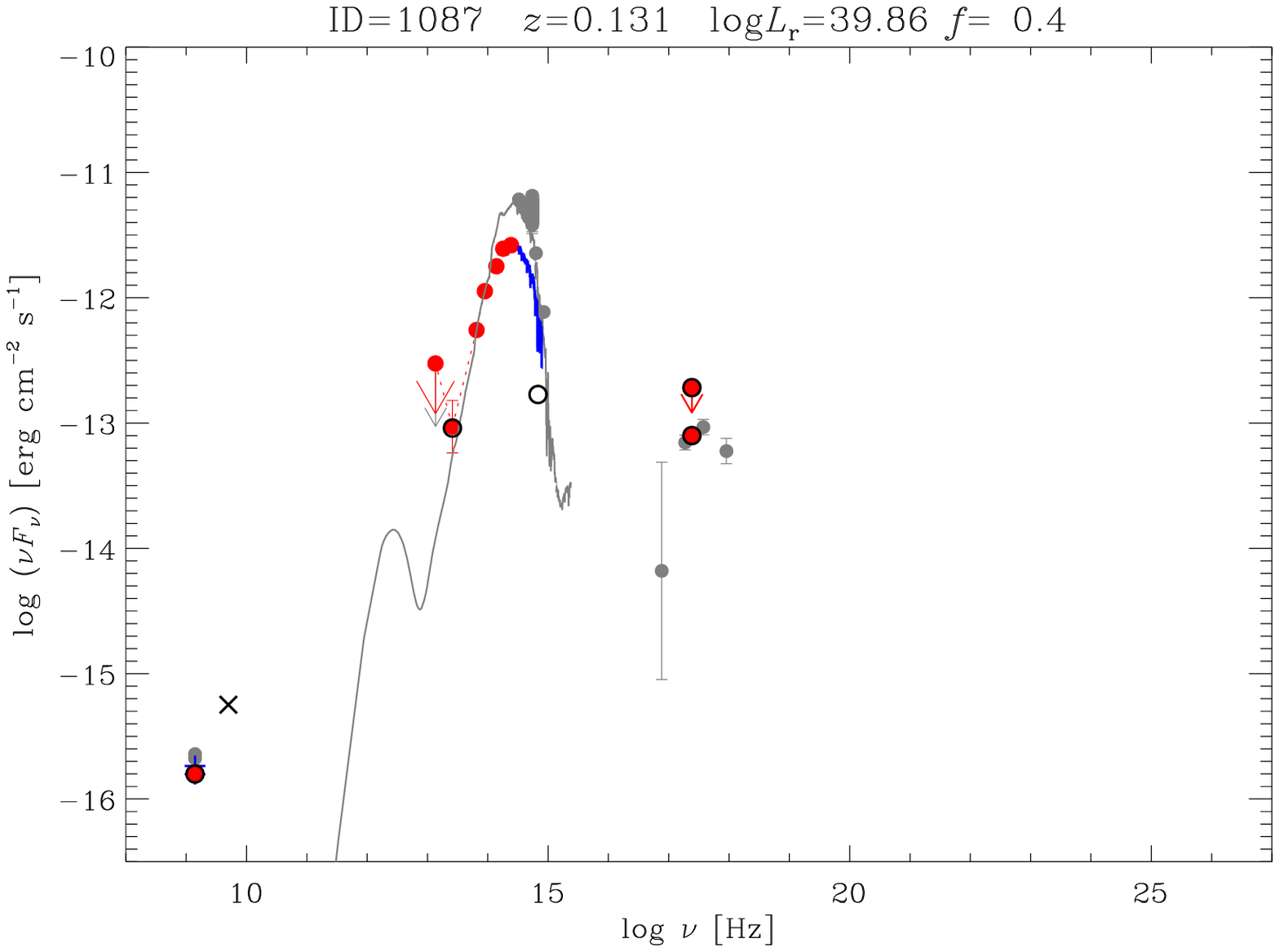,width=0.50\linewidth}
\psfig{figure=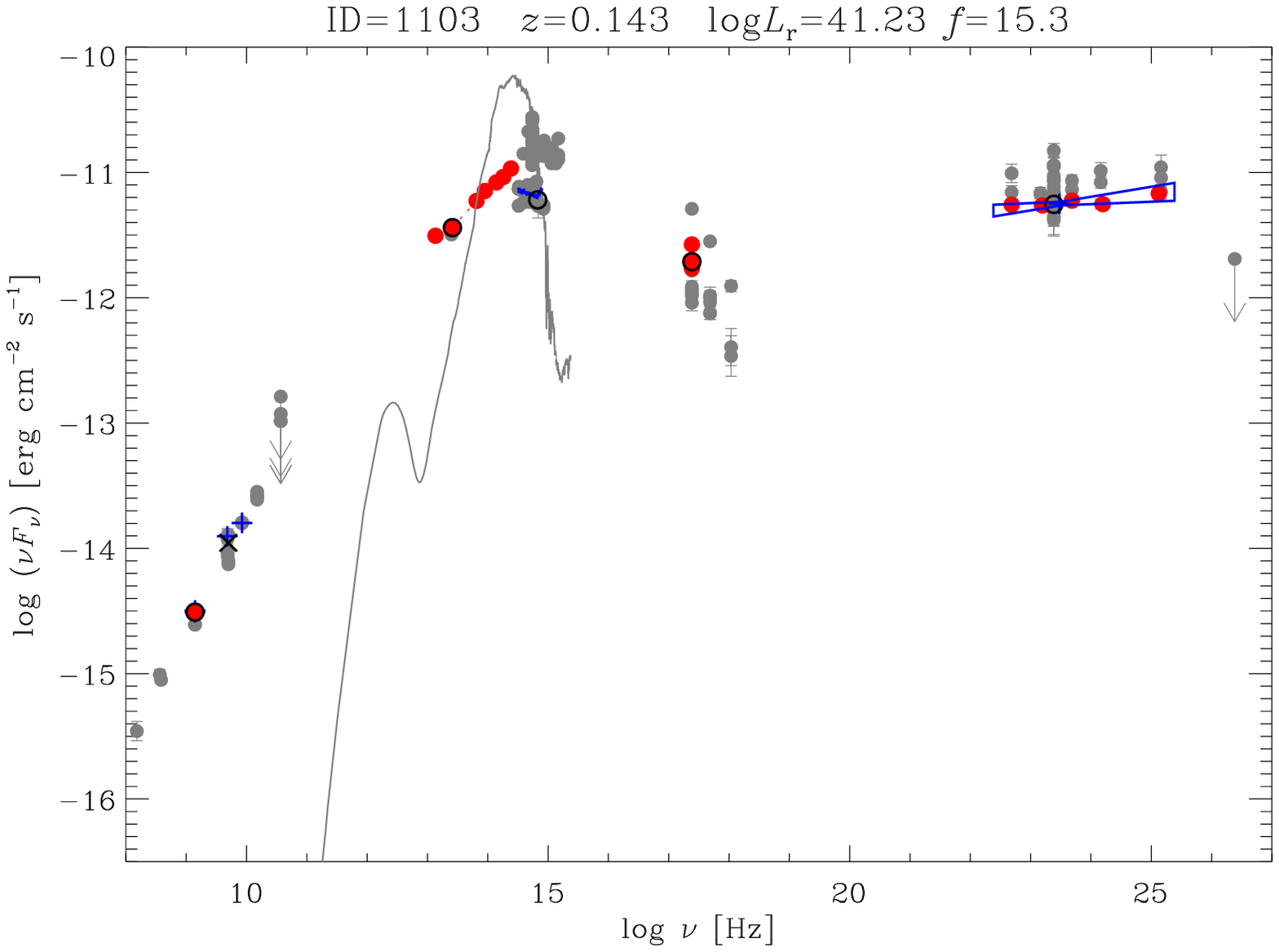,width=0.50\linewidth}}
\vspace{0.2cm}
\centerline{\psfig{figure=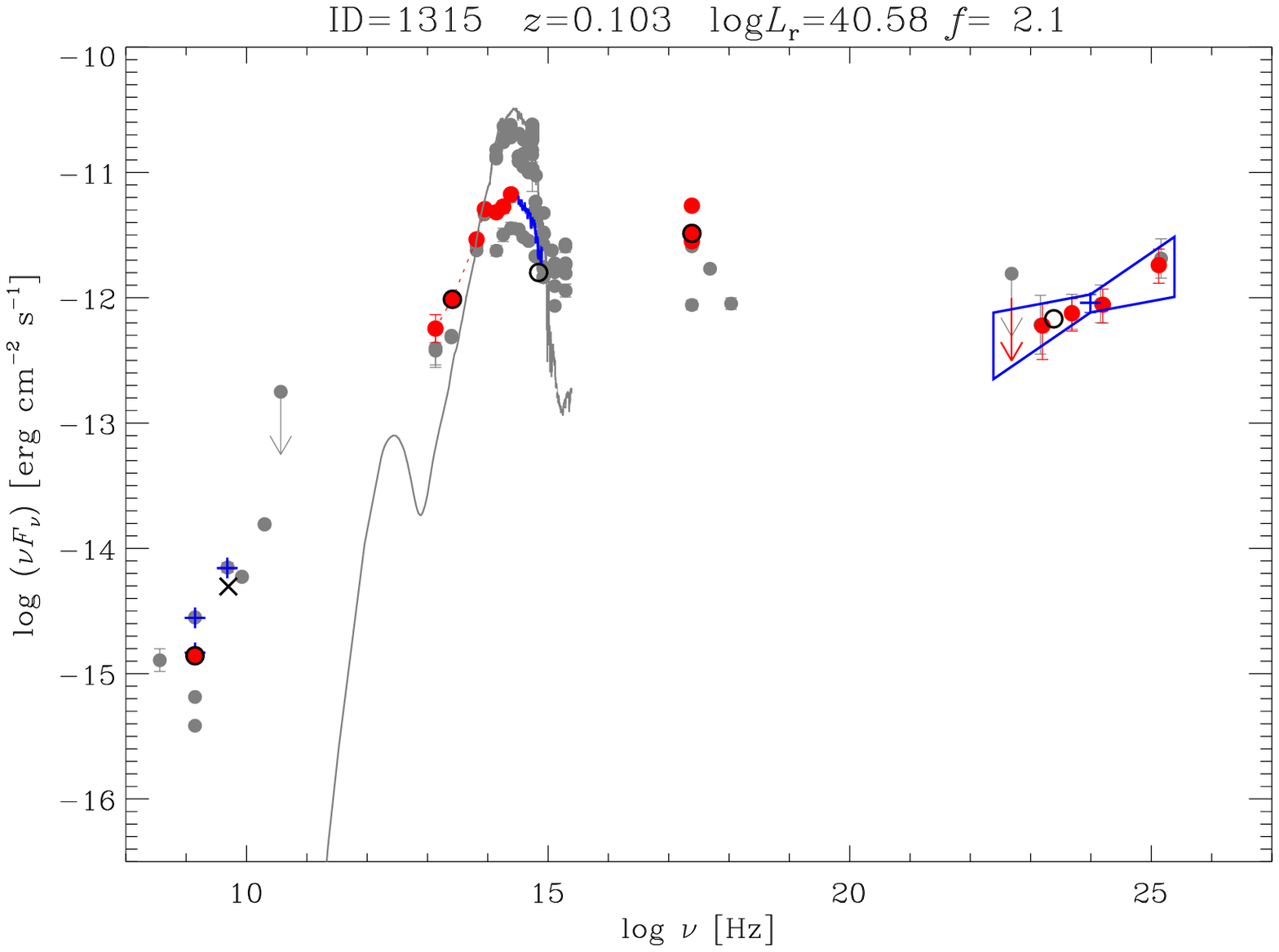,width=0.50\linewidth}
\psfig{figure=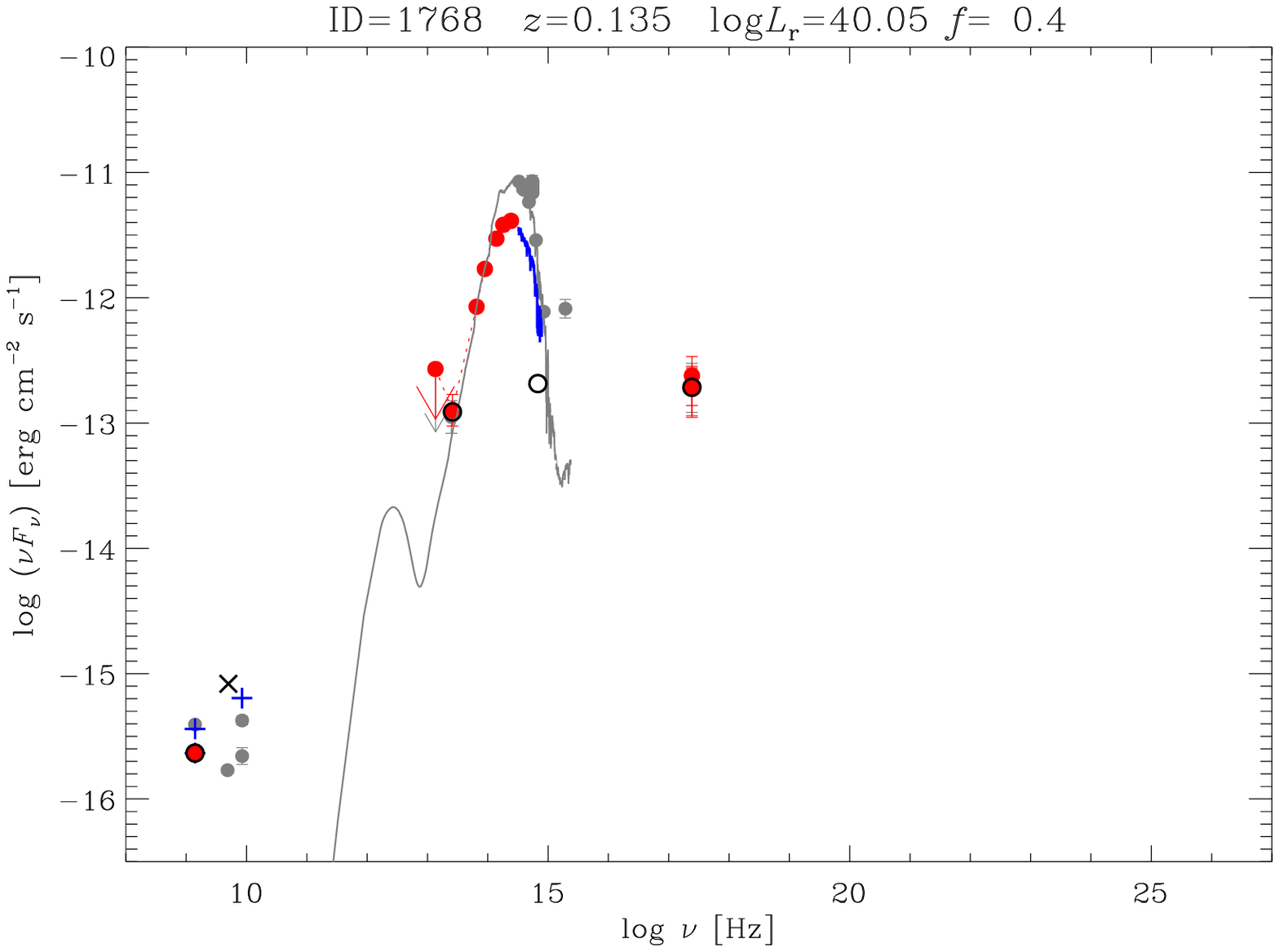,width=0.50\linewidth}}
\vspace{0.2cm}
    \caption{Broad-band SEDs of the 34 LPBLs analysed in this paper. 
      Red dots represent flux densities derived from the 
      FIRST, AllWISE, 2MASS, RASS, 3XMM, and 3FGL catalogs. 
      Blue plus signs show other radio data and the $\gamma$-ray flux density at the pivot energy; 
      the black crosses represent the 1.4 GHz FIRST data converted into 5 GHz ones with
      $\alpha_{\rm r}=0$.
      The SDSS spectrum and the {\em Fermi} power-law fit butterfly are plotted in blue.
      Near-infrared, optical, and X-ray data have been corrected for Galactic absorption. 
      The black circles highlight the reference radio (1.4 GHz),
      infrared (W3 band), optical (3900 \AA\ rest-frame), X-ray (1 keV), and
      $\gamma$-ray (1 GeV) jet flux densities. 
      In the W3 band, the jet flux density has been corrected for the host contamination
      estimated from an elliptical galaxy template (grey line) normalized in
      the W2 band. In the optical, the jet contribution has been derived from the Dn(4000).
      Grey dots show data downloaded from the ASDC.}
    \label{sed_z1a}
   \end{figure*}

    \addtocounter{figure}{-1}
\begin{figure*}
\centerline{\psfig{figure=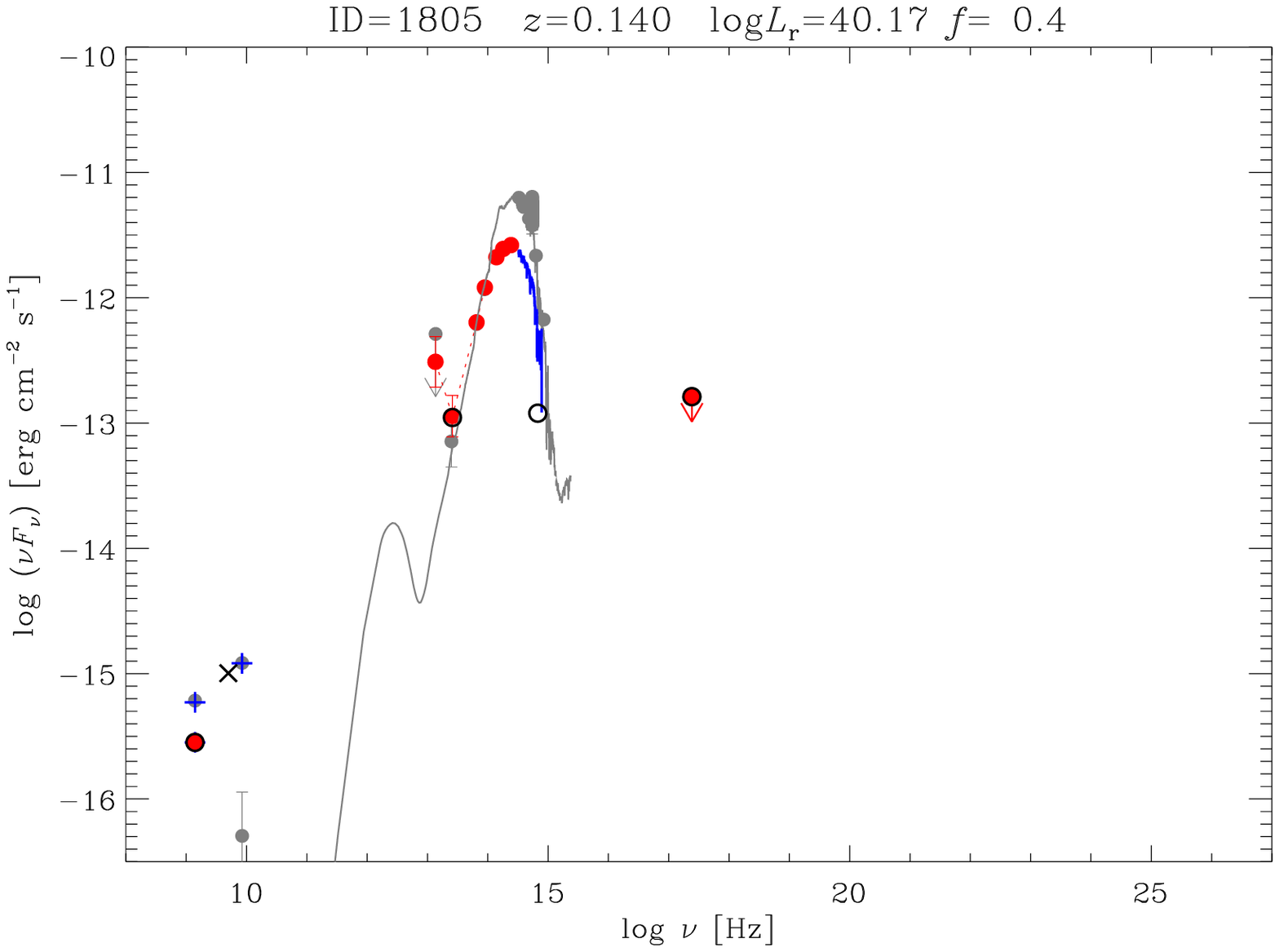,width=0.50\linewidth}
\psfig{figure=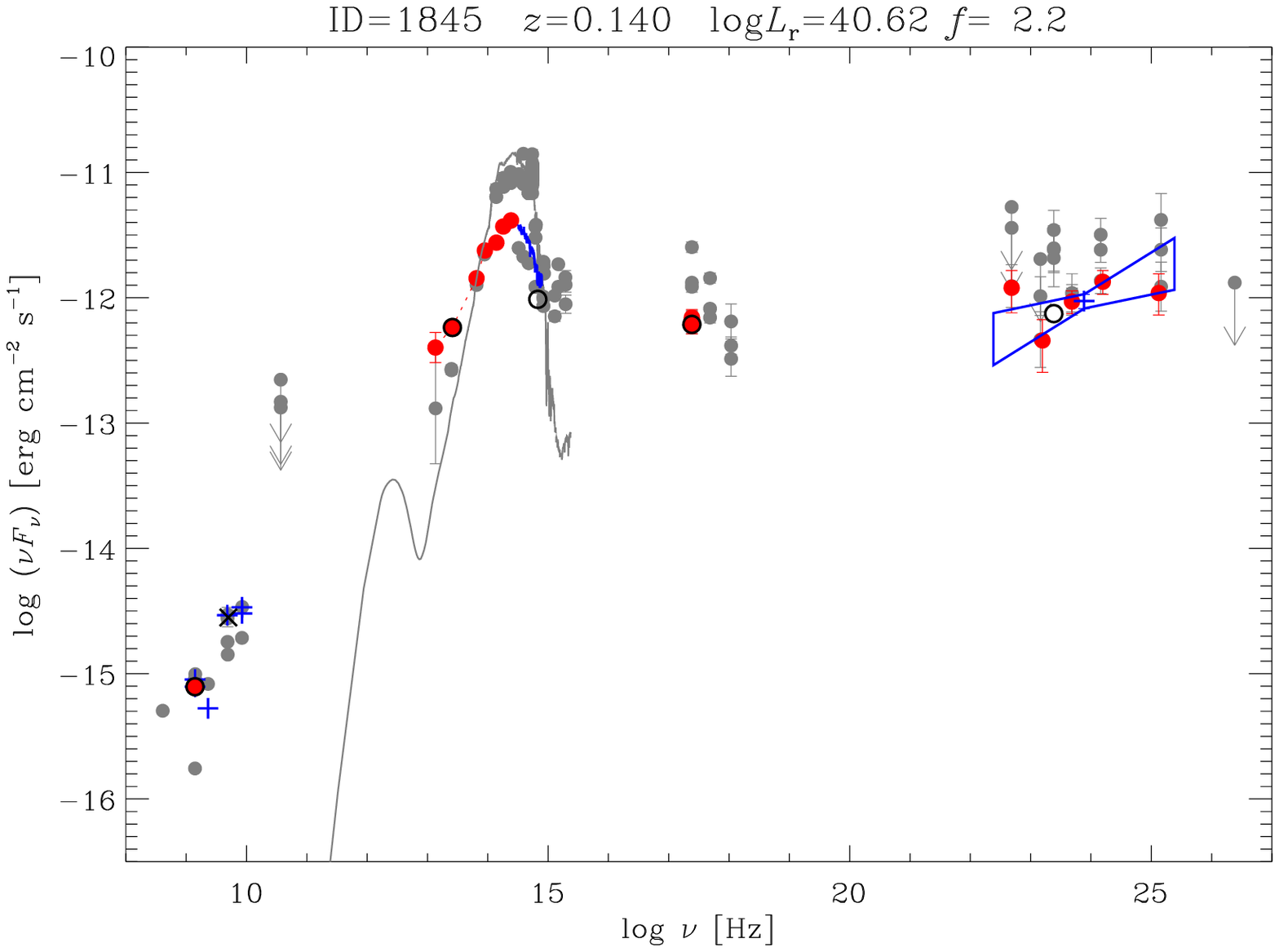,width=0.50\linewidth}}
\vspace{0.2cm}
\centerline{\psfig{figure=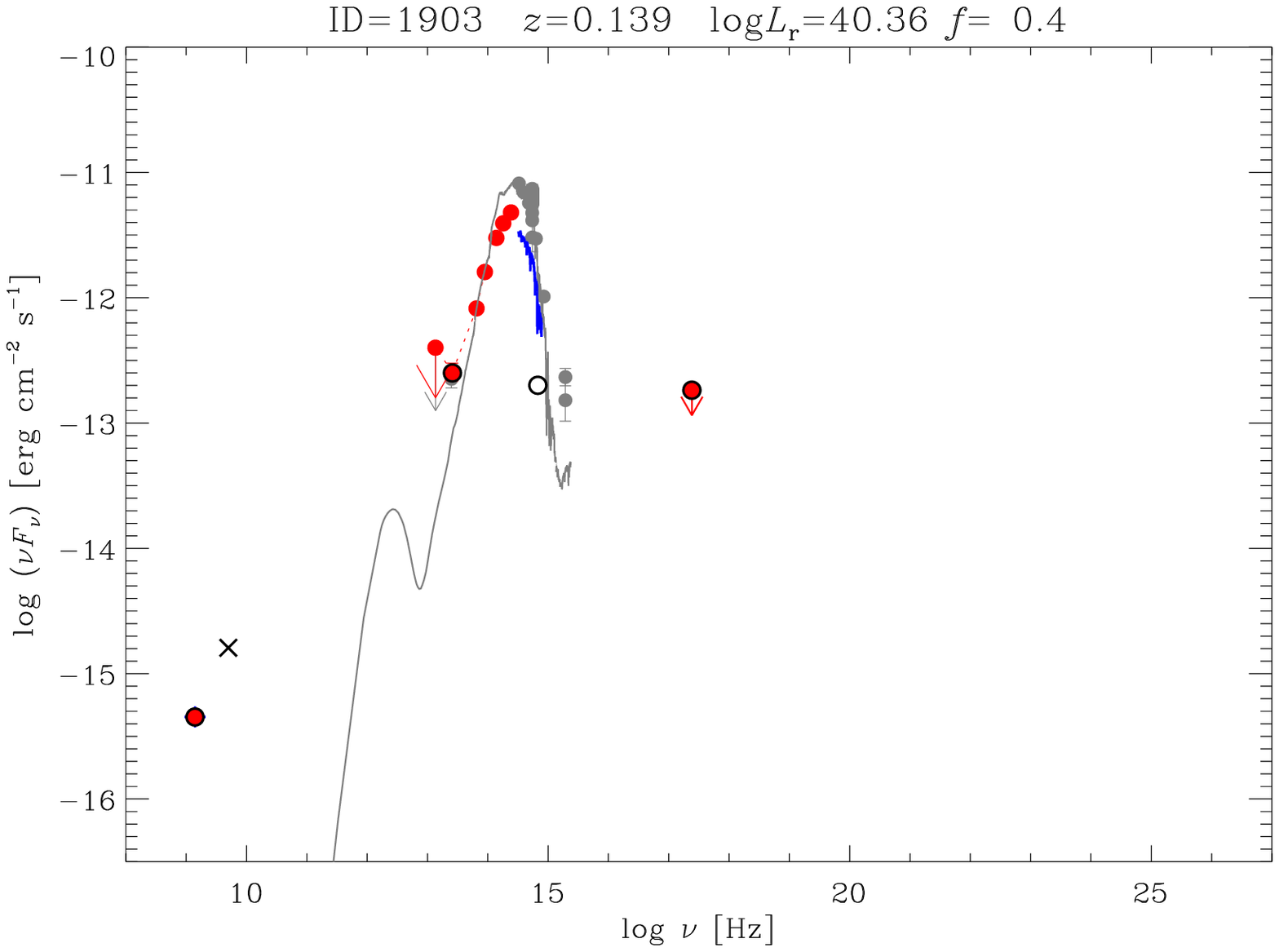,width=0.50\linewidth}
\psfig{figure=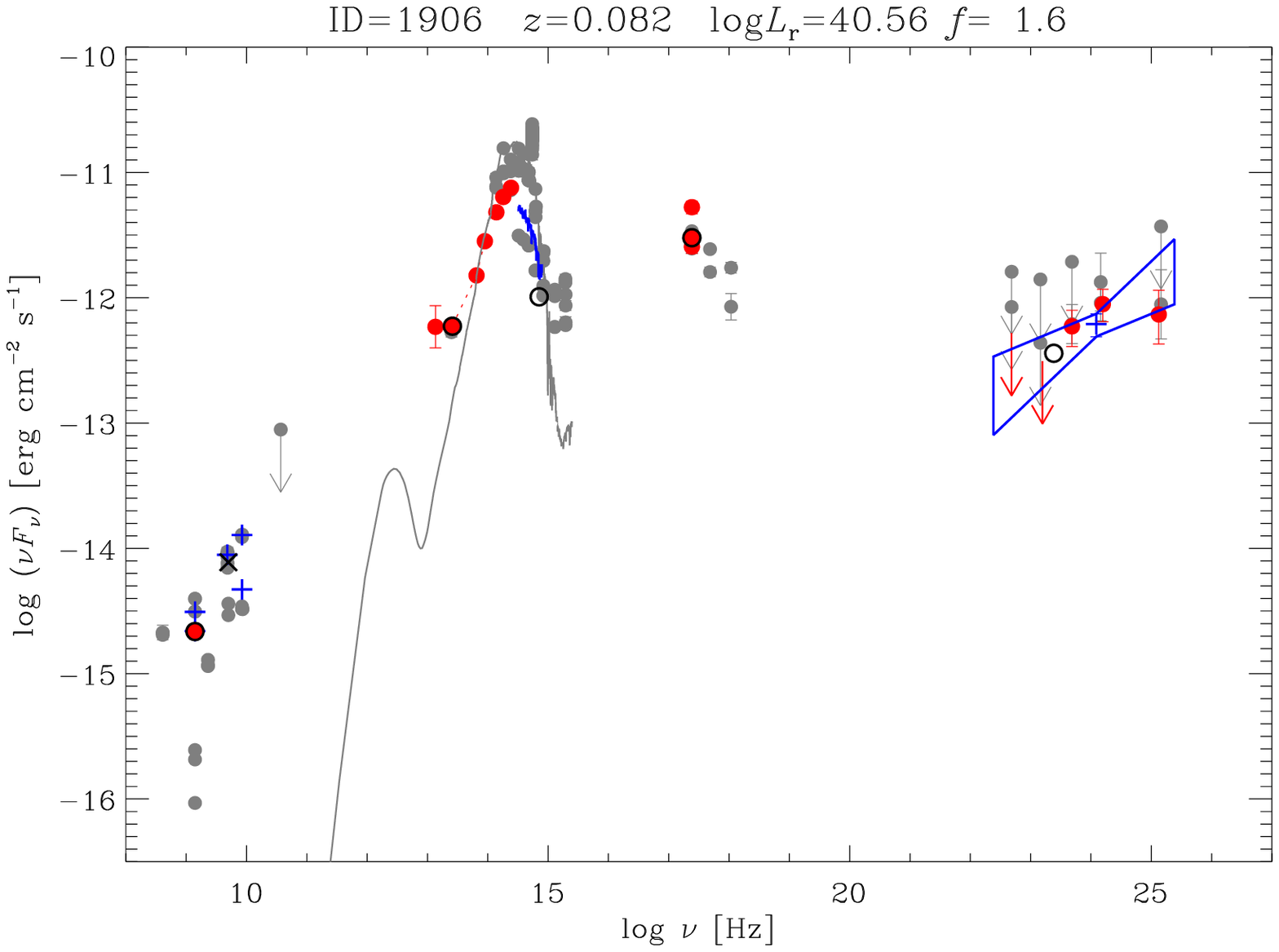,width=0.50\linewidth}}
\vspace{0.2cm}
\centerline{\psfig{figure=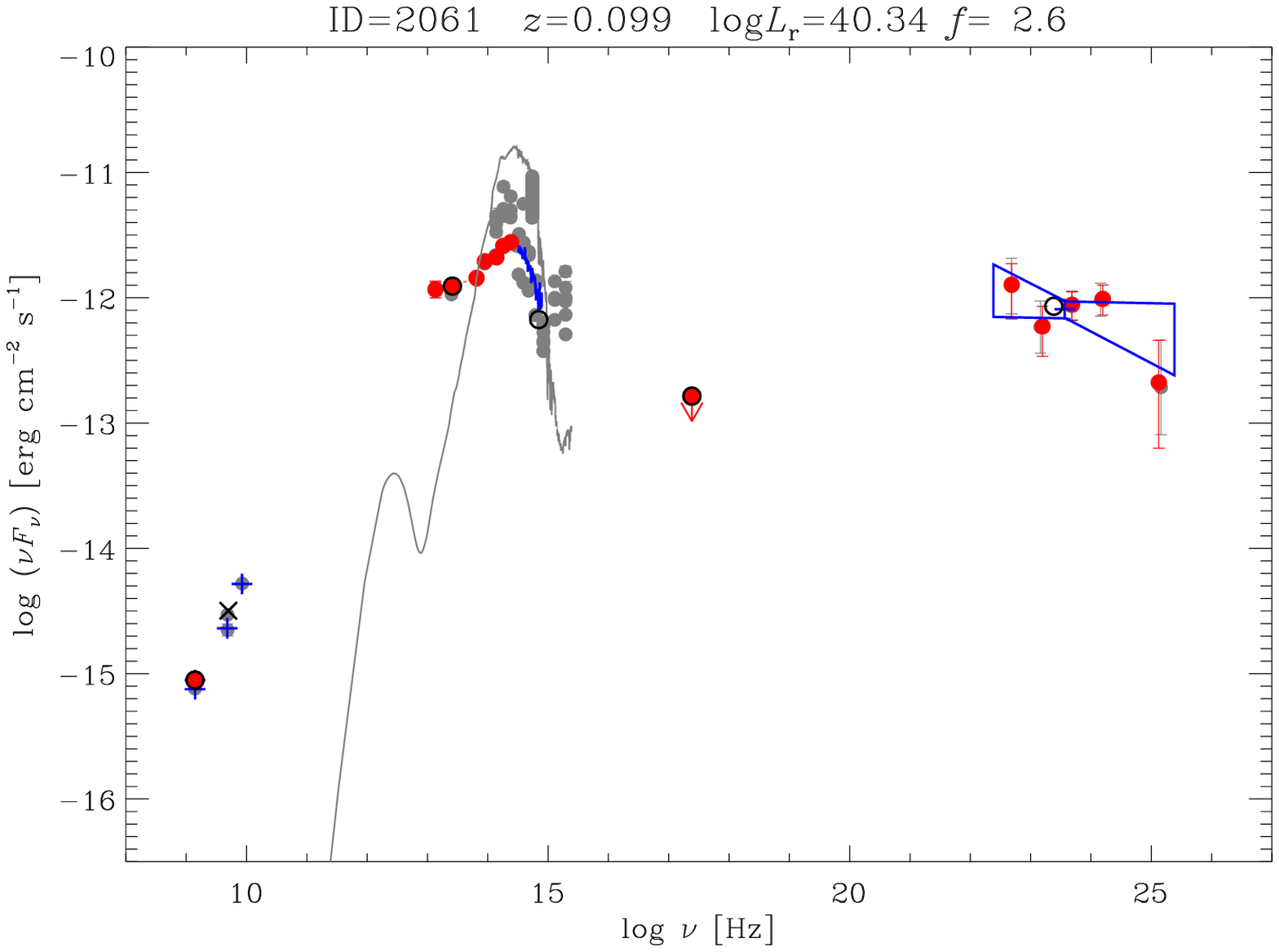,width=0.50\linewidth}
\psfig{figure=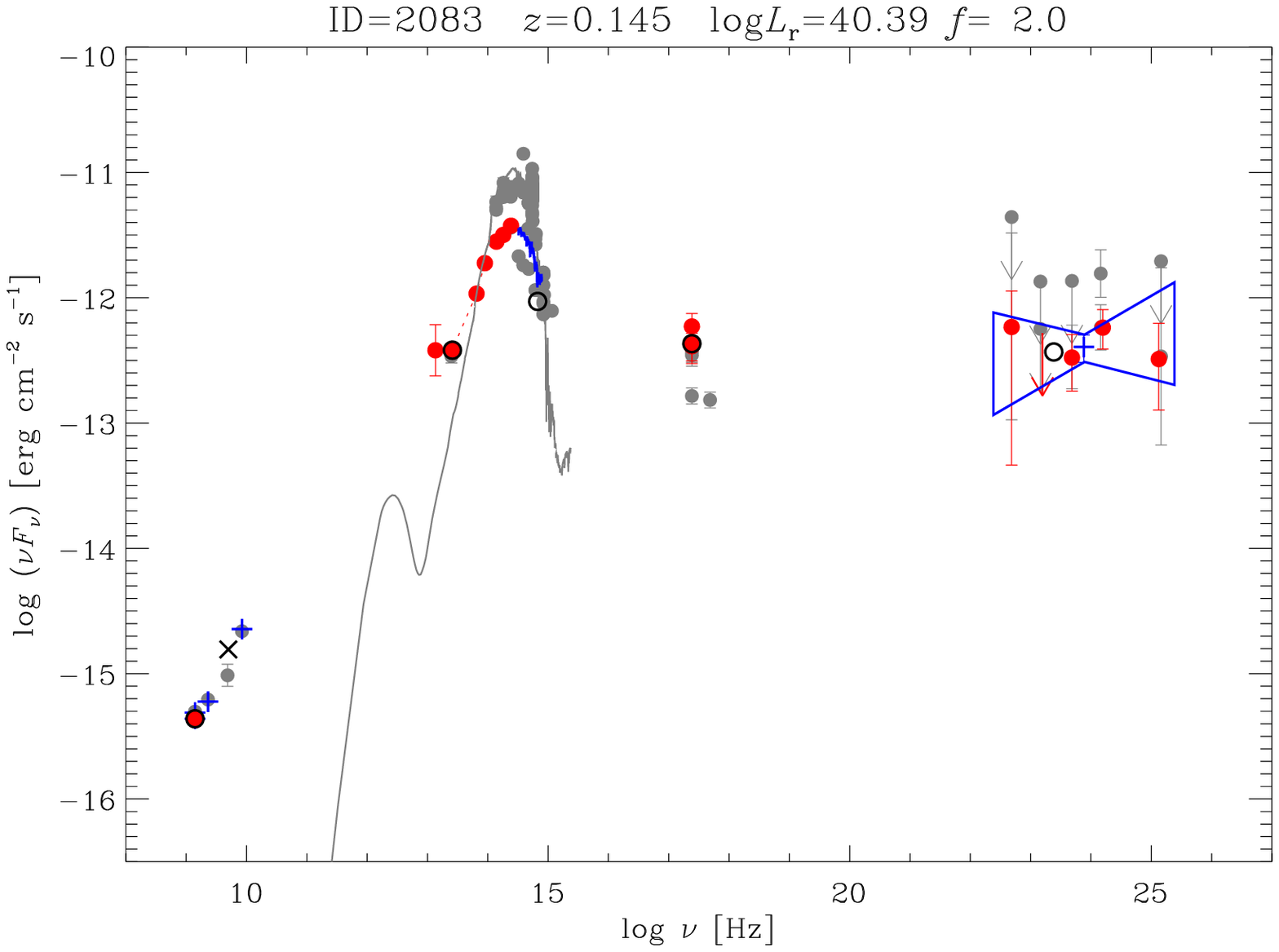,width=0.50\linewidth}}
\vspace{0.2cm}
\centerline{\psfig{figure=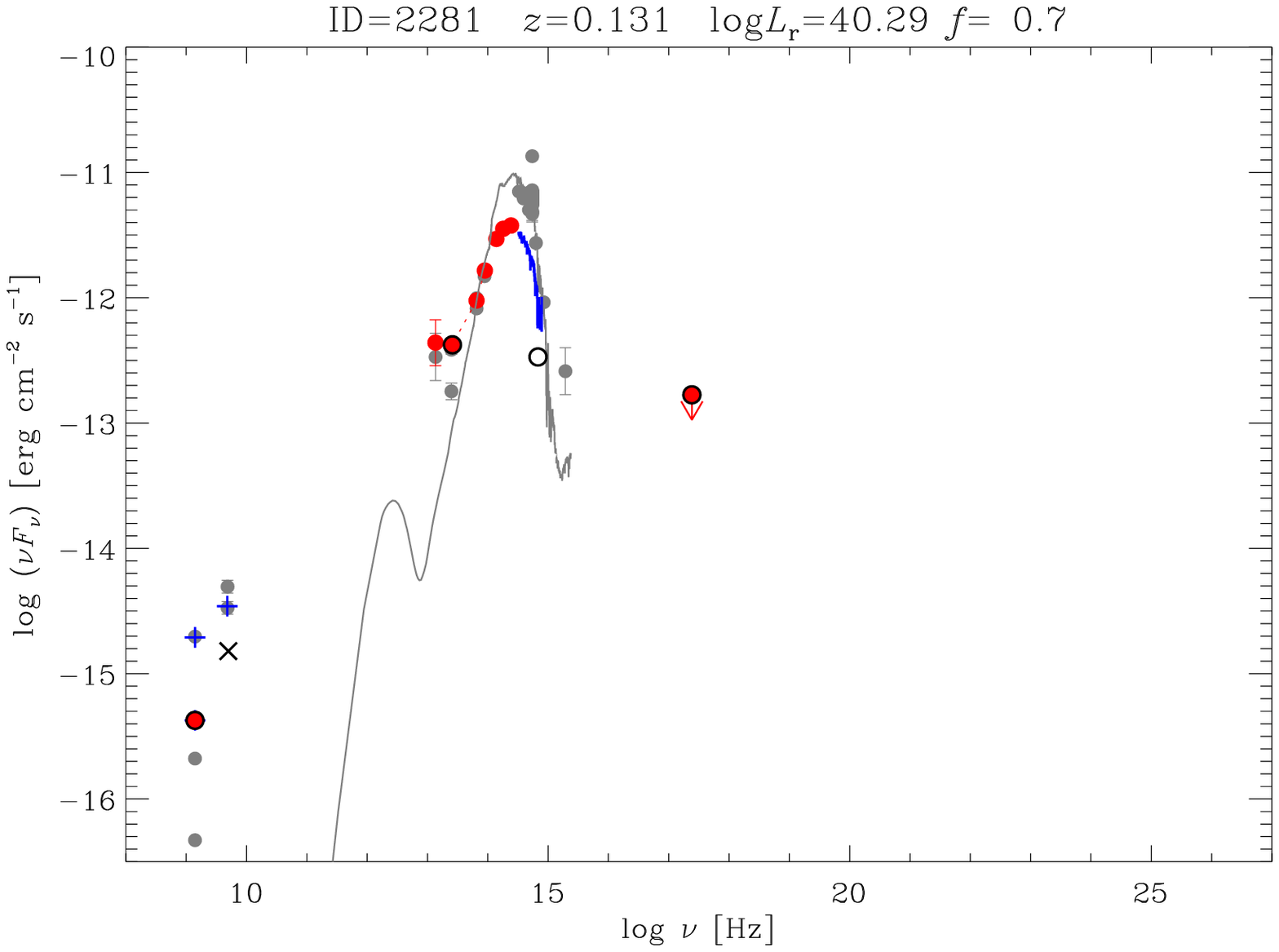,width=0.50\linewidth}
\psfig{figure=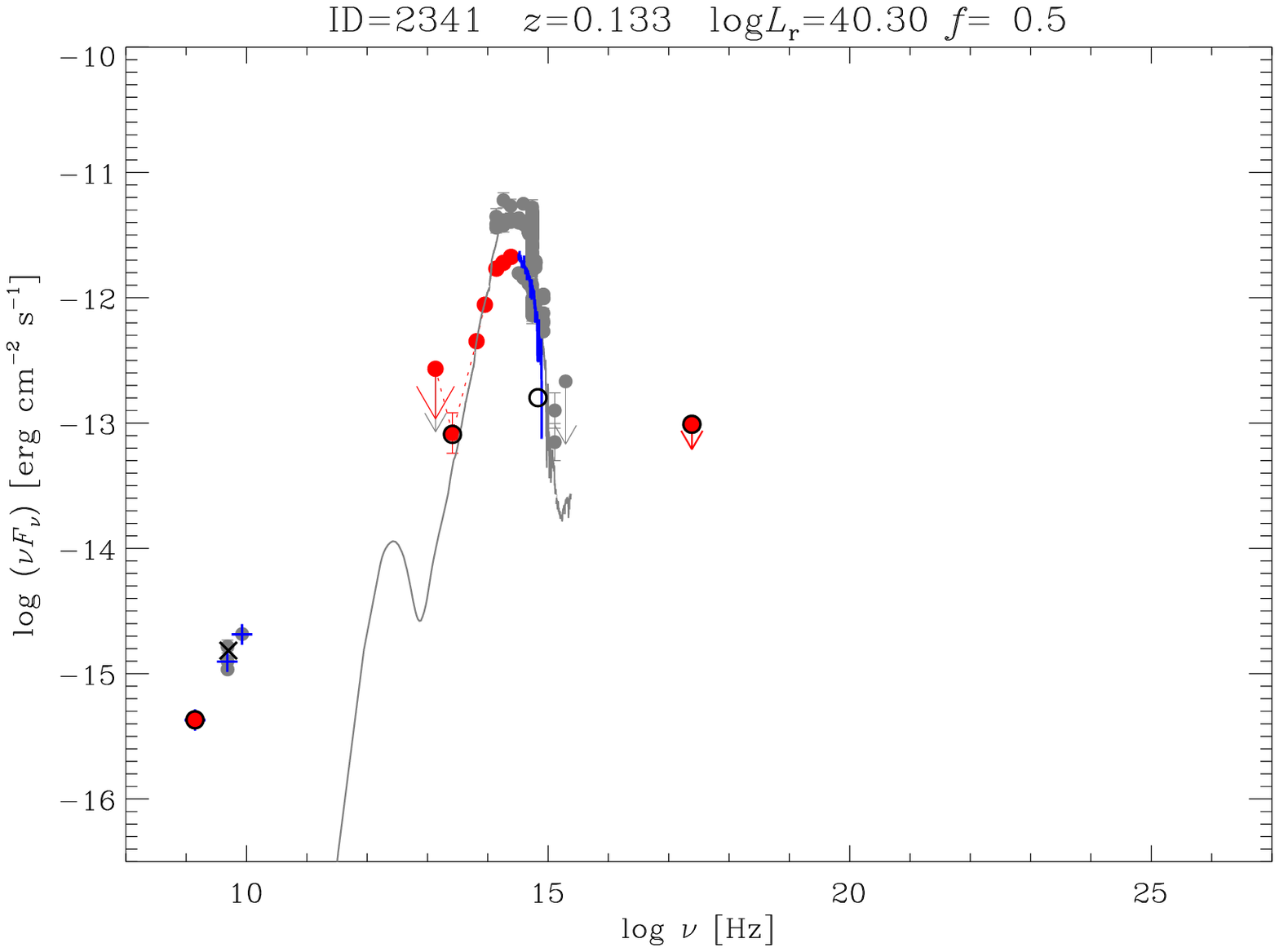,width=0.50\linewidth}}
    \caption{(continued)}
   \end{figure*}

    \addtocounter{figure}{-1}
\begin{figure*}
\centerline{\psfig{figure=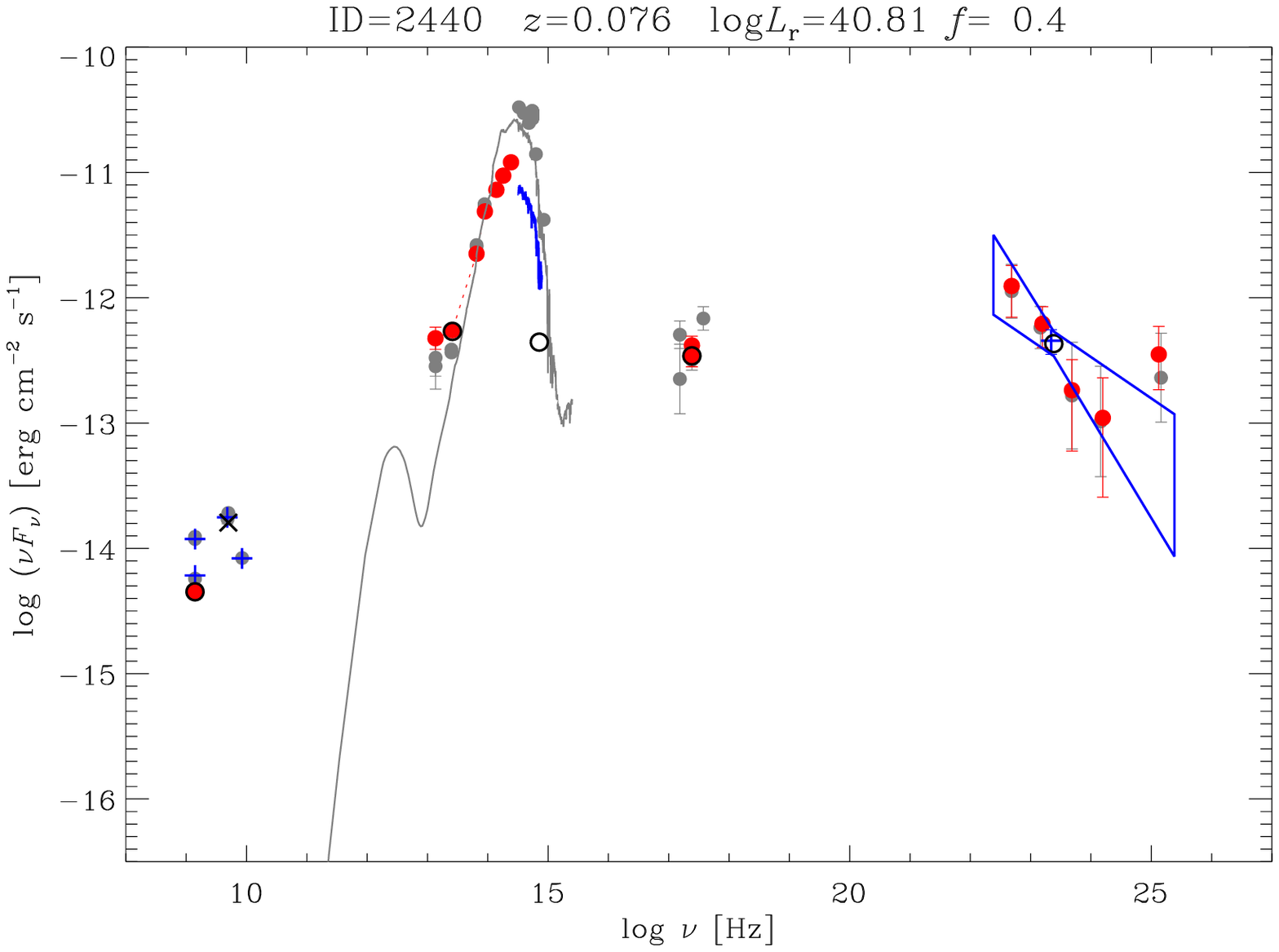,width=0.50\linewidth}
\psfig{figure=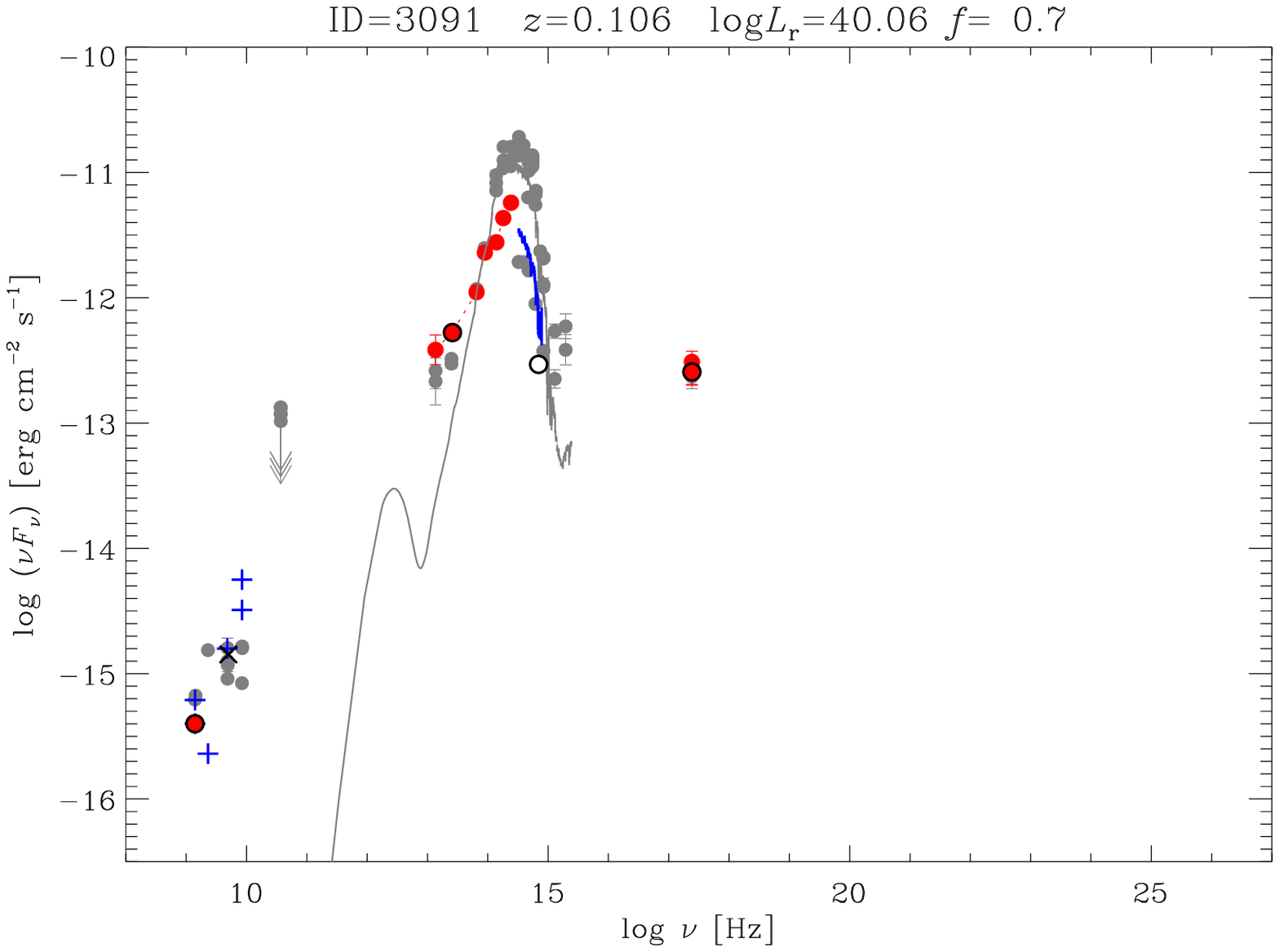,width=0.50\linewidth}}
\vspace{0.2cm}
\centerline{\psfig{figure=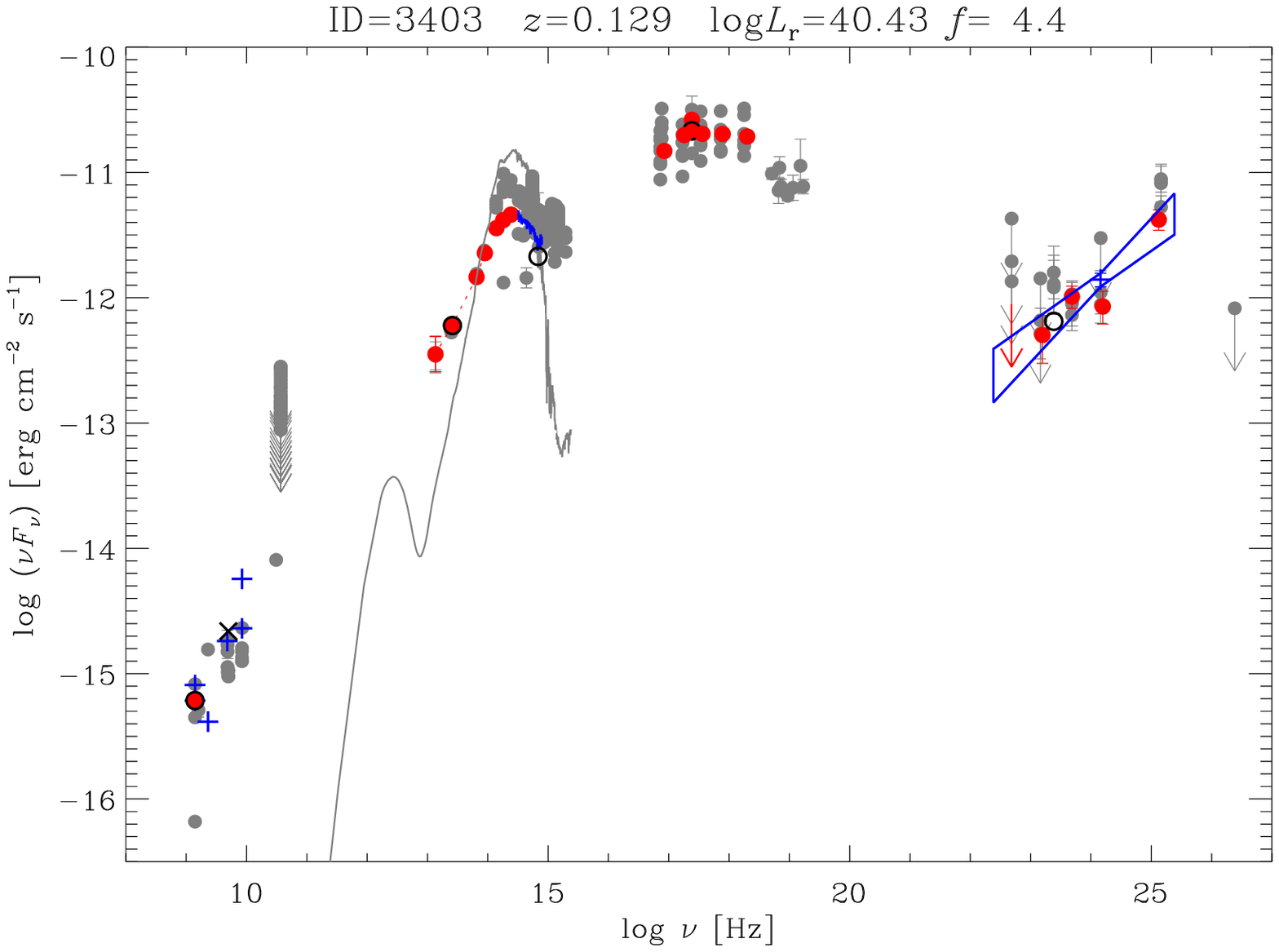,width=0.50\linewidth}
\psfig{figure=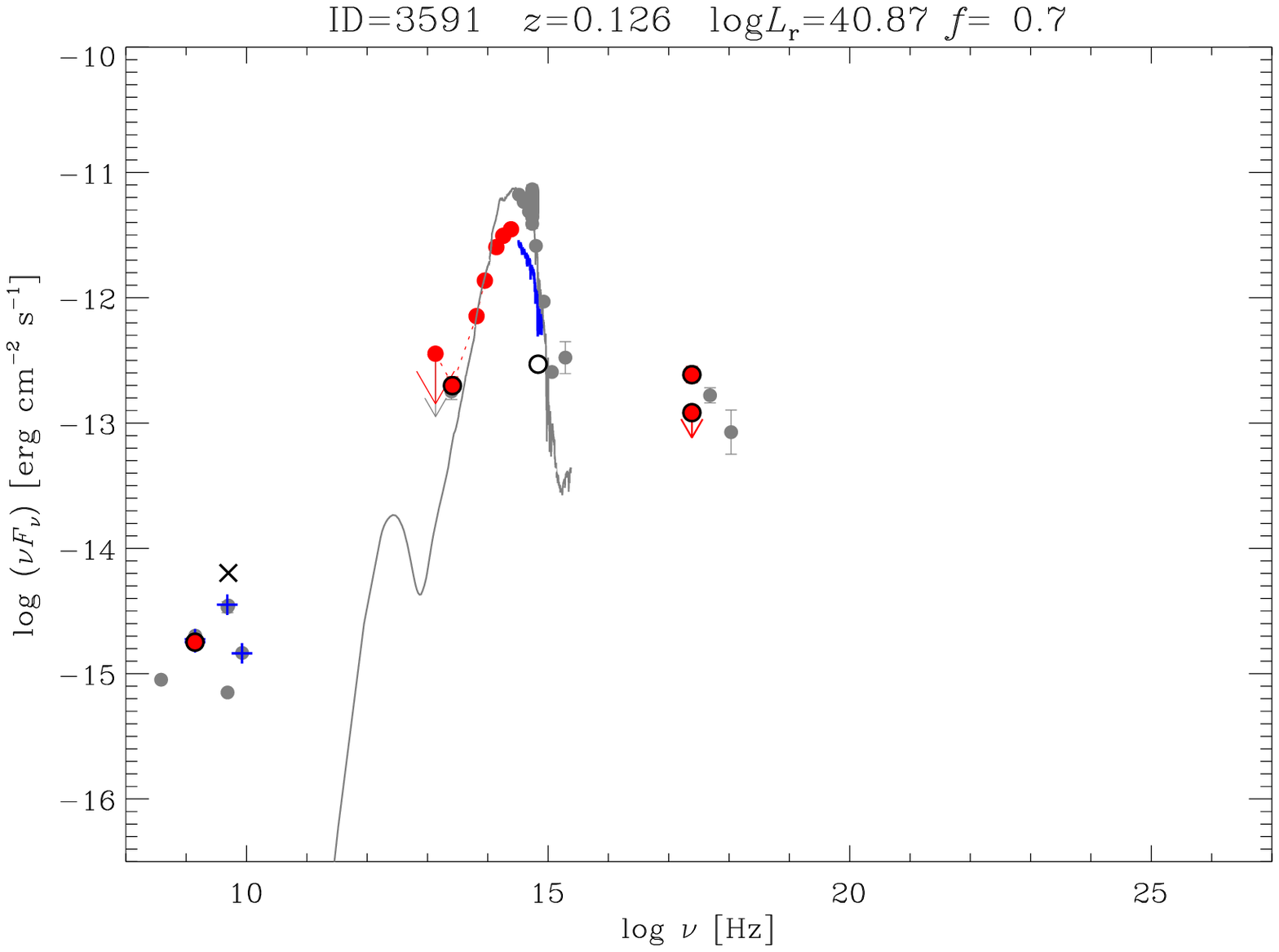,width=0.50\linewidth}}
\vspace{0.2cm}
\centerline{\psfig{figure=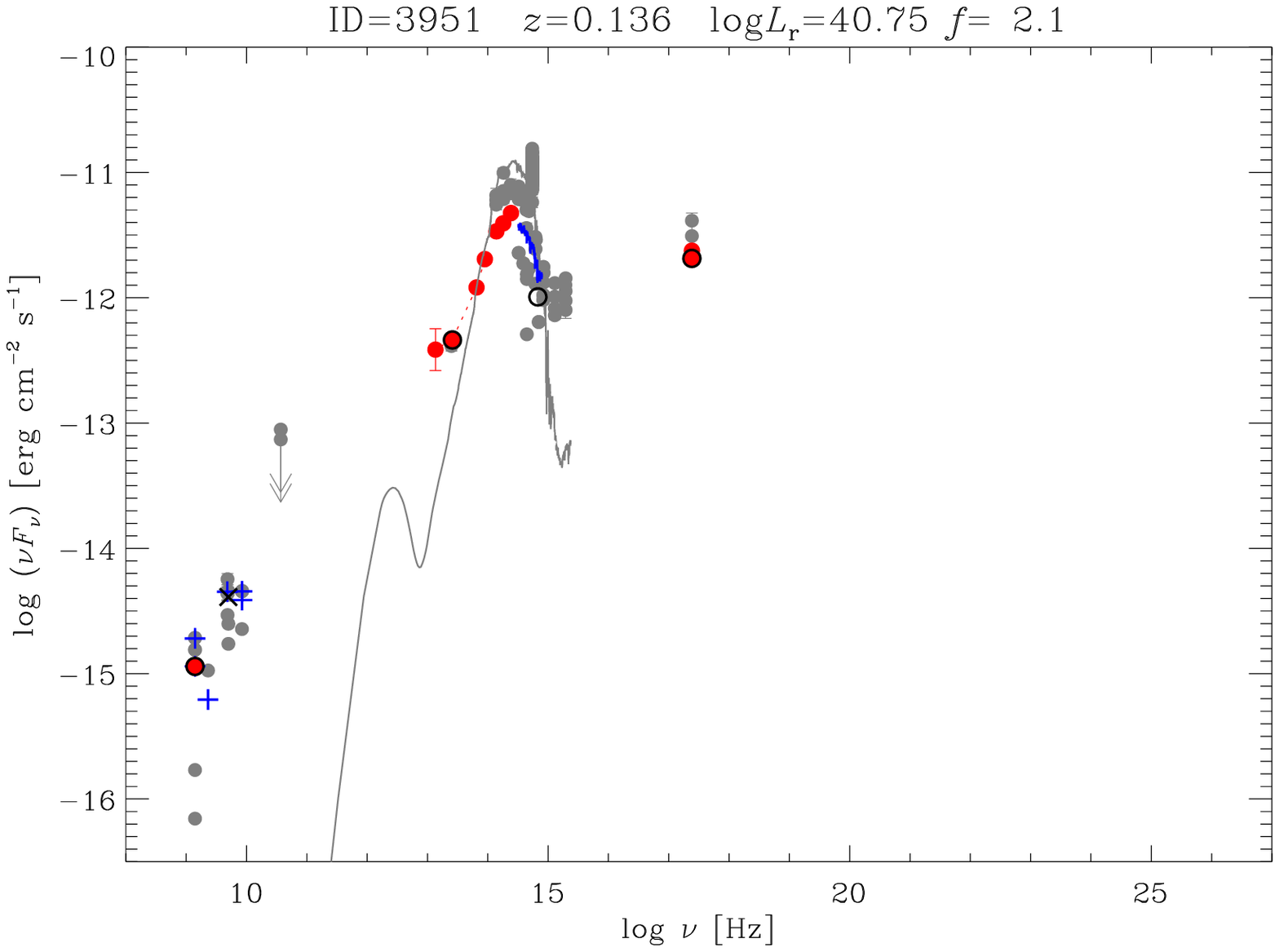,width=0.50\linewidth}
\psfig{figure=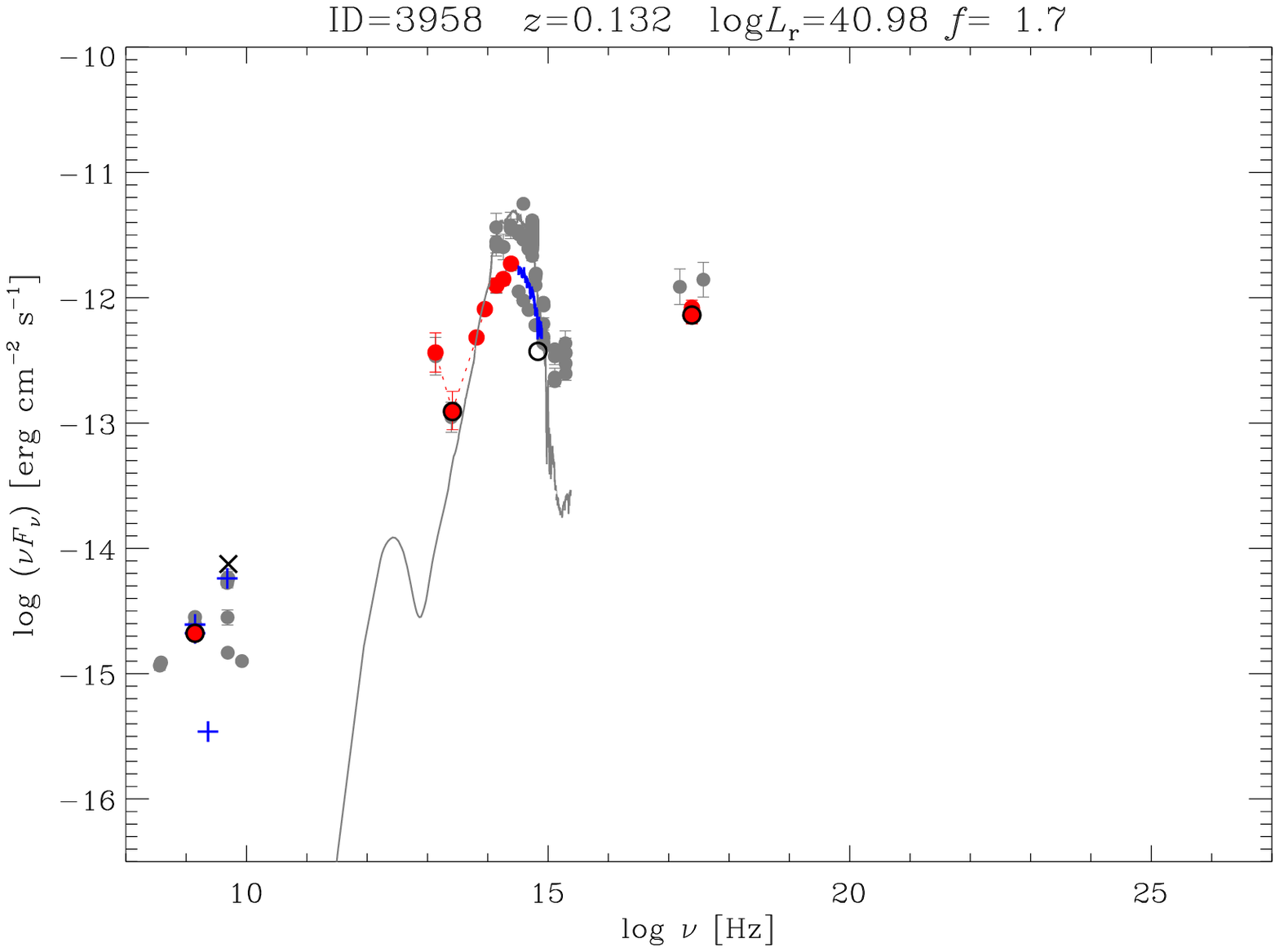,width=0.50\linewidth}}
\vspace{0.2cm}
\centerline{\psfig{figure=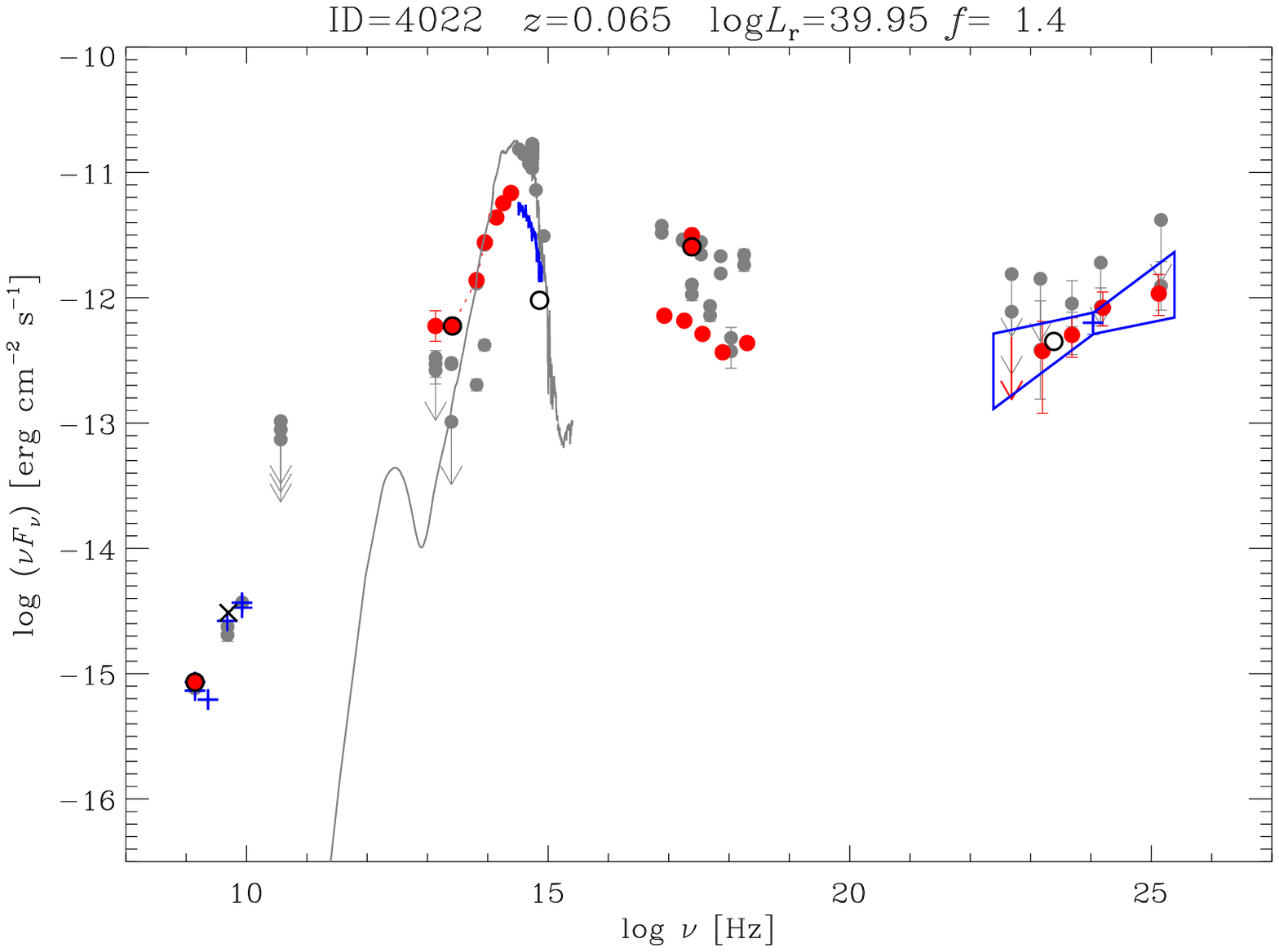,width=0.50\linewidth}
\psfig{figure=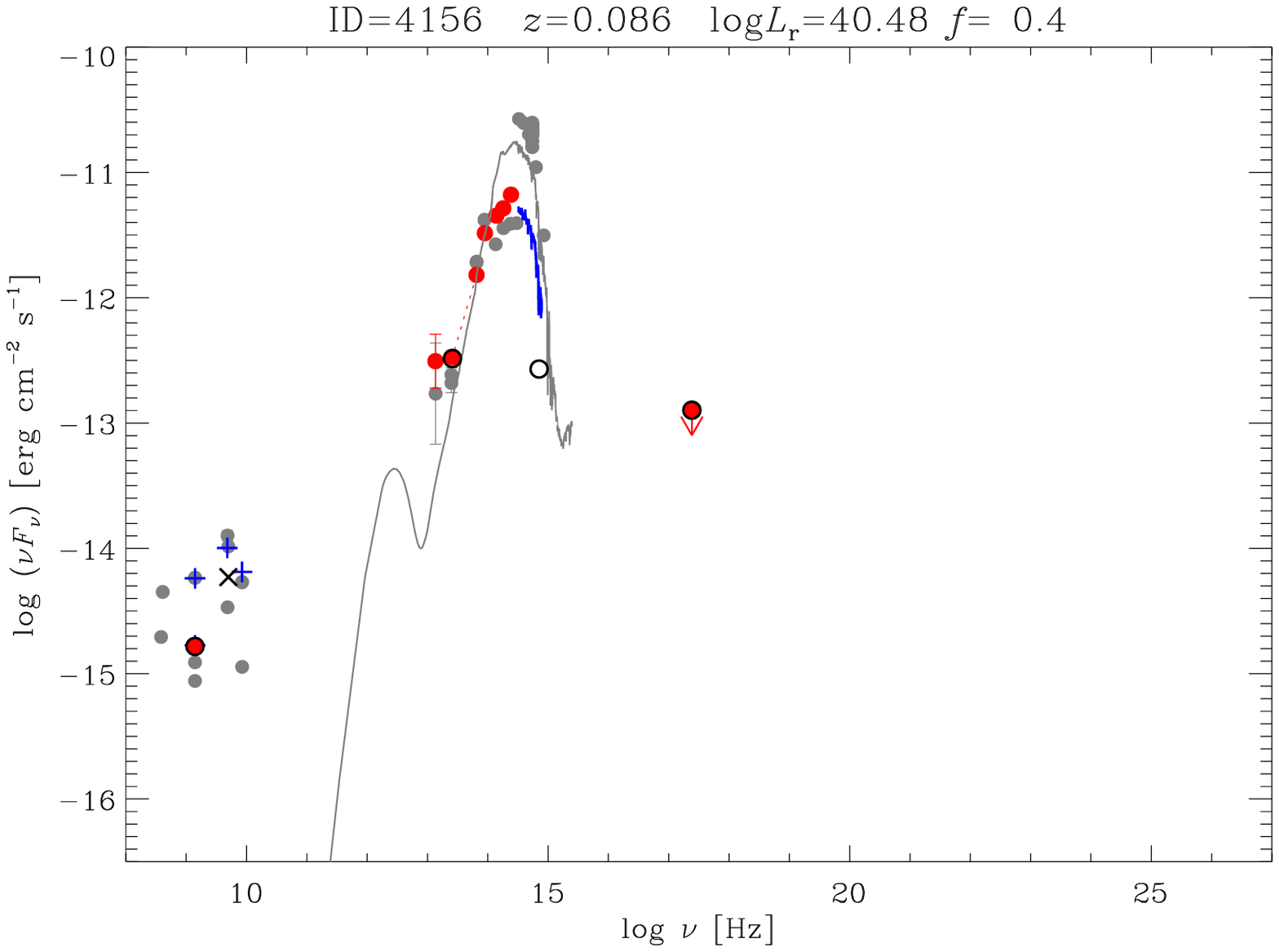,width=0.50\linewidth}}
    \caption{(continued)}
   \end{figure*}

    \addtocounter{figure}{-1}
\begin{figure*}
\centerline{\psfig{figure=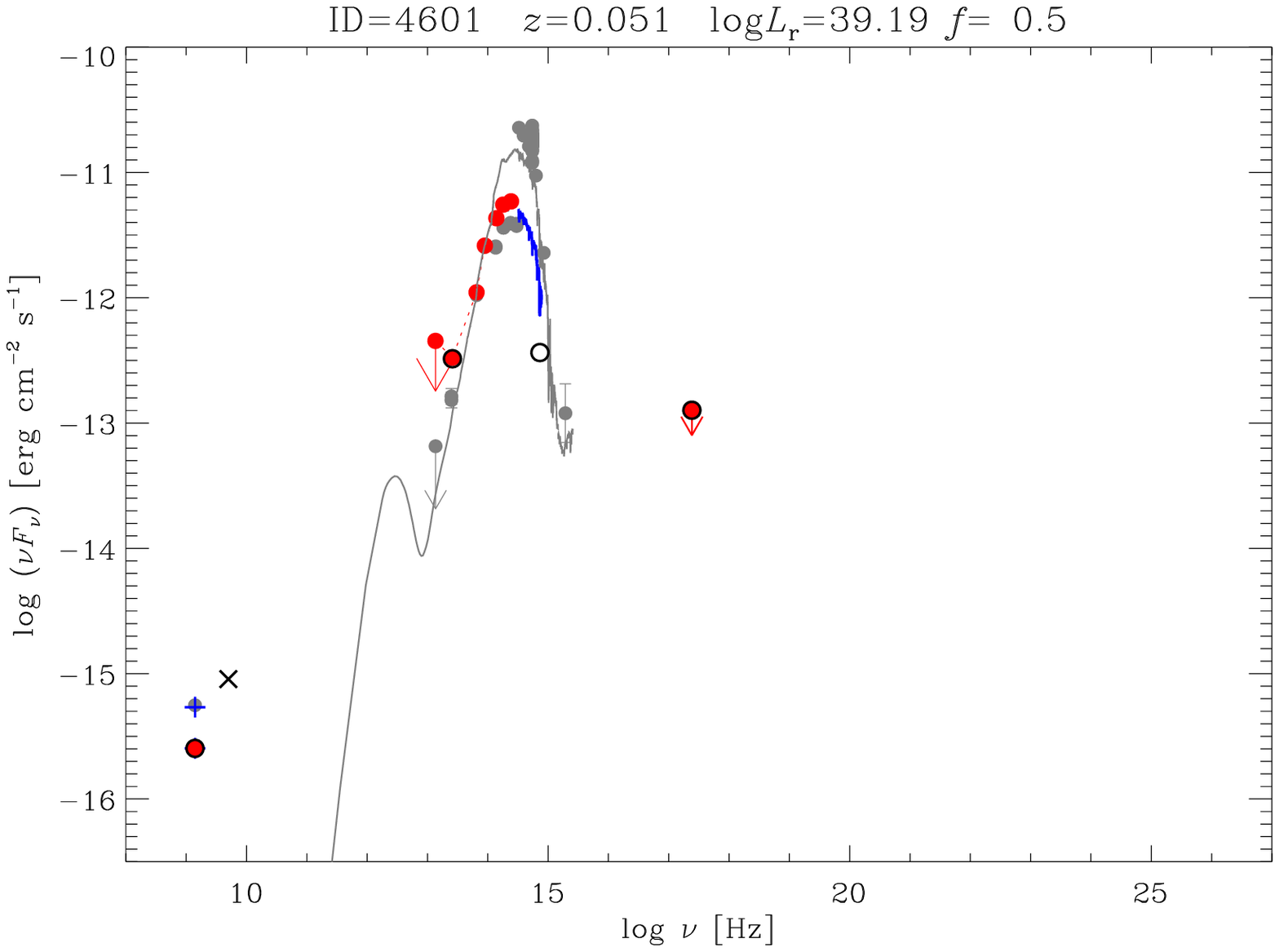,width=0.50\linewidth}
\psfig{figure=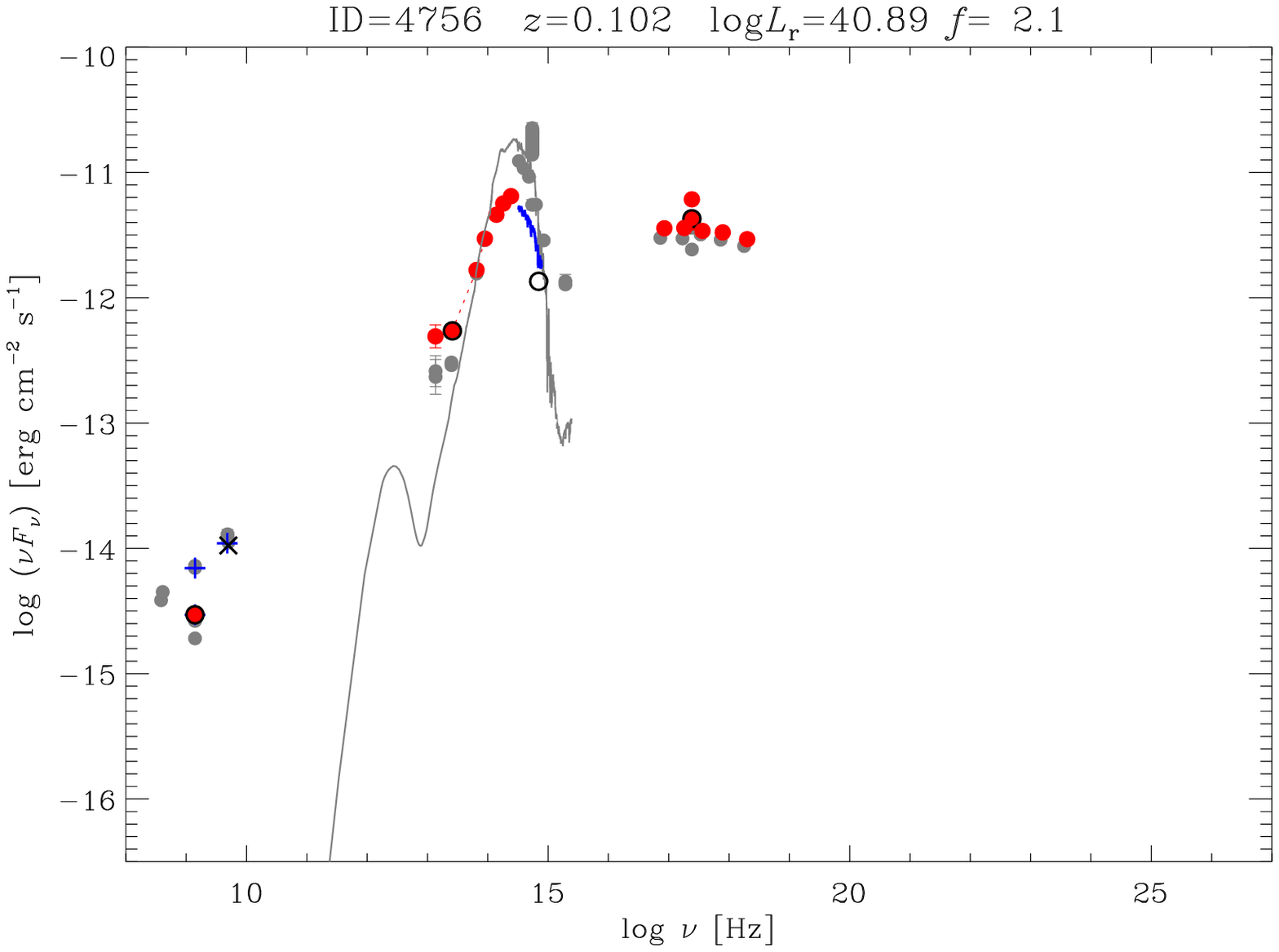,width=0.50\linewidth}}
\vspace{0.2cm}
\centerline{\psfig{figure=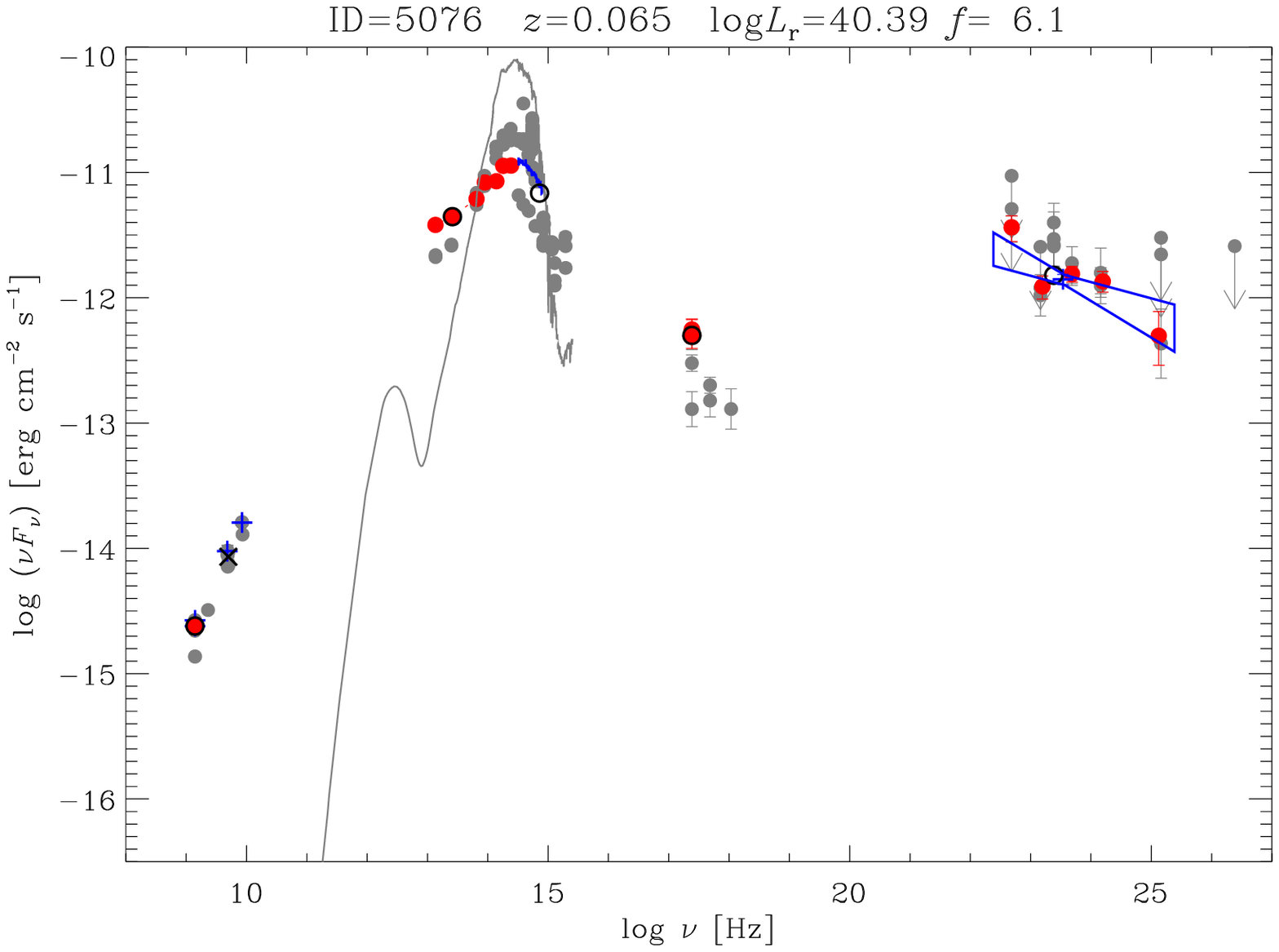,width=0.50\linewidth}
\psfig{figure=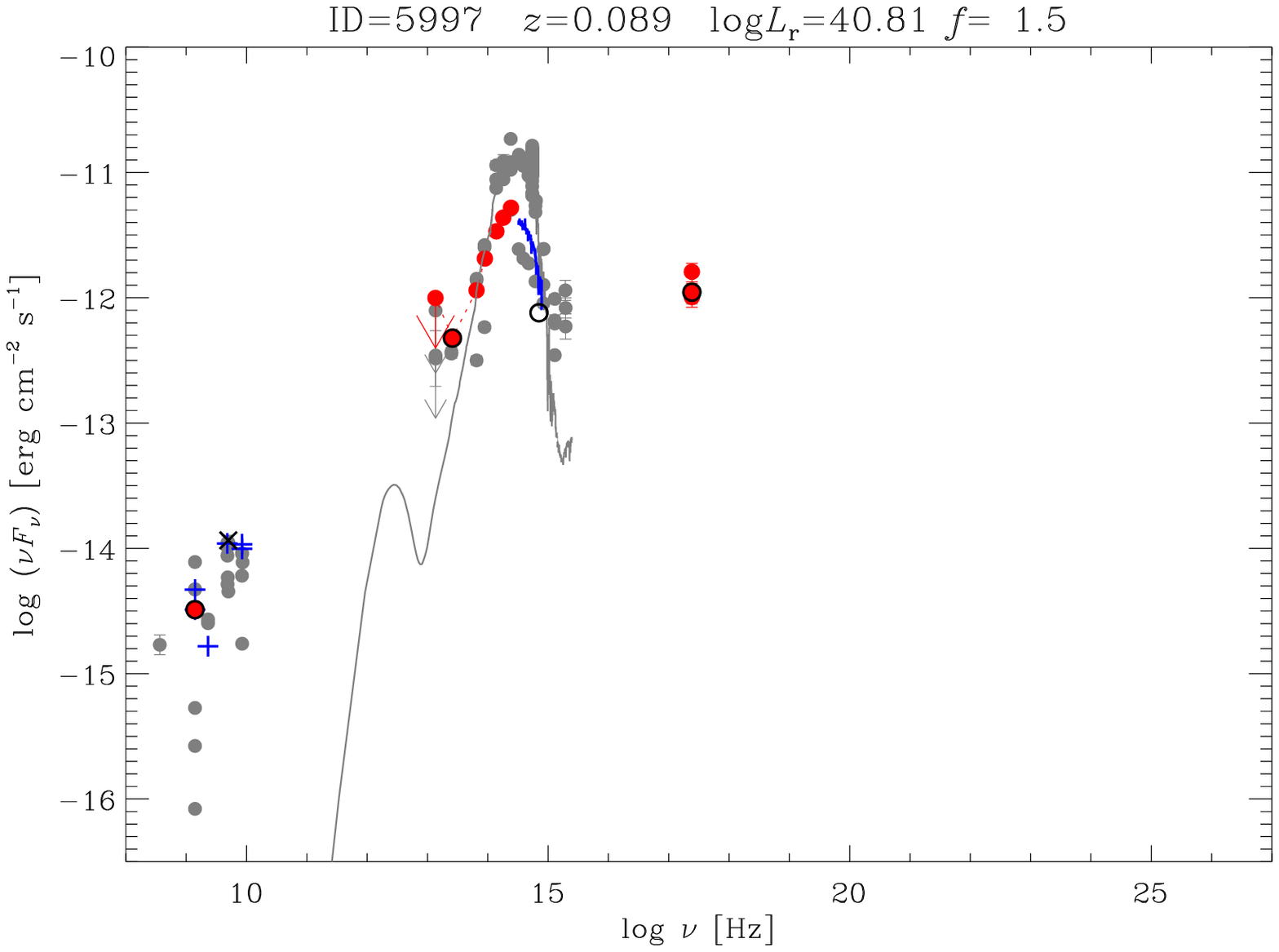,width=0.50\linewidth}}
\vspace{0.2cm}
\centerline{\psfig{figure=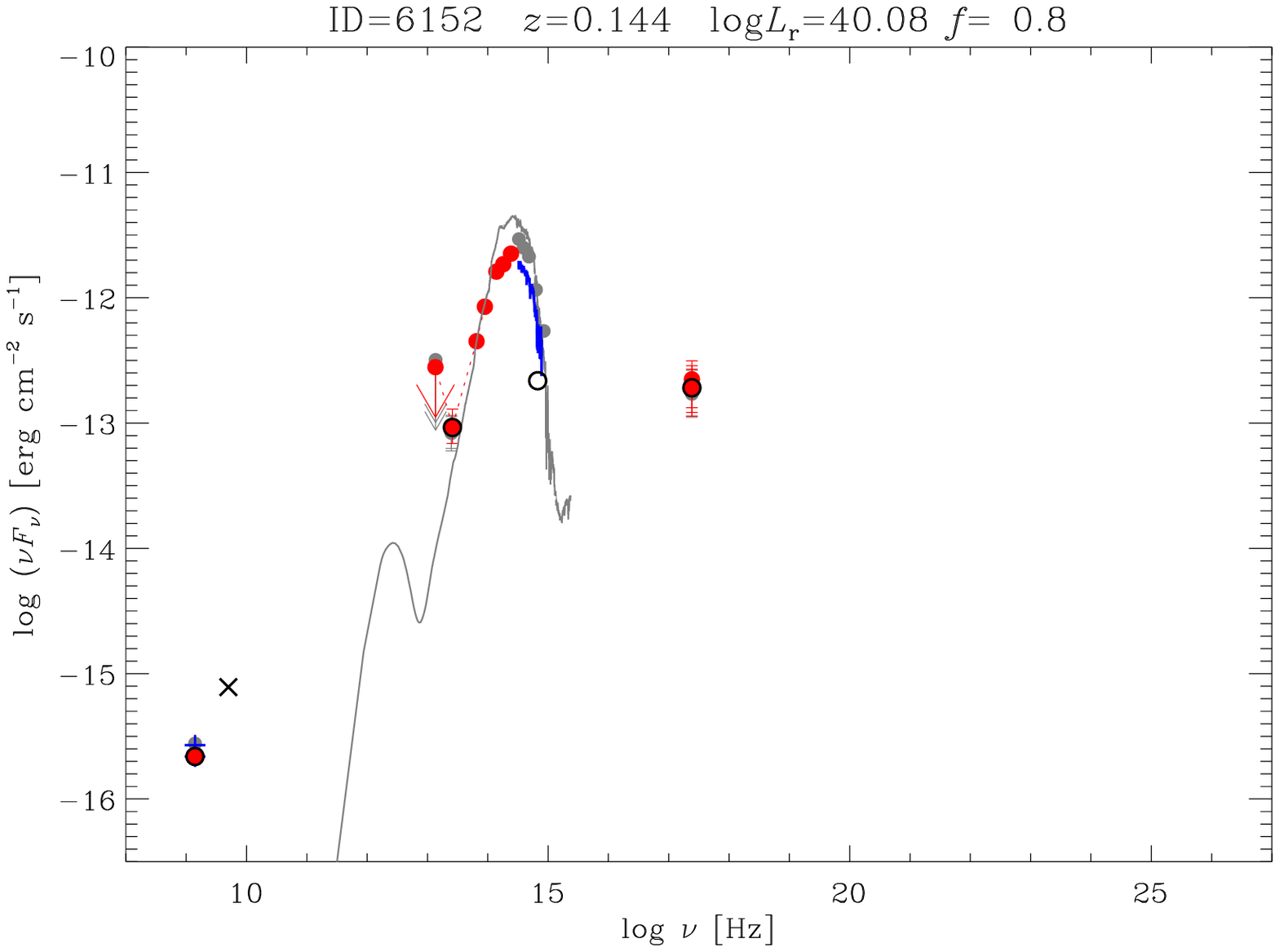,width=0.50\linewidth}
\psfig{figure=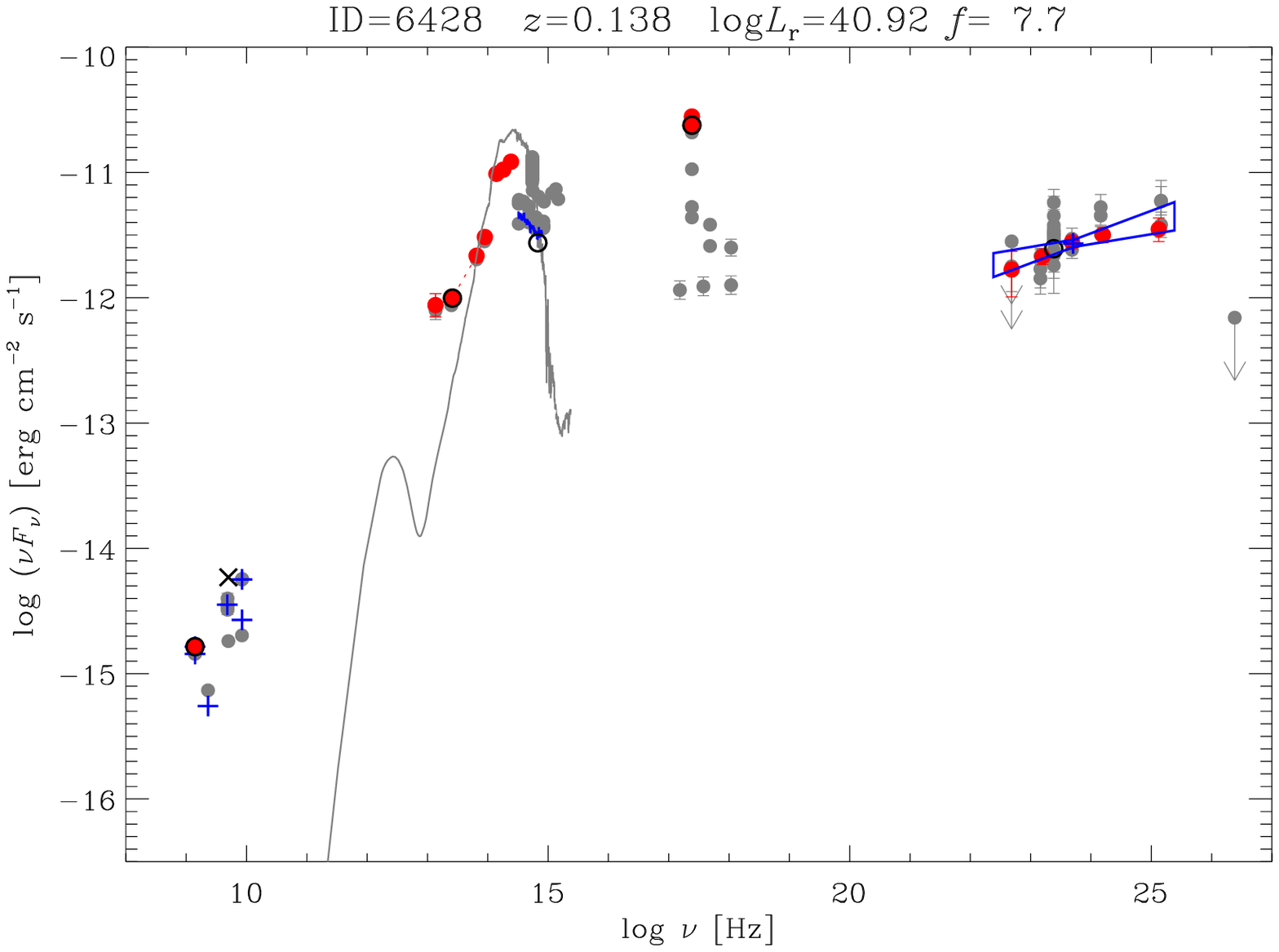,width=0.50\linewidth}}
\vspace{0.2cm}
\centerline{\psfig{figure=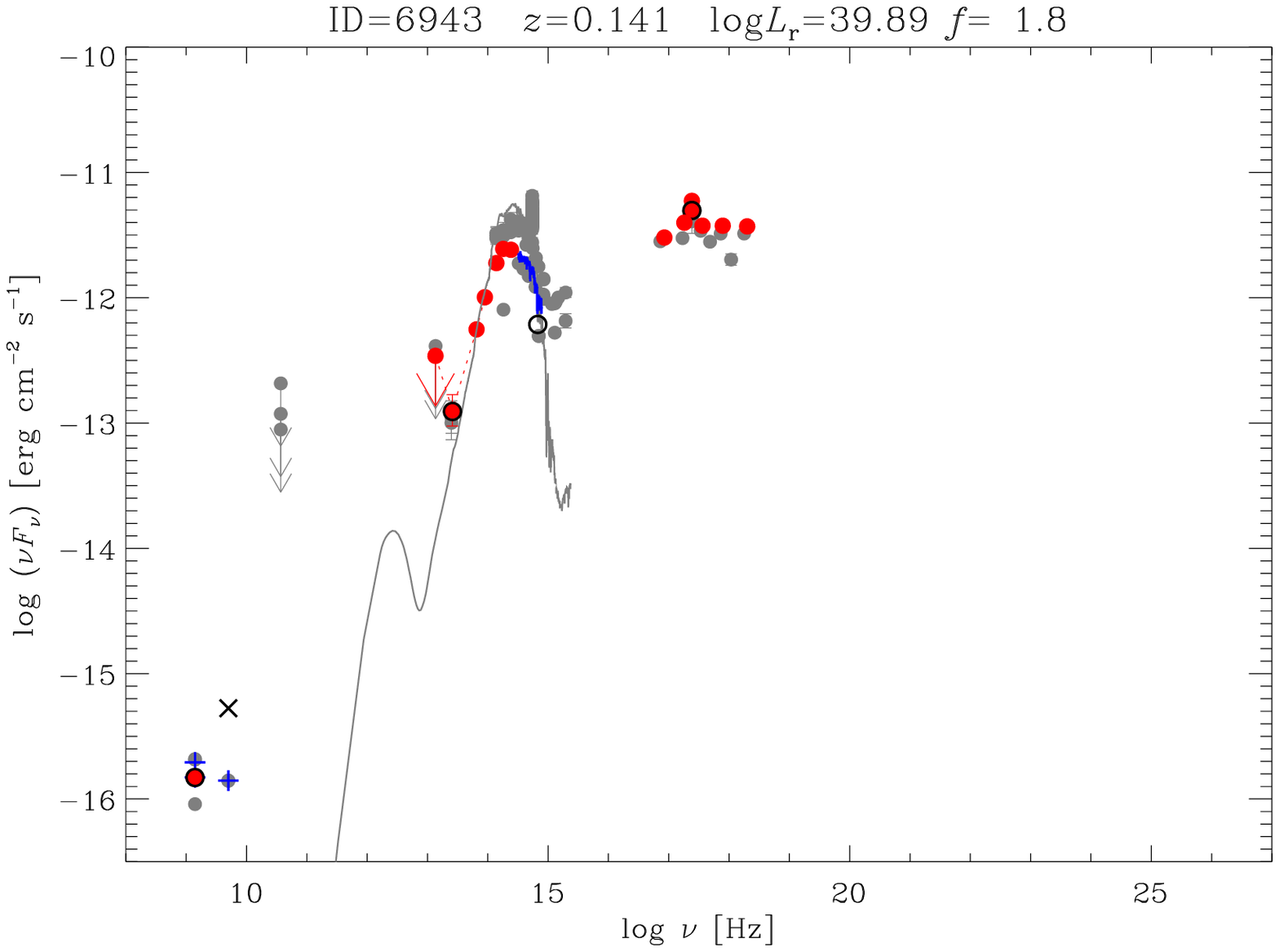,width=0.50\linewidth}
\psfig{figure=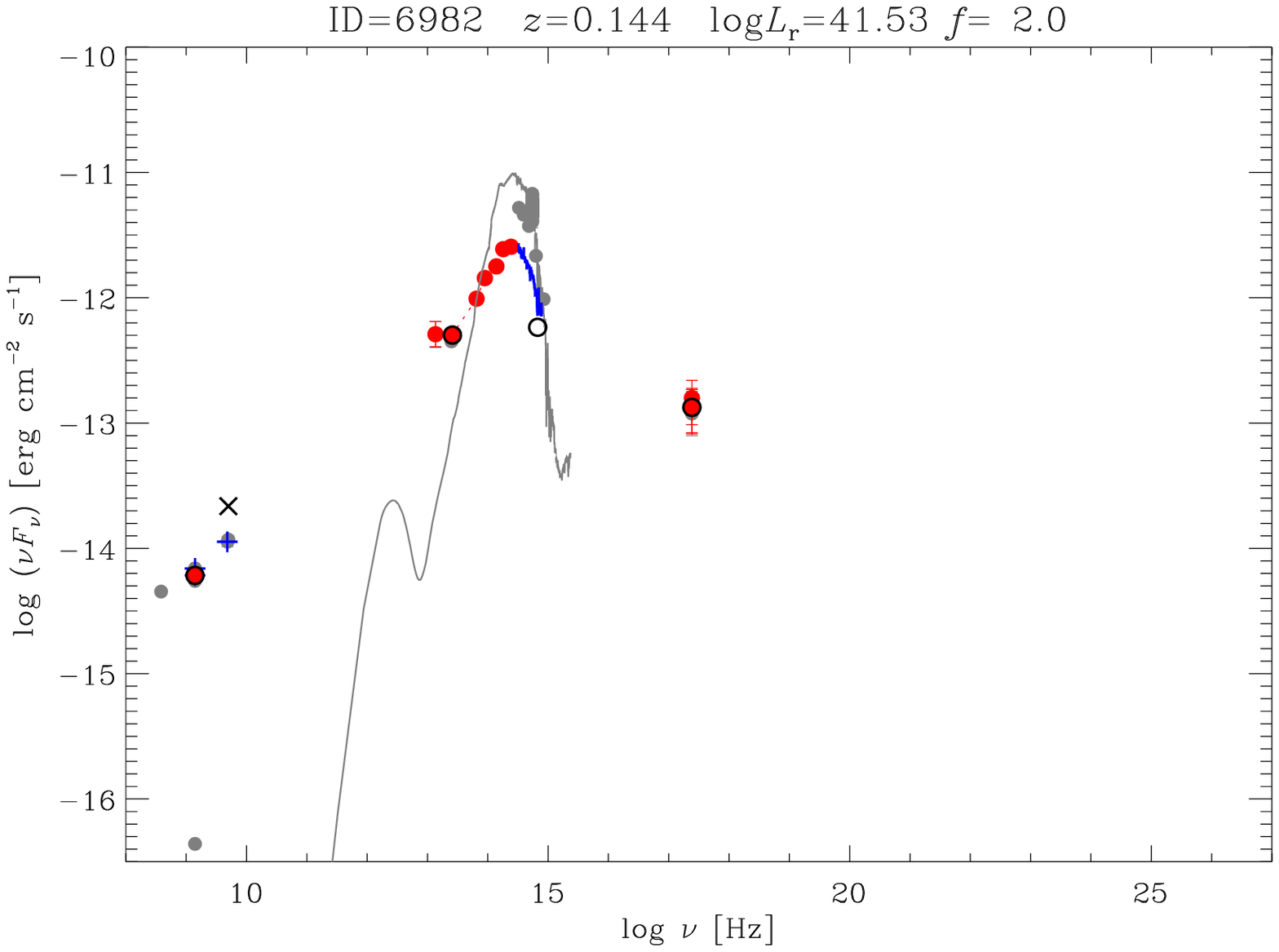,width=0.50\linewidth}}
    \caption{(continued)}
   \end{figure*}

    \addtocounter{figure}{-1}
\begin{figure*}
\centerline{\psfig{figure=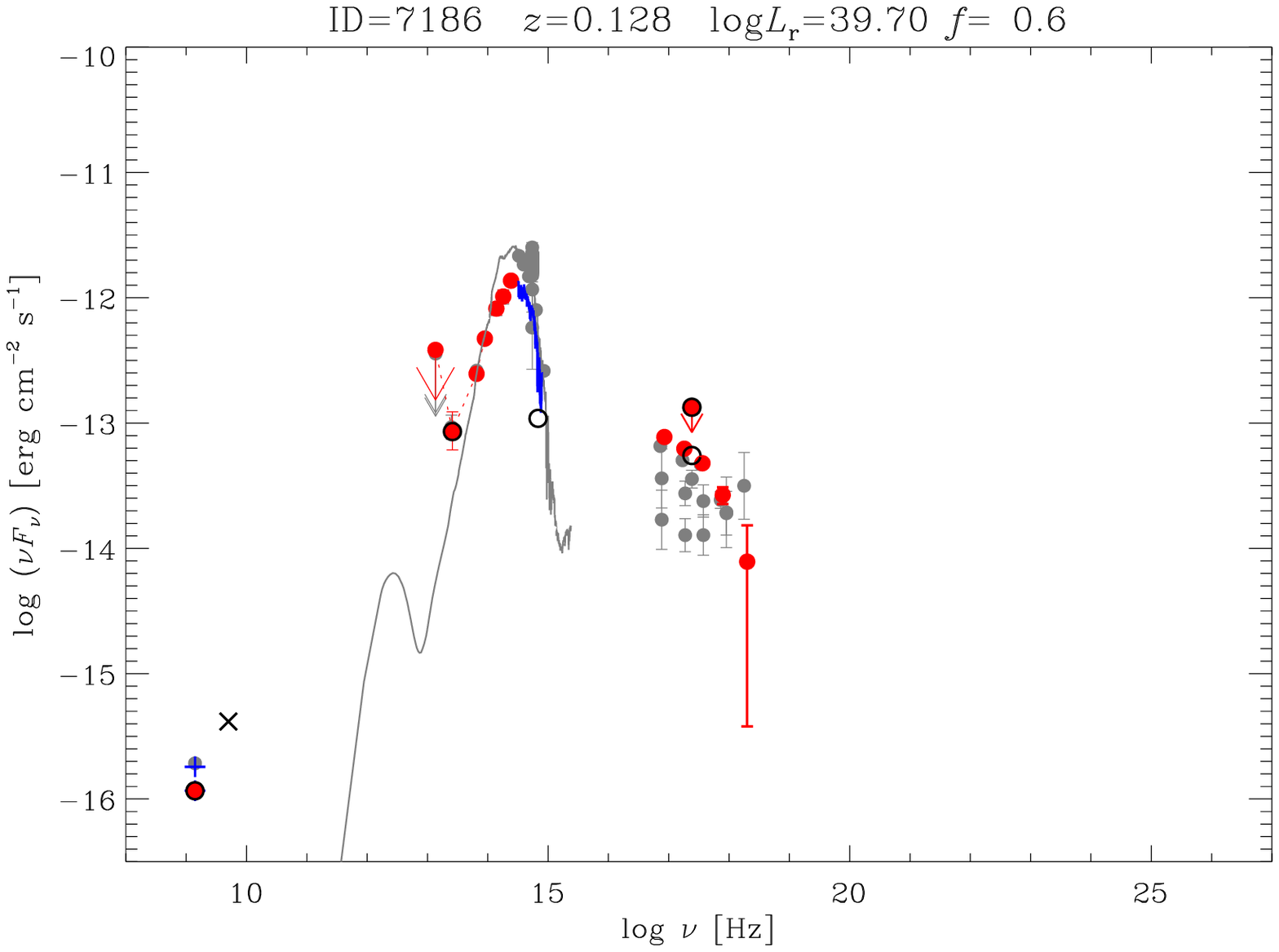,width=0.50\linewidth}
\psfig{figure=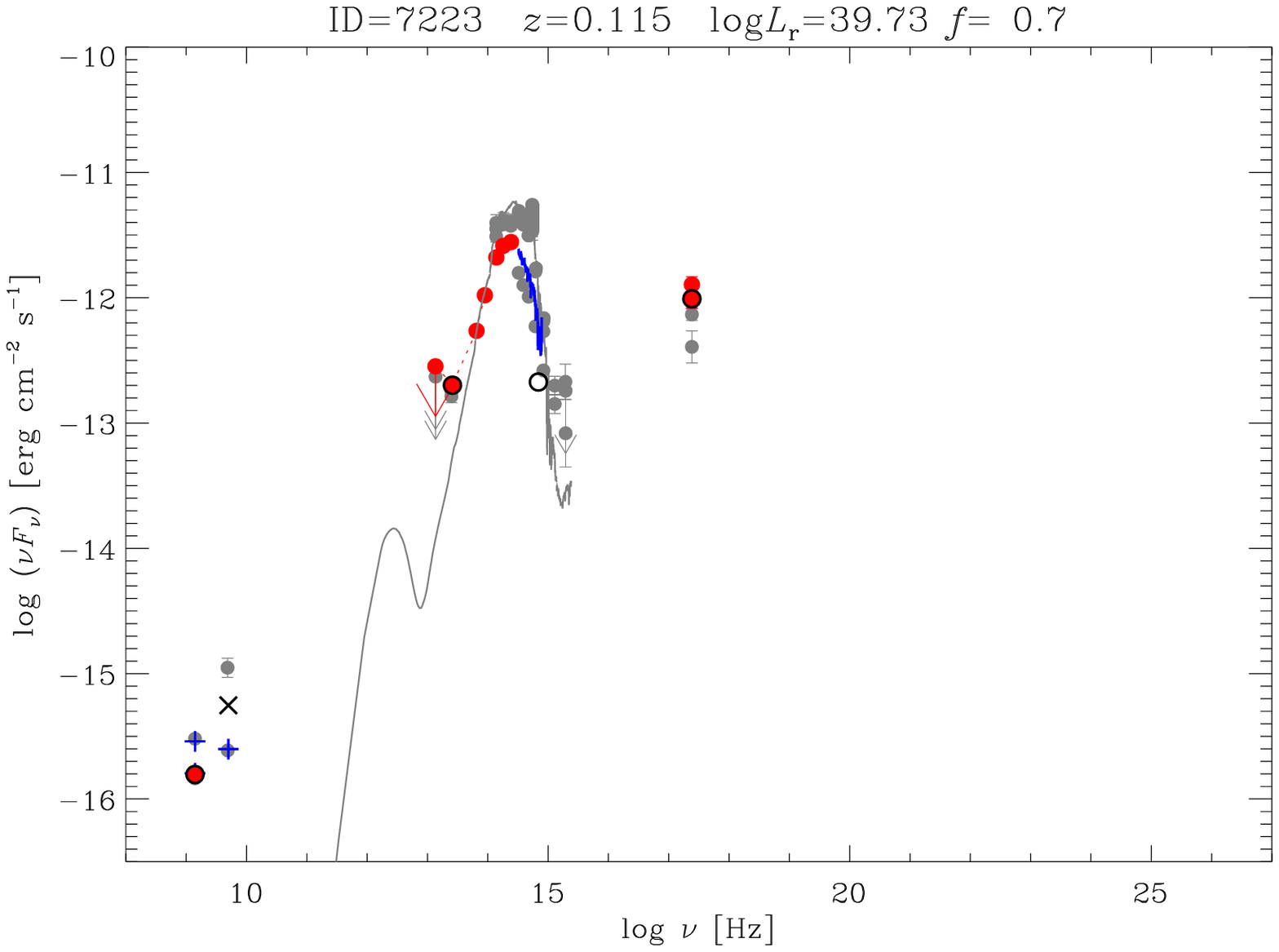,width=0.50\linewidth}}
\vspace{0.2cm}
\centerline{\psfig{figure=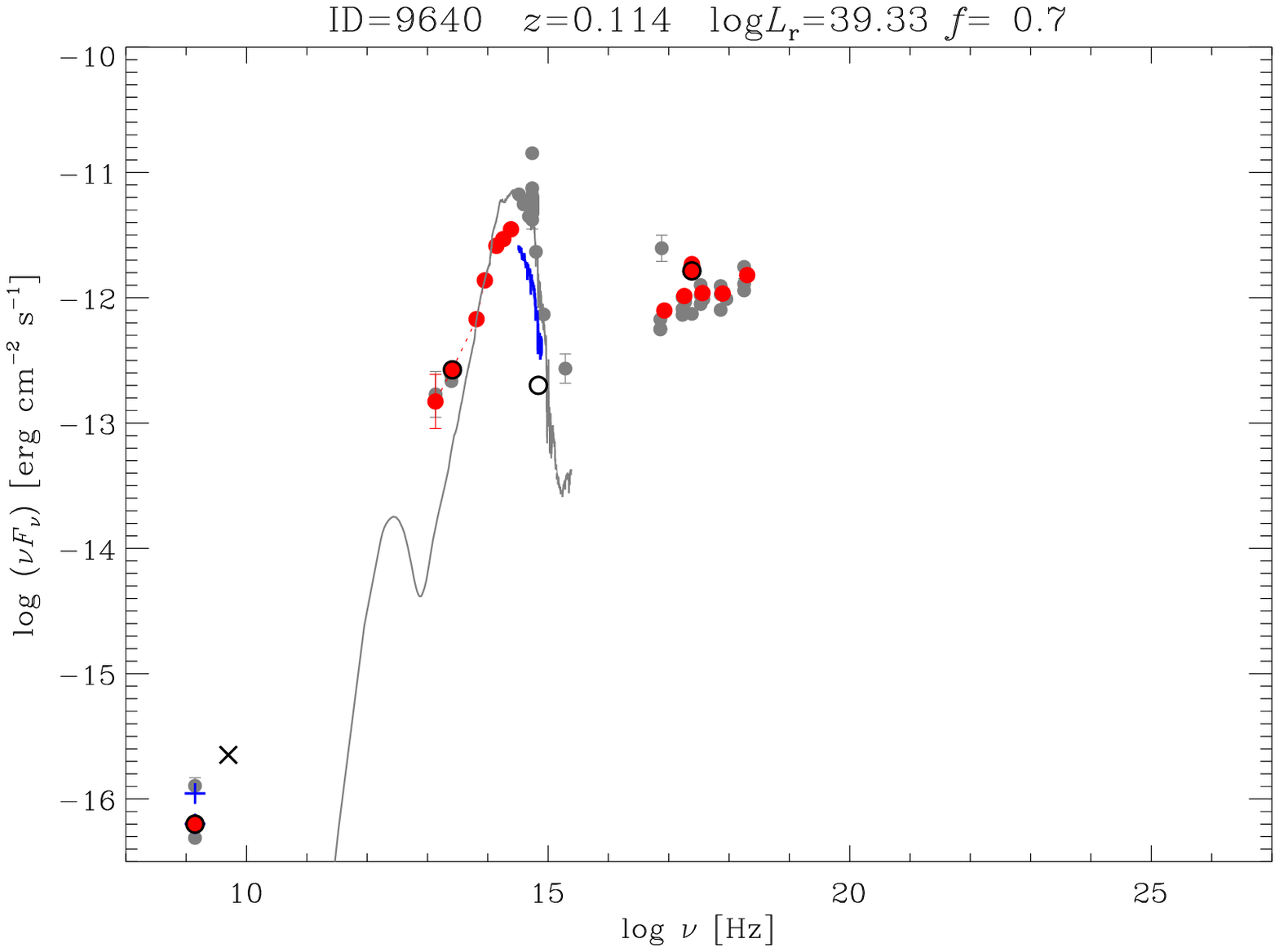,width=0.50\linewidth}
\psfig{figure=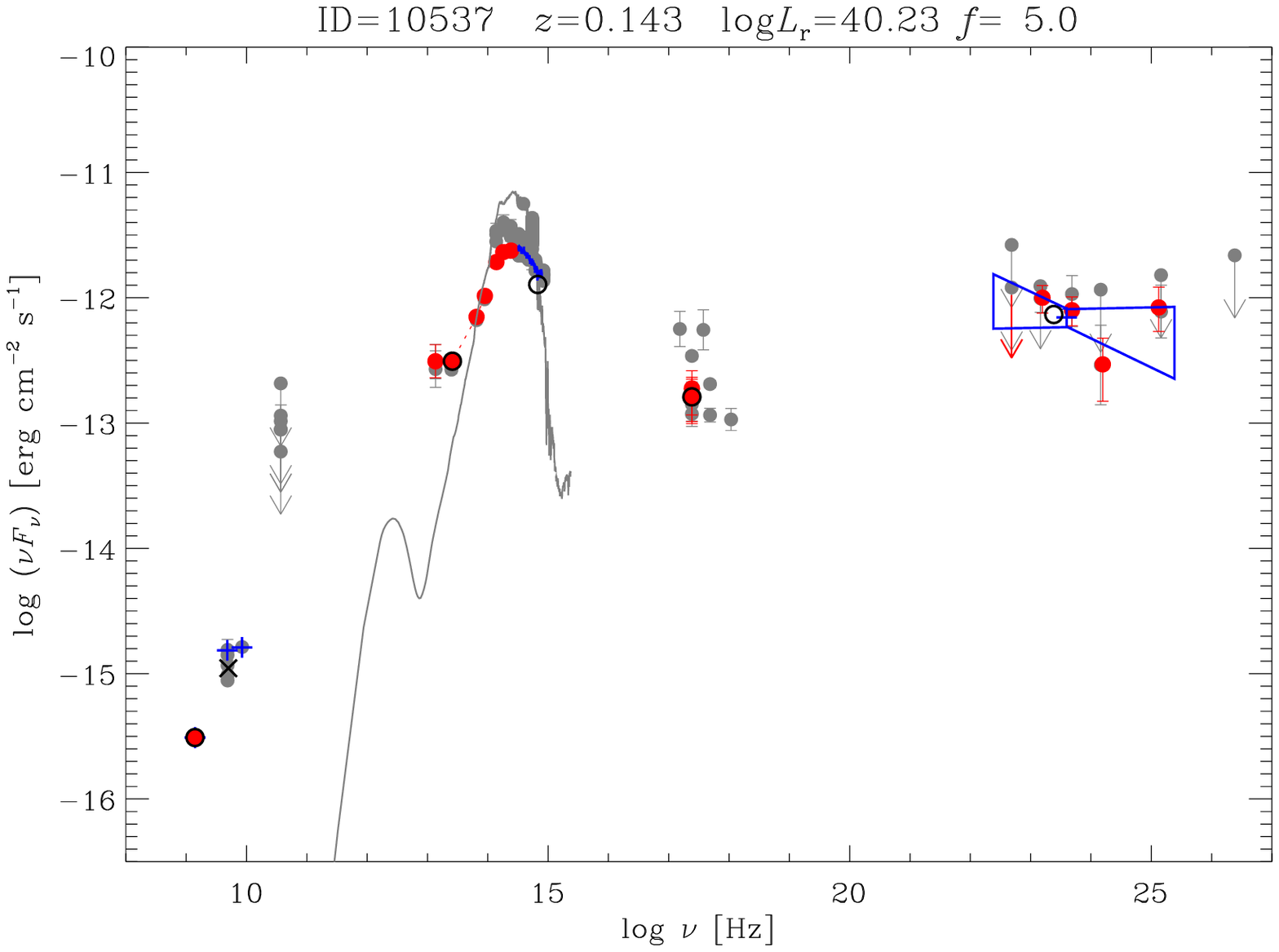,width=0.50\linewidth}}
    \caption{(continued)}
   \end{figure*}

\end{document}